\documentclass[letterpaper,12pt]{article}
\usepackage{newtxtext,newtxmath}
\usepackage[utf8]{inputenc}
\usepackage[T1]{fontenc}
\DeclareUnicodeCharacter{2500}{\textemdash{}}
\DeclareUnicodeCharacter{00A0}{~}
\usepackage{textcomp}
\usepackage[margin=1in]{geometry}
\usepackage{amsmath}
\usepackage{graphicx}
\usepackage{hyperref}
\usepackage{tabularx}
\usepackage{newunicodechar}
\newunicodechar{≈}{\approx}
\let\bibhang\relax
\usepackage{listings}
\usepackage{enumitem}
\usepackage{booktabs}
\usepackage{varwidth}
\usepackage{pdflscape}
\usepackage{tcolorbox}

\usepackage[justification=centering]{caption}

\usepackage{titlesec}
\titlespacing*{\section}
  {0pt}    
  {0.8ex}  
  {0.5ex}  
\titlespacing*{\subsection}
  {0pt}
  {0.6ex}
  {0.4ex}
\titlespacing*{\subsubsection}
  {0pt}
  {0.4ex}
  {0.3ex}

\usepackage{mdframed}

\newmdenv[
  linewidth=0.4pt,
  linecolor=black,
  topline=true,
  bottomline=true,
  leftline=false,
  rightline=false,
  innerleftmargin=0pt,
  innerrightmargin=0pt,
  innertopmargin=4pt,
  innerbottommargin=4pt,
  skipabove=\baselineskip,
  skipbelow=\baselineskip
]{checklistbox}

\newlist{todolist}{itemize}{1}
\setlist[todolist]{
  label=$\square$,
  leftmargin=*,
  topsep=2pt,
  itemsep=1pt
}

\usepackage[dvipsnames]{xcolor}

\usepackage{pgfplots}
\pgfplotsset{compat=1.18}
\usepgfplotslibrary{statistics,groupplots}

\definecolor{yesno}{HTML}{4E79A7}
\definecolor{likert}{HTML}{59A14F}
\definecolor{pct}{HTML}{F28E2B}

\definecolor{oiBlue}{RGB}{0,114,178}
\definecolor{oiOrange}{RGB}{230,159,0}
\definecolor{oiGreen}{RGB}{0,158,115}
\definecolor{oiSky}{RGB}{86,180,233}
\definecolor{oiGrey}{gray}{0.25}

\usepackage[style=apa, backend=biber, maxcitenames=2, mincitenames=1, uniquelist=false]{biblatex}
\DeclareFieldFormat{journaltitle}{#1}
\DeclareFieldFormat[article]{title}{\mkbibitalic{#1}}
\DeclareFieldFormat[book]{title}{\mkbibitalic{#1}}
\DeclareFieldFormat[article]{volume}{{#1}}

\renewcommand{\arraystretch}{1.05}
\usepackage{float}
\usepackage{setspace}
\usepackage{tikz}
\usetikzlibrary{arrows.meta,positioning,fit,backgrounds,shapes.multipart,decorations.pathreplacing}
\usepackage{adjustbox}
\usepackage{verbatim}
\usepackage{array}     
\usepackage{longtable} 

\usepackage{siunitx}
\hypersetup{hidelinks}
\setlist[itemize]{leftmargin=1.2em}
\setlist[enumerate]{leftmargin=1.4em}
\begin{document}

\title{\LARGE{Variance-Aware LLM Annotation for Strategy Research: Sources, Diagnostics, and a Protocol for Reliable Measurement}}

\author{%
  Arnaldo Camuffo\footnotemark[1] \and
  Alfonso Gambardella\footnotemark[1] \and
  Saeid Kazemi\footnotemark[1] \and
  Jakub Malachowski\footnotemark[1] \and
  Abhinav Pandey\footnotemark[1]
}

\footnotetext[1]{Bocconi University, ION Management Science Lab.}
\footnotetext[2]{Emails: \texttt{arnaldo.camuffo@unibocconi.it},
\texttt{alfonso.gambardella@unibocconi.it},
\texttt{saeid.kazemi@unibocconi.it},
\texttt{jakub.malachowski@studbocconi.it},
\texttt{abhinav.pandey@unibocconi.it}.}

\date{\today}

\maketitle

\vspace{5cm}

\begin{center}
PRELIMINARY DRAFT.\\
\end{center}
\newpage

\begin{abstract}
Large language models (LLMs) offer strategy researchers powerful tools for annotating text at scale, but treating LLM-generated labels as deterministic overlooks substantial instability. Grounded in content analysis and generalizability theory, we diagnose five variance sources: construct specification, interface effects, model preferences, output extraction, and system-level aggregation. Empirical demonstrations show that minor design choices—prompt phrasing, model selection—can shift outcomes by 12–85 percentage points. Such variance threatens not only reproducibility but econometric identification: annotation errors correlated with covariates bias parameter estimates regardless of average accuracy. We develop a variance-aware protocol specifying sampling budgets, aggregation rules, and reporting standards, and delineate scope conditions where LLM annotation should not be used. These contributions transform LLM-based annotation from ad hoc practice into auditable measurement infrastructure. \\

\textbf{Keywords:} large language models, text annotation, measurement, reliability, research methods
\end{abstract}

\begin{center}
\textbf{Managerial Summary}
\end{center}
\begin{center}
\begin{minipage}{0.87\textwidth}
\small

Strategy researchers increasingly use LLMs to annotate text corpora, such as classifying disclosures, scoring strategic initiatives, and extracting signals at scale. This study shows that such annotations can be highly unstable: small variations in prompts, model choice, or extraction procedures may shift results by 12–85 percentage points. We introduce a variance-aware annotation protocol that adapts established measurement principles to LLM-based analysis. The protocol specifies clear coding rubrics, systematic sampling across prompts and models, and transparent reporting standards. By following these practices, researchers can produce annotations that are auditable and replicable. For practitioners who rely on research-based evidence, this reduces the risk that strategic insights reflect measurement artifacts rather than underlying organizational phenomena. 

\end{minipage}
\end{center}
\newpage
\doublespacing
\section*{Introduction}
\label{sec:introduction}
\setcounter{page}{1}
Large language models (LLMs) now power research workflows in strategy that previously depended on teams of human coders. Researchers use LLMs to annotate, score, rank, and classify text corpora---from business plans and investor pitches to executive communications and corporate disclosures. These annotations become data in empirical studies. LLMs have been used not only to assess entrepreneurial business plans \parencite{Csaszar2024AIStrategicDecision, Lane2024NarrativeAIAdvantage, Doshi2025GenAIEvaluatingStrategicDecisions} and M\&As' announcements \parencite{Mirzayev2025ArtificialAgentsMAs}, but also to annotate entrepreneurial pitches along multiple criteria \parencite{Boussioux2024CrowdlessFuture,Lane2024NarrativeAIAdvantage}, code sustainability claims in crowdfunding campaigns \parencite{Carlson2025LLMsAnnotate}, and score the ``scientificness'' of founders' decision making in pitch blurbs and interviews \parencite{Camuffo2025DP20300}.

However, LLM annotations are not deterministic or unbiased. As in any field of application \parencite{Chang2024SurveyLLMEval}, also in strategy, LLM annotation output can be characterized by undesired variance that must be understood and mitigated if it is to be used as data in strategy studies. This study frames the use of LLMs in strategy research as a \emph{measurement task} and investigates the sources of unwanted variation in their output and what protocol strategy researchers should adopt to acknowledge, document and mitigate it. We extend the foundational framework of \textcite{Carlson2025LLMsAnnotate} by systematically diagnosing the sources of annotation variance and proposing an operationalized protocol. Where they establish principles, we provide an implementable workflow with specific diagnostics, sampling budgets, aggregation rules, and reporting standards.

Our research questions are: (i) How large is annotation output variance in strategy-typical tasks? (ii) What are its sources, and how do research design choices interact with LLM annotation? (iii) How can we design a usable, variance-aware protocol for strategy research?

Our contributions are fourfold. First, we provide a \textit{five-source variance framework} that maps failure modes to diagnostics and controls, grounded in content analysis and generalizability theory. Second, we provide \textit{empirical demonstrations} on strategy-typical annotation tasks showing that construct wording, context framing, prompt phrasing, and model choice can shift annotations by 12--85 percentage points. Third, we develop a \textit{normative, variance-aware protocol} with level-specific defaults, sampling budgets ($P$, $S$, $M$), aggregation rules, and a checklist for authors and reviewers. Fourth, we specify \textit{scope conditions}---when LLM annotation should not be used---that protect against misapplication.

Annotation variance matters not only for reliability but also for inference. \textcite{Ludwig2025LLMsEconometric} show that, without validation data, seemingly minor choices—such as model selection or prompt wording—can lead to substantially different empirical results, including changes in coefficient size, direction, and statistical significance. When errors in LLM-generated annotations are systematically related to key variables, estimates become biased even if average error is small. This means that LLM-based annotation raises an identification concern, not merely a reproducibility one: empirical conclusions may be systematically misleading. Our protocol addresses this problem by reducing annotation variance at the measurement stage, while their validation-based methods correct remaining bias. Together, these approaches support the credible use of LLM-generated data in strategy research.

After defining the measurement problem (Section 2), we develop the five-source variance framework and review the technical mechanisms documented in the AI literature (Section 3). Section 4 provides controlled demonstrations quantifying the effects of various sources of LLM annotation output variance in strategy research. We show that variance is material even in low-complexity tasks and argue that, if problems exist at this level, they compound in more complex settings. These findings inform a comprehensive normative protocol (Section 5). The final section discusses the findings and points to future research. Five appendices complement the paper. They provide:  (A) a full operational runbook for implementing variance-aware LLM annotation, (B) complete empirical materials and robustness results, (C) an expanded technical taxonomy of variance mechanisms grounded in the AI literature, (D) a formal mathematical treatment of the LLM annotation process and aggregation, and (E) applied case studies demonstrating how the protocol performs in realistic organizational and educational settings, illustrating boundary conditions and practical trade-offs.

\section*{LLM Annotation in Strategy Research}
\label{sec:llm-annotation}

\subsection*{LLM Annotation and the Text-as-Data Tradition}

Strategy researchers have long recognized the value of transforming unstructured text into structured variables for empirical analysis. The text-as-data approach treats documents---annual reports, press releases, earnings calls, business plans, patents---as repositories of strategic signals that can be systematically extracted and quantified \parencite{Short2010Research,Hannigan2019Topic}. This tradition encompasses multiple generations of methods, each with characteristic strengths and limitations. Dictionary and keyword methods count the occurrences of predefined word lists to measure constructs such as uncertainty, optimism, or innovation orientation \parencite{Tetlock2007GivingContent,Loughran2011LiabilityMisleading}. Supervised machine learning classifiers learn construct-relevant patterns from human-labeled training data \parencite{Choudhury2021ML}. Topic models and embeddings discover a latent semantic structure without predefined categories \parencite{Hannigan2019Topic}.

Large language models represent the latest evolution in this trajectory. LLMs can perform complex annotation tasks---classification, ranking, extraction, scoring---without custom training data, guided instead by natural language instructions. Early experiments demonstrate that LLMs can match or exceed the performance of both crowd workers and expert coders on many annotation tasks, with substantial cost and time savings \parencite{Gilardi2023GPT,Tornberg2023ChatGPT,Rathje2023GPT}. Recent methodological work confirms these capabilities while highlighting sensitivity to implementation choices.

\textcite{Carlson2025LLMsAnnotate} provide the foundational framework for LLM annotation in strategy research, organizing implementation decisions across five stages: method selection, model selection, prompt engineering, cost and scale considerations, and validation. Their framework emphasizes sensitivity analysis---systematically varying prompts and models to assess result robustness---rather than searching for a single ``best'' configuration. Using sustainability claims in crowdfunding campaigns as an empirical application, they demonstrate that LLMs can match traditional methods while also showing that prompt variations can significantly affect downstream analyses. 
\textcite{Doshi2025GenAIEvaluatingStrategicDecisions} examine LLMs annotating business model descriptions along evaluative dimensions such as ``likelihood of success.'' They find that individual LLM annotations are often inconsistent and biased, but that aggregating across multiple models, prompts, and assumed roles produces annotations that align more closely with human expert judgments. \textcite{Csaszar2024AIStrategicDecision} compare LLM-generated and LLM-annotated entrepreneurial strategies with those of human entrepreneurs and investors, highlighting the potential for LLMs to augment strategic decision-making while noting reliability concerns. These studies represent important initial steps in raising awareness about LLM annotation reliability. However, to transform annotation variance from a methodological weakness into robustness testing for strategy research conclusions, a more systematic approach is needed. First, it is necessary to ground the multiple sources of annotation variance on how LLM annotation works in producing output and how research design choices interact with LLM annotation use. Second, it is important to leverage recent advances from specialized AI literature to thoroughly understand what drives annotation variance and how the problem evolves as LLMs improve. Third, once the multiple drivers of variance are contextualized in strategy research, it is important to develop actionable best practices.

\subsection*{Annotation as Measurement}

LLM annotation should be subject to the same measurement discipline as human coding. The content analysis literature provides the conceptual foundation. \textcite{Krippendorff2018Content} defines content analysis as ``a research technique for making replicable and valid inferences from texts.'' Reliability---the extent to which a measuring procedure yields the same results on repeated trials---is a prerequisite for validity \parencite{Neuendorf2017Content}. When human coders annotate text, researchers assess interrater reliability using statistics such as Cohen's $\kappa$ \parencite{Cohen1960Kappa}, Fleiss's $\kappa$ \parencite{fleiss1971measuring}, or Krippendorff's $\alpha$ \parencite{Krippendorff2011ComputingAlpha}. LLM annotation should meet the same standards.

The consequences of annotation variance extend beyond reliability to parameter identification. \textcite{Ludwig2025LLMsEconometric} formalize the measurement error $\Delta_r = \hat{V}_r - V_r$ and demonstrate that if this error correlates with economic covariates $W_r$, plug-in regression using LLM annotations produces inconsistent parameter estimates---regardless of how small the average annotation error may be. Even an LLM that achieves 95\% accuracy can induce arbitrary bias in downstream estimation if its errors are systematically related to the variables under study. Their framework establishes that absent validation data, seemingly innocuous choices---which model, which prompt---can produce dramatically different coefficient estimates, varying in magnitude, sign, and statistical significance. Our framework (Section 3) decomposes this measurement error into tractable components, enabling researchers to diagnose and mitigate specific failure modes before annotations enter downstream analysis.

When an LLM labels a text, it functions as a synthetic coder whose output is probabilistic, context-dependent, and potentially biased. Treating these outputs as deterministic data---as if the LLM were an infallible oracle---ignores the stochastic nature of the annotation process and risks propagating measurement error into substantive conclusions. The interrater reliability tradition assumes multiple raters annotating the same items. With LLMs, we instantiate this design through multiple models that annotate the same texts (cross-model reliability), multiple prompts that elicit annotations from the same model (cross-prompt reliability), and multiple samples from the same model-prompt combination (within-configuration reliability). Our protocol operationalizes all three through sampling budgets for prompt ensembles ($P$), sampling replicates ($S$), and model panels ($M$).

While LLMs could also be used as "judges" \parencite{Zheng2023MTBench}, in this study we posit that normativity resides in the construct and rubric designed by the researcher, not in the model itself. The LLM is viewed as an instrument that applies researcher-specified criteria to text; its role is analogous to that of a trained coder following a codebook, not an autonomous judge exercising independent discretion \parencite{haaland2025understanding}.

LLM annotations produce \textit{authoritative output}---structured data usable in empirical analysis. We classify the annotation output into three levels that determine appropriate variance controls and aggregation procedures. Level~1 (single-label annotation) yields one categorical label from a predefined set, such as binary categories, yes/no responses, or Likert-scale points; only the label token is authoritative. Level~2 (multi-field annotation) yields a vector of labels across multiple schema-typed criteria, where criteria may be independent or dependent, as when percentage allocations must sum to 100. Level~3 (extractive annotation) treats open text---such as extracted spans, summaries, or verbatim quotes---as itself the measure. These levels are orthogonal to decision scope, which may be pointwise (annotating each text independently), pairwise (comparing two texts), listwise (ranking or scoring a set), or setwise (assigning texts to groups). More complex annotation tasks often combine levels in pipelines---for example, extracting evidence spans at Level~3, verifying relevance at Level~1, then scoring criteria at Level~2. We treat pipelines as a meta-design that inherits variance from each component level.

The protocol we propose provides the annotation ``hygiene'' needed by empirical studies that use LLM annotation output from textual corpora in strategy research. By documenting precisely how annotations are produced and how stable they are, the application of this protocol can also reduce the aversion of authors', reviewers' and editors' \parencite{Dietvorst2018AlgorithmAversion} -- or their complacency \parencite{Harbarth2025OvertrustingAI}-- towards studies that use data generated by LLM annotations. Furthermore, because the costs of annotating through LLMs have fallen dramatically, our protocol helps prevent that LLMs are not used to test numerous prompts until the desired research output is obtained, a practice similar to ``HARK-ing'' or ``p-hacking'' through prompt tweaking \parencite{Carlson2025LLMsAnnotate,Kosch2025PromptHacking}. Unreliable LLM-based annotations would lead---via unreliable data and analyses---to biased research output and potentially generate misleading advice for researchers and practitioners who use research-based insights \textcite{Ludwig2025LLMsEconometric}. Although we are unaware of retractions or major corrections in strategy research due to unreliable LLM-based annotations to date, our goal is to set standards \textit{before} problems scale.

\subsection*{When Not to Use LLM Annotation}

There are settings where LLM annotation should be avoided or used only as a non-decisional advisory signal with strict human oversight. LLM annotation is inappropriate for high-stakes individual assessments where decisions have legal, ethical, or material consequences that require human accountability, such as hiring, performance evaluation, or clinical diagnosis. It should also be avoided for culturally sensitive annotations where human judgment, contextual understanding, and cultural competence are irreplaceable. Tasks requiring domain expertise that exceeds model training data present another boundary condition---for instance, coding highly technical or proprietary content where the model cannot have learned relevant patterns. LLM annotation is also ill-suited for constructs with contested validity, where the ground truth is genuinely uncertain, and human expert disagreement is substantively meaningful rather than mere noise. Finally, researchers should avoid LLM annotation in situations where annotation error costs exceed efficiency gains.

\section*{Conceptual Framework: Sources of Annotation Variance}
\label{sec:variance-sources}

This section formalizes how an LLM maps a construct, rubric, prompt, context, and text item to an \emph{annotation} \parencite{haaland2025understanding}. Making this mechanism explicit surfaces potential sources of unreliability and clarifies why the variance-aware procedures we propose in Section 5 are necessary. We first describe how LLMs compute annotations at a high level. Then, building on measurement traditions and drawing on recent advances in the AI literature, we develop a five-source framework for understanding annotation variance. For each source, we specify signatures (what failure looks like), diagnostics (how to detect it), and controls (how to mitigate it). Section 3  develops this framework in detail; an expanded taxonomy with detailed subdimensions, diagnostics, and controls for each source is provided in Appendix C, which also provides a technical review of the AI literature on each mechanism. The complete mathematical formalization, including tokenization, embedding, transformer equations, and decoding details, is provided in Appendix D.

\subsection*{The Annotation Mechanism}

LLMs do not possess task-specific domain knowledge by default. They rely on patterns extracted from training data, unless additional information is provided at the prompt. Consequently, before querying an LLM, the researcher must clearly specify the research construct under investigation, have a valid measure/rubric of such construct, assess whether their own understanding of the construct is more precise than what the model could infer unaided, and provide sufficient context to ensure that the annotation is meaningful.

Formally, let $r$ denote the system or rubric instruction; let $k$ represent the broader context, including the role assigned to the LLM, details about the annotation task, or documents providing necessary information; let $p$ be the prompt specifying the construct of interest and the desired response format; and let $x$ denote the item to be coded, such as a pitch blurb or strategic document. Define $\mathcal{Y}$ as the set of admissible labels. The complete input to the LLM is:\[
s = \mathrm{concat}\big(r,k,p,x,\mathrm{options}(\mathcal{Y})\big).
\]
The LLM maps this sequence to a probability distribution over the vocabulary through a multi-stage process. Tokenization converts text to token IDs; embedding and positional encoding produce initial hidden states; transformer blocks refine representations through attention and feed-forward computations; a readout layer produces logits; and softmax normalization yields probabilities:
\[p_\theta(t \mid s) 
= \frac{\exp\!\big(z_t/T\big)}
       {\sum_{t' \in \mathcal V} \exp\!\big(z_{t'}/T\big)},\]
where $T>0$ is the temperature parameter that controls the sharpness of the distribution. Decoding strategies (greedy, top-$k$, top-$p$, or grammar-constrained) then sample the next token.

The key insight is that the LLM produces annotations as autoregressive next-token prediction. What appears as reasoning is the reuse of this prediction mechanism over structured prompts. Higher-level behaviors such as tool use, scratchpad reasoning, planning, and validation are protocol layers built on top of this basic mechanism. Small textual or layout changes can shift token patterns, alter probability distributions, and potentially flipping output near decision boundaries. This is why treating the annotator as a \emph{stochastic instrument} rather than a deterministic oracle is essential.

\subsection*{Five Sources of Variance: The Framework}

Building on the measurement foundations established in Section 2, we now detail how five sources of variance manifest in the annotation mechanism. This framework serves as the organizing structure for diagnostics and controls throughout the protocol. Figure~\ref{fig:variance-framework} provides a visual overview of the framework, showing how each source contributes to annotation output variance.

\begingroup
\setlength{\fboxsep}{0pt}
\setlength{\fboxrule}{0pt}

\definecolor{oiBlue}{RGB}{0,114,178}
\definecolor{oiOrange}{RGB}{230,159,0}
\definecolor{oiGreen}{RGB}{0,158,115}
\definecolor{oiSky}{RGB}{86,180,233}
\definecolor{oiGrey}{gray}{0.25}

\tikzset{
  box/.style={
    draw=oiBlue!60!black, fill=oiBlue!6,
    rounded corners=2pt, align=left, inner sep=5pt,
    text width=\BoxW, line width=0.5pt
  },
  core/.style={
    draw=oiOrange!70!black, fill=oiOrange!12,
    rounded corners=2pt, align=center, inner sep=7pt,
    text width=0.32\textwidth, line width=0.6pt
  },
  arrow/.style={-Stealth, line width=0.6pt, draw=oiGrey},
}
\begin{figure}[htbp]
\centering
\begin{adjustbox}{max width=\linewidth,center,scale=0.76, max height=0.8\textheight, keepaspectratio}
\begin{tikzpicture}[
  font=\footnotesize,
  >=Stealth,
  every node/.style={inner sep=5pt}
]
\def\BoxW{0.36\textwidth}

\node[core] (core) at (0,0)
{\textbf{LLM Annotation: Output Variance}
\\[2pt]\emph{Targets: reliability, validity, and transparency}
\\[4pt]\footnotesize For each variance source: \emph{signatures} $\rightarrow$ \emph{diagnostics} $\rightarrow$ \emph{controls}};

\node[box] (d1) at (0,6.5)
{\textbf{1.\ Construct \& Rubric Specification}
\begin{itemize}\itemsep2pt
\item Construct map \& BARS anchors
\item Select L1/L2/L3; declare L2 independence/dependence
\item Decision scope (point/pair/list/set)
\item Authoritative vs.\ non-authoritative fields; evidence binding
\item Item integrity (ASR/OCR fidelity; anonymize/redact)
\end{itemize}
\textit{Mechanisms:} construct ambiguity; level mis-specification; authority confusion; schema dependence errors};

\node[box] (d2) at (-6.8,3.6)
{\textbf{2.\ Interface \& Context Effects}
\begin{itemize}\itemsep2pt
\item Fixed layout; one-question prompts
\item Order/position \& length/verbosity effects
\item Paraphrase brittleness; template offsets
\item Long-context placement; ``lost in the middle''
\item Identity/prestige/persona cues
\end{itemize}
\textit{Mechanisms:} salience--order misattention (SOM); prompt brittleness};

\node[box] (d3) at (6.8,3.6)
{\textbf{3.\ Model Preferences \& Composition}
\begin{itemize}\itemsep2pt
\item Alignment-induced deference/sycophancy
\item Persona \& framing dependence
\item Family/self-preference (generator--coder coupling)
\item CoT unfaithfulness; rationale--label mismatch
\item Tool-gating/validator misses
\end{itemize}
\textit{Mechanisms:} deferential confirmation (DC); unfaithful composition (UC)};

\node[box] (d4) at (-6.8,-3.6)
{\textbf{4.\ Output Constraining \& Extraction}
\begin{itemize}\itemsep2pt
\item Grammar-guided decoding; typed JSON schemas
\item Deterministic token$\rightarrow$label maps; \texttt{max\_tokens=1}
\item Schema validation; reject-on-fail \& bounded retries
\item Probability/logit logging; calibration (temperature/isotonic)
\item Thresholding \& tie rules
\end{itemize}
\textit{Mechanisms:} extractor fragility; calibration gaps};

\node[box] (d5) at (6.8,-3.6)
{\textbf{5.\ System \& Aggregation}
\begin{itemize}\itemsep2pt
\item Cross-model disagreement; ensemble diversity
\item Provider/version drift; pinned audit sets
\item Precision/hardware/batch/seed effects
\item Aggregation (majority vs.\ Dawid--Skene/GLAD)
\item CIs, preregistration, and drift checks
\end{itemize}
\textit{Mechanisms:} drift; precision sensitivity; aggregation error};

\draw[arrow] (d2.south east) .. controls +(+1.5,-0.4) and +(-1.5,0.4) .. (core.west);
\draw[arrow] (d4.north east) .. controls +(+1.5,0.4)  and +(-1.5,-0.4) .. (core.west);
\draw[arrow] (d3.south west) .. controls +(-1.5,-0.4) and +(+1.5,0.4) .. (core.east);
\draw[arrow] (d5.north west) .. controls +(-1.5,0.4)  and +(+1.5,-0.4) .. (core.east);
\draw[arrow] (d1.south) -- ([yshift=2pt]core.north);

\end{tikzpicture}
\end{adjustbox}\
\caption{The five sources of variance generation for LLM annotation output, with associated mechanisms. Arrows indicate how each source contributes variance to annotation output; diagnostics and controls operate within each box.}
\label{fig:variance-framework}
\end{figure}
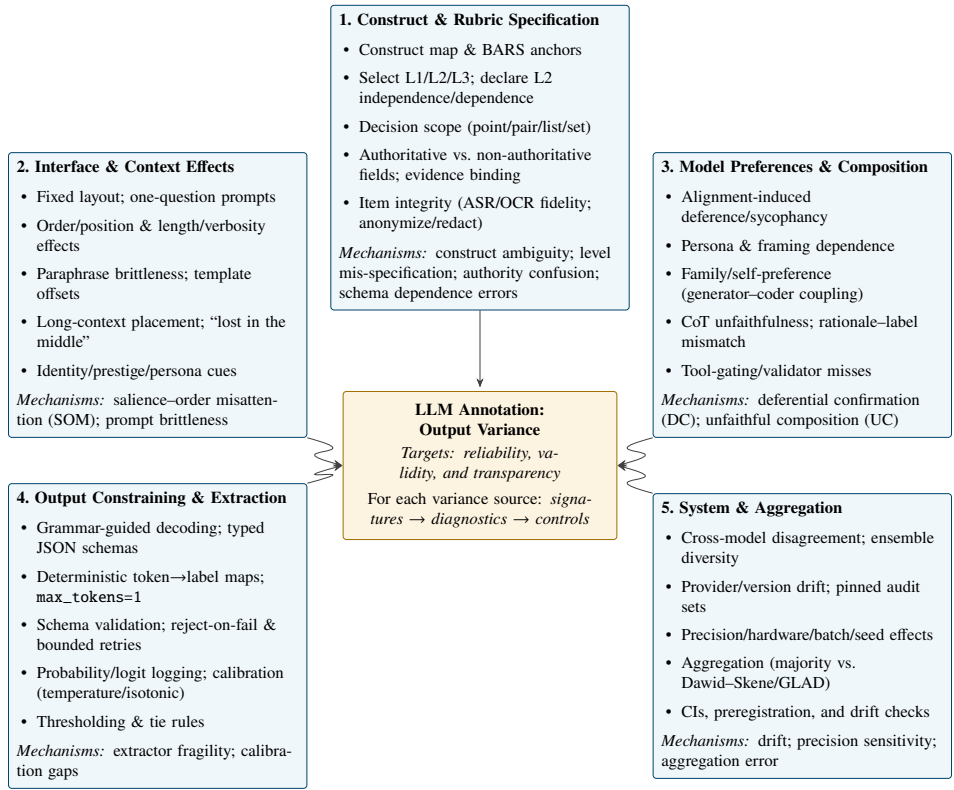
\endgroup

The first source of instability is construct and rubric specification\parencite{boyd2005construct}. When the construct is underspecified or the rubric lacks behavioral anchors, the model substitutes surface proxies---such as fluency or promotional language---for the target criterion. This source of variance connects directly to foundational work on construct validity \textcite{Cronbach1955ConstructValidity} according to which a measure must capture the theoretical construct it purports to assess, not merely surface features correlated with that construct. In human coding, behaviorally anchored rating scales (BARS) address this challenge by providing concrete examples that define each scale point \parencite{Smith1963RetranslationAnchors}. LLM annotation faces an analogous challenge: without explicit behavioral anchors, the model may substitute statistical regularities from training data for the researcher's intended construct. Ambiguous constructs push attention toward salient but irrelevant cues. Level mis-specification---treating multi-field annotation (L2) as single-label (L1), or parsing scores from extractive text (L3)---introduces additional fragility. The mechanisms through which this source operates include construct ambiguity, level mis-specification, authority confusion when the model cannot distinguish authoritative from explanatory output, and schema dependence errors when annotation formats interact with content.

The second source of instability is interface and context effects. Prompt and layout choices directly affect the token sequence $s$ that the model processes. This source of variance parallels established findings on questionnaire design and order effects in survey methodology \parencite{Schuman1981QuestionOrder,Krosnick1987ResponseOrder}. Just as human respondents are influenced by question wording, response option ordering, and context effects, LLM annotators exhibit systematic sensitivity to interface characteristics that should be irrelevant to the construct being measured. 
Order and position effects (salience-order misattention) cause the model to favor items in certain slots. Brittleness means that semantically equivalent prompts produce different token patterns and thus different probability distributions. Long contexts suffer from ``lost-in-the-middle'' effects, where evidence distant from the query receives less attention \parencite{liu2024lost}. Identity or prestige cues in the context or item leak world knowledge into coding decisions. The mechanisms here include salience-order misattention and prompt brittleness.

The third source of instability is model preferences and composition. Different models encode different annotation biases, analogous to rater bias in human coding. The content analysis literature has long recognized that human coders exhibit systematic biases, including coder drift (changing standards over time), halo effects (allowing impressions on one dimension to influence ratings on others), and central tendency bias \parencite{Krippendorff2018Content,Neuendorf2017Content}. LLM annotators exhibit their own characteristic biases that vary across model families and versions. Alignment-induced deference (sycophancy) causes models to favor responses matching stated beliefs or personas. Family or self-preference means models reward outputs stylistically similar to their own generations, creating systematic bias when the same model family both generates and codes content. Chain-of-thought unfaithfulness occurs when rationales contradict final labels---the model generates plausible-sounding reasoning that does not faithfully reflect its decision process. The key mechanisms are deferential confirmation and unfaithful composition.

The fourth source of instability is output constraining and extraction. How reliably annotations can be extracted from model output determines whether the measured variable corresponds to the model's actual classification. This source parallels research on response format effects in psychometrics, where the choice between forced-choice and Likert scales, or between open-ended and closed-ended questions, systematically affects measured values \parencite{Krosnick1999SurveyResearch}. Without constraints, extracting labels from free text is fragile---a phenomenon described as ``right answer, wrong score'' \parencite{Molfese2025RightAnswerWrongScore}. Decoding without grammar constraints or schema validation allows for invalid tokens. Probability distributions may be poorly calibrated: high softmax probability does not guarantee correctness. Near decision boundaries, small probability shifts flip outcomes. Mechanisms include extractor fragility and calibration gaps between stated confidence and actual accuracy.

The fifth source of instability is system and aggregation. How annotations are combined across replicates, prompts, and models determines the final measurement properties. This source connects to the multi-rater aggregation literature in psychometrics, where combining ratings from multiple judges with varying reliability requires statistical models that account for rater quality \parencite{Dawid1979ObserverErrorEM,Passonneau2014BenefitsAnnotation}. The challenge is amplified for LLM annotation because the ``raters'' (model-prompt combinations) may exhibit correlated errors rather than independent noise. Different model families encode different priors and reward different features (cross-model disagreement). Provider and version drift means the ``same'' API endpoint changes behavior over time. Numerical precision (FP16/BF16/TF32) and hardware, batch, and seed configurations introduce inference non-determinism even at temperature zero. Aggregating across noisy annotators without accounting for reliability differences compounds errors. The mechanisms are model drift, precision sensitivity, and aggregation error.

\subsection*{Connecting Variance Sources to Measurement Error}

These five variance sources collectively generate the non-classical measurement error that \textcite{Ludwig2025LLMsEconometric} show can invalidate downstream inference. Each source contributes to the measurement error $\Delta_r = \hat{V}_r - V_r$ in ways that are systematically related to observable characteristics of the data and research design. This correlation structure is precisely what makes LLM measurement error ``non-classical'' in the econometric sense. Classical measurement error---random noise uncorrelated with true values or covariates---attenuates regression coefficients but does not bias them toward particular values. Non-classical measurement error, on the contrary, can induce bias in any direction, including sign reversals. Our diagnostics enable researchers to identify which sources dominate in their specific application and apply targeted controls. By reducing the correlation between annotation error and study variables, the protocol moves measurement error closer to the classical case where its consequences are more predictable and manageable.

\subsection*{From Framework to Protocol}

The above described five sources of variance motivate the protocol design we describe in Section 5. 
Table~\ref{tab:levels} summarizes the instantiation of the framework across output levels. By explicitly designing for variance measurement and reduction at each source, we transform LLM annotation from an \textit{ad hoc }tool into a reliable measurement instrument. The next section provides empirical demonstrations quantifying the effects of these sources in strategy-typical tasks, serving as empirical support for the normative protocol.

\begin{table}[htbp]
\centering
\tiny
\caption{LLM annotation output types/levels: definitions, signatures, diagnostics, controls, and defaults}
\label{tab:levels}
\setlength{\tabcolsep}{3pt}
\renewcommand{\arraystretch}{1.2}
\begin{adjustbox}{max width=\linewidth, max totalheight=\textheight, keepaspectratio, center}
\tiny
\setlength{\tabcolsep}{2pt}
\renewcommand{\arraystretch}{1.05}

\begin{tabularx}{\textwidth}{
  >{\raggedright\arraybackslash}p{1.8cm}
  >{\raggedright\arraybackslash}p{2.6cm}
  >{\raggedright\arraybackslash}p{2.6cm}
  >{\raggedright\arraybackslash}p{2.6cm}
  >{\raggedright\arraybackslash}p{2.6cm}
  >{\raggedright\arraybackslash}p{5.0cm}}
\toprule
\textbf{Level} & \textbf{Definition \& typical use} & \textbf{Authoritative output} & \textbf{Common failure signatures} & \textbf{Diagnostics to run} & \textbf{Controls \& defaults (incl.\ budgets/metrics)} \\
\midrule
\textbf{L1} \newline single label
&
One query $\rightarrow$ one analyzable label (A/B, Yes/No, Likert token). Use when the construct can be coded as a single category or digit.
&
Single token in finite set $Y$ (\texttt{max\_tokens=1}); deterministic token$\rightarrow$label map. Any note is non-authoritative.
&
Order/position flips; surface-form leakage when using verbal anchors; ``right answer, wrong score'' from parsing free text; threshold instability near ties.
&
A/B order swap; paraphrase ensemble; replicate $S$ draws at $T\!\approx\!1$; record per-label log-probs; off-schema counts.
&
Fixed layout; randomized option order; grammar/logit bias; reject-on-fail. \newline
\emph{Budgets:} minimal $P{=}3,S{=}10,M{=}1$; recommended $P{=}3$--$5,S{=}20,M{=}2$. \newline
\emph{Aggregation:} majority; upgrade to Dawid--Skene/GLAD if $\kappa$ low. \newline
\emph{Agreement:} EM, Cohen/Fleiss $\kappa$ (bootstrap CIs). \newline
\emph{Calibration:} Brier/log-loss, reliability curves (temp./isotonic). \\
\addlinespace
\textbf{L2} \newline multi-label
&
Multiple criteria scored together or separately; declare independence vs.\ dependence; pointwise/numeric slots when needed.
&
Typed schema with per-slot tokens or bounded numerics; only slots are authoritative; short notes optional and non-authoritative.
&
Undeclared dependence causes spillovers; scale drift across prompts; unit/format inconsistencies; anchor ambiguity.
&
Separate vs.\ joint elicitation A/B; criterion re-order test; unit/format checks; inter-prompt agreement (weighted $\kappa$/ICC).
&
JSON schema with ranges/types; validator $+$ bounded retries; declare independence/dependence; z-score by prompt for numeric slots; median/trimmed mean per slot. \newline
\emph{Budgets:} minimal $P{=}3,S{=}5,M{=}1$; recommended $P{=}3$--$5,S{=}10$--$15,M{=}2$. \newline
\emph{Aggregation:} slot-wise median/trim; then across prompts/models. \newline
\emph{Agreement:} ICC(2,1)/ICC(3,$k$), Kendall's $W$; CIs via bootstrap. \newline
\emph{Calibration:} cut-point alignment to human anchors; show spread. \\
\addlinespace
\textbf{L3} \newline authoritative text
&
Text itself is the measure (e.g., exact span extraction; bounded summary used downstream). Often feeds L2 in evidence$\rightarrow$score pipelines.
&
Bounded text subject to structure/length or substring-of-source constraint; text is authoritative \emph{only} when declared.
&
Hallucinated evidence; layout sensitivity; sample-to-sample variability; downstream parsing for numbers reintroduces extraction errors.
&
Exact-match/F1 against a small human set; replicate ensembles; position tests; rationale--label alignment checks when L3 feeds L2.
&
Span binding or bounded-structure targets; evidence-then-score pipeline when numbers are needed; keep scores separate from rationales; show dispersion across samples. \newline
\emph{Budgets:} minimal $P{=}3,S{=}5,M{=}1$; recommended $P{=}3$--$4,S{=}10,M{=}2$. \newline
\emph{Aggregation:} slot/text-level EM/F1 or clustering for summaries. \newline
\emph{Agreement:} EM/F1 per slot; $\kappa$/ICC by slot type. \newline
\emph{Uncertainty:} report variance across summaries/extractions. \\
\addlinespace
\textbf{Pipeline} \newline meta (L3$\rightarrow$L1/L2)
&
DAG: extract $\rightarrow$ verify $\rightarrow$ score $\rightarrow$ synthesize. Use for complex constructs needing evidence binding and staged decisions.
&
Stage-wise schema fields are authoritative; aggregator $g(z_{1:J})$ is preregistered and separates stages from synthesis.
&
Early-stage noise leaking downstream; inconsistent schemas; silent verifier failures; end-to-end results drift across versions.
&
Per-stage metrics and stop rates; end-to-end bootstrap CIs; drift checks on a pinned audit set pre/post version/precision change.
&
Stage gates (\emph{no evidence} $\Rightarrow$ NA/triage); per-stage sampling $+$ aggregation then compose; propagate uncertainty to final output. \newline
\emph{Budgets:} per stage minimal $P{=}2,S{=}3,M{=}1$; recommended $P{=}3,S{=}5$--$10,M{=}2$. \newline
\emph{Agreement:} stage-wise (EM/F1/ICC) $+$ end-to-end. \newline
\emph{Drift:} pin provider/name/version/date/precision; audit set cadence. \\
\bottomrule
\end{tabularx}
\end{adjustbox}
\end{table}

\section*{Empirical Demonstrations of Annotation Variance}
\label{sec:empirical-evidence}

Sections 2 and 3 established that na\"ive use of LLMs as annotators is fragile even in the simplest coding tasks and materially shift annotation output. Since LLM-based annotation output is produced by autoregressive next-token prediction under constraints, protocol choices are an integral part of the \emph{measurement model}, not cosmetic options. This section provides controlled demonstrations quantifying these effects in strategy-typical tasks, serving as empirical support for the normative protocol in Section 5.

\subsection*{Analytic Framework}

Following \textcite{Doshi2025GenAIEvaluatingStrategicDecisions}, we annotate the same pair of business model descriptions, keeping these artifacts fixed across all analyses and varying only the \emph{process} (prompts, instructions, models). We cast the task as a pairwise, forced-choice problem with admissible labels $\mathcal{Y}=\{\mathrm{A},\mathrm{B}\}$ and also consider format variants (binary, pointwise token, probability). Unless noted otherwise, all conditions use controlled sampling with $n{=}20$ (Level~1) and $n{=}30$ (Level~2) repeated annotations per manipulation, single-token constrained outputs, and temperature $T{=}1$ to expose variance. For binary labels we extract log-probabilities and compare induced Bernoulli distributions using $t$-tests and ANOVA; for percentage/probability responses we use Welch's $t$-test and the 1-Wasserstein distance \parencite{cai2022distances}. For omnibus tests, we use ANOVA with Tukey's HSD ($\alpha{=}0.05$); for cross-model agreement we report Cohen's $\kappa$ to capture agreement beyond chance. Verbatim stimuli, $\kappa$ definitions, and detailed implementation settings are provided in Appendices A and B.

\subsection*{Construct and Rubric Instability}

\subsubsection*{Construct Ambiguity Effects}

Ambiguous constructs push the model to substitute surface proxies for the target criterion---a failure mode well documented in human coding when codebooks lack behavioral anchors \parencite{Krippendorff2018Content}. We contrasted three prompts intended to measure ostensibly similar constructs while holding all artifacts constant: (1) \emph{Success likelihood}: ``From your viewpoint as an investor in this startup, which business model is more likely to succeed? Answer only `A' or `B';" (2)  \emph{Idea quality}: ``From your viewpoint as an investor in this startup, which business model is a better idea? Answer only `A' or `B';" and (3) \emph{Writing quality}: ``From your viewpoint as an investor in this startup, which business model is better written? Answer only `A' or `B'."

Outcomes differ across these formulations (one-way ANOVA $F(2,57)=10.54$, $p=0.0001$). Post-hoc tests indicate that success and idea quality are statistically indistinguishable (mean difference $=-0.339$, $p=0.994$), while both diverge significantly from writing quality (mean differences $-12.688$ and $-12.349$ respectively, both $p<0.001$). A 12--13 point shift on otherwise identical inputs is a construct-failure signature: without an explicit construct map and behaviorally anchored rubric, the annotator substitutes whichever surrogate the prompt invites. This pattern mirrors findings in human content analysis, where underspecified constructs produce unreliable coding \parencite{Neuendorf2017Content}.

\subsubsection*{Context Provision Sensitivity}

Even simply changing the role assigned to an LLM annotator fundamentally shapes its coding stance. We tested three different annotator perspectives while holding all other factors constant: (1) Investor perspective: ``From your viewpoint as an investor in this startup, which business model is more likely to succeed? Answer only `A' or `B';" (2) Founder perspective: ``From your viewpoint as the founder of this startup, which business model is more likely to succeed? Answer only `A' or `B';" and (3) Environmental activist perspective: ``From your viewpoint as an environmental activist, which business model is more likely to succeed? Answer only `A' or `B'."

Even a minimally different context produces large differences (ANOVA $F(2,57) = 1849.36$, $p < 0.001$). The environmental activist perspective differs from the investor perspective by 69.36 percentage points ($p < 0.001$) and from the founder perspective by 85.12 percentage points ($p < 0.001$). Even the investor and founder perspectives differ by 15.76 percentage points ($p < 0.001$). These swings---up to 85 percentage points---reveal that LLM annotations adopt distinct coding frameworks shaped by the assigned role, analogous to how human coders with different training or disciplinary backgrounds apply constructs differently \parencite{Krippendorff2018Content}.

\subsection*{Interface and Prompt Instability}

\subsubsection*{Prompt Wording Sensitivity}

Consistent with the questionnaire design literature, which documents that semantically equivalent question wordings can produce different response distributions \parencite{Schuman1981QuestionOrder}, LLM annotations are highly sensitive to prompt wording and formatting (\textit{prompt brittleness}). To test whether semantic-preserving reformulations maintain annotation stability, we applied multiple manipulations to the base investor prompt while preserving its intended meaning (see Appendix B for full variants). These supposedly equivalent formulations produce significantly different outcomes ($F(4,95) = 32.32$, $p < 0.001$). Restructuring clauses alone shifts annotations from 8.38 to 28.10 percentage points compared to other phrasings (all $p < 0.05$). Nominalization differs from the original by 16.50 percentage points ($p < 0.001$), while synonym substitution shows a minimal difference (3.23 percentage points, $p = 0.818$).

This sensitivity parallels findings on question wording effects in survey methodology: small changes in phrasing can shift responses substantially, even when the intended meaning is preserved \parencite{Krosnick1987ResponseOrder}. Just as survey researchers must pilot-test question wordings for stability, LLM annotation protocols should assess robustness to paraphrase.

\subsubsection*{Prestige Cue Contamination (Halo Effects)}

LLM annotators exhibit halo effects from prestige cues---a well-documented form of rater bias in human performance appraisal \parencite{DeNisi2017PerformanceAppraisal}. Because LLMs encode knowledge about real-world entities, their annotations can be contaminated by irrelevant prestige associations. We test this by adding three ``endorsement cues'' to the prompt: (1) Prestigious tech investor: ``Peter Thiel backs this model;" (2) Prestigious tech entrepreneur: ``Elon Musk backs this model;" and (3) Unknown control: ``Saeid Kazemi backs this model."

Each endorsement was appended to the standard prompt. The presence of identity/prestige cues significantly alters annotations ($F(2,57) = 19.25$, $p < 0.001$). Peter Thiel's and Elon Musk's endorsements produce similar boosts (mean difference = 0.558, $p = 0.971$), while both differ markedly from the unknown name control (mean differences of $-12.683$ and $-13.241$ respectively, both $p < 0.001$). This 12--13 percentage point effect confirms systematic contamination from the model's latent real-world knowledge---a form of halo effect where prestige on one dimension (fame, success) influences coding on an unrelated dimension (business model quality). In human coding, halo effects occur when a rater's overall impression of a target influences ratings on specific dimensions \parencite{Neuendorf2017Content}. LLM annotators exhibit an analogous bias: the knowledge that a prestigious individual endorses an option inflates ratings for that option, regardless of its substantive merits.

\subsection*{System and Cross-Model Instability}

\subsubsection*{Sampling Variance and Probabilistic Uncertainty}
\label{subsubsec:sampling-variance}

Single-token outputs are draws from an underlying categorical probability distribution. Using log-probability extraction with GPT-4o, we document the uncertainty masked by single-token output. We repeatedly sample ($N=100$) annotations using identical prompts and context. The probability of selecting business model A ranges from 0.68 to 0.92. This 24-percentage-point range persists even at temperature $T=0$ due to hardware-induced non-determinism and floating-point precision effects. What appears as a confident annotation at $P(A)=0.55$ actually represents a draw from a broad, nearly uniform distribution. Without repeated sampling and probability logging, researchers might think LLMs provide deterministic classifications instead of probabilistic predictions.

\subsubsection*{Cross-Model Disagreement}
\label{subsubsec:cross-model}

Different LLM ``coders'' interpret constructs differently, requiring reliability assessment---just as human coding studies assess intercoder reliability before treating annotations as valid data \parencite{Krippendorff2018Content}. We examine whether ostensibly similar frontier models reach consistent annotations by comparing six LLMs---GPT-4o, o4-mini, Claude-3.5-Sonnet, Claude-Sonnet-4, Gemini-2.0-Flash, and Gemini-2.5-Pro---on identical business model comparisons. Each model annotated 20 binary choices, yielding 120 observations after excluding four API failures.

Cross-model agreement analysis reveals significant annotation inconsistency ($F = 13.847$, $p < 0.001$). Most critically, in 90.0\% of annotations, the choice of model determined the outcome---meaning research inferences would differ based solely on model selection. Pairwise $t$-tests revealed significant differences in 7 of 15 model comparisons. Claude-3.5-Sonnet showed the most divergent annotation pattern, differing significantly from all other models (all $p < 0.001$). Within-provider consistency varied substantially: OpenAI models showed identical response patterns ($p = 1.000$), while Anthropic models ($p < 0.001$) and Google models ($p = 0.016$) differed significantly between versions. Reasoning models did not demonstrate superior consistency compared to non-reasoning models ($p = 0.100$). The response distributions (Table~\ref{tab:model-responses}) expose systematic biases across model families---analogous to how different human coders may exhibit systematic tendencies toward leniency or severity \parencite{Neuendorf2017Content}. Gemini-2.0-Flash exhibited complete uniformity (100\% selecting A), while Claude-3.5-Sonnet showed opposite preferences (85\% selecting B).

\begin{table}[htbp]
\centering
\tiny
\begin{tabular}{lcc}
\hline
\textbf{Model (Annotator)} & \textbf{Selected A (\%)} & \textbf{Selected B (\%)} \\
\hline
Gemini-2.0-Flash & 100.0 & 0.0 \\
GPT-4o & 85.0 & 15.0 \\
o4-mini & 85.0 & 15.0 \\
Claude-Sonnet-4 & 80.0 & 20.0 \\
Gemini-2.5-Pro & 75.0 & 25.0 \\
Claude-3.5-Sonnet & 15.0 & 85.0 \\
\hline
\end{tabular}
\caption{Response distributions across LLM annotators coding identical business model pairs. Different model families exhibit systematic coding biases analogous to rater effects in human content analysis.}
\label{tab:model-responses}
\end{table}

Cohen's $\kappa$\footnote{Strictly speaking, Cohen's $\kappa$ is defined for agreement on distinct items where the item-wise pairing is substantively meaningful. In our setting, all models annotate the same business model comparisons, and the ordering of those comparisons is not itself of substantive interest. We nevertheless use $\kappa$ here both as a conservative diagnostic and as an illustration of the methodology researchers should apply when comparing model annotations across \emph{different} texts. If models fail to exhibit high $\kappa$ even when coding identical text pairs under a favorable setup, one should expect at least as much disagreement---and likely more---when they are applied to heterogeneous corpora.} averaged only 0.026, indicating agreement barely distinguishable from chance despite 58.7\% raw agreement. Within-provider consistency varies dramatically (Table~\ref{tab:model-agreement}). OpenAI models demonstrated moderate agreement ($\kappa=0.608$), while Anthropic models showed poor consistency ($\kappa=0.085$), and Google models exhibited no agreement beyond chance ($\kappa=0.000$) despite 75\% raw agreement---reflecting Gemini-2.0-Flash's complete bias toward A. Surprisingly, reasoning-enhanced models showed \textit{worse} internal consistency than their non-reasoning counterparts ($\kappa=-0.043$ vs.\ $-0.025$), consistent with recent evidence of inverse scaling in test-time compute \parencite{Gema2025InverseScalingTestTimeCompute}.\

\begin{table}[htbp]
\tiny
\centering
\begin{tabular}{lcc}
\toprule
\textbf{Annotator Pairing} & \textbf{Cohen's $\kappa$} & \textbf{\% Agreement} \\
\midrule
\multicolumn{3}{l}{\textit{Within-Provider}} \\
\quad OpenAI (GPT-4o vs o4-mini) & 0.608 & 90.0 \\
\quad Anthropic (Claude-3.5 vs Claude-4) & 0.085 & 35.0 \\
\quad Google (Gemini-2.0 vs Gemini-2.5) & 0.000 & 75.0 \\
\midrule
\multicolumn{3}{l}{\textit{By Reasoning Capability}} \\
\quad Within Reasoning Models & -0.043 & 66.7 \\
\quad Within Non-Reasoning Models & -0.025 & 44.2 \\
\quad Cross-Type Comparison & 0.067 & 61.9 \\
\midrule
\textbf{Overall Mean} & \textbf{0.026} & \textbf{58.7} \\
\bottomrule
\end{tabular}
\caption{Inter-annotator agreement and within-provider consistency for LLM coders. The low $\kappa$ values indicate that different LLM annotators interpret the same construct differently, analogous to intercoder disagreement in human content analysis.}
\label{tab:model-agreement}
\end{table}

With 90\% of outcomes determined by model choice rather than substantive factors, single-model studies risk producing artifacts of model selection rather than insights about strategic phenomena \parencite{Doshi2025GenAIEvaluatingStrategicDecisions}. This finding underscores the importance of multi-rater reliability assessment: just as human coding studies would not rely on a single coder without assessing intercoder agreement, LLM annotation studies should employ multiple models and report cross-model reliability.

\subsection*{Level-Specific Patterns}

\subsubsection*{Level 1: Response Format Changes Outcomes}

\begin{figure}[htbp]
\centering
\includegraphics[width=0.49\textwidth]{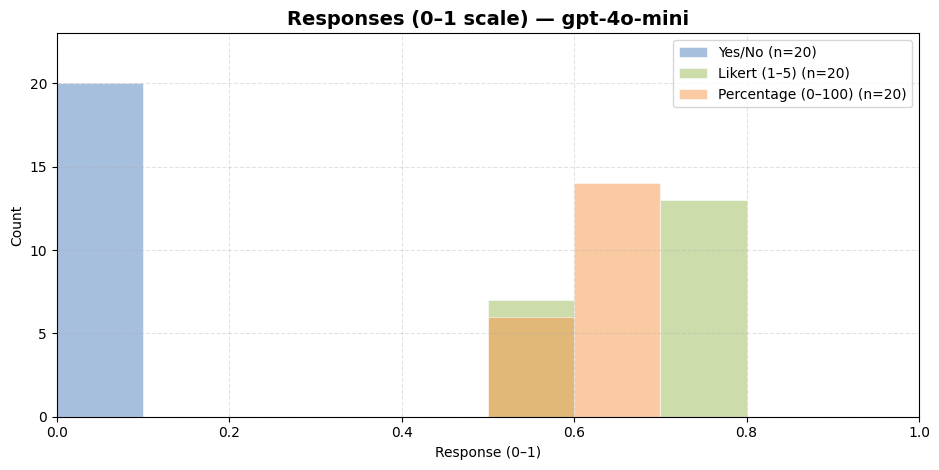}
\includegraphics[width=0.49\textwidth]{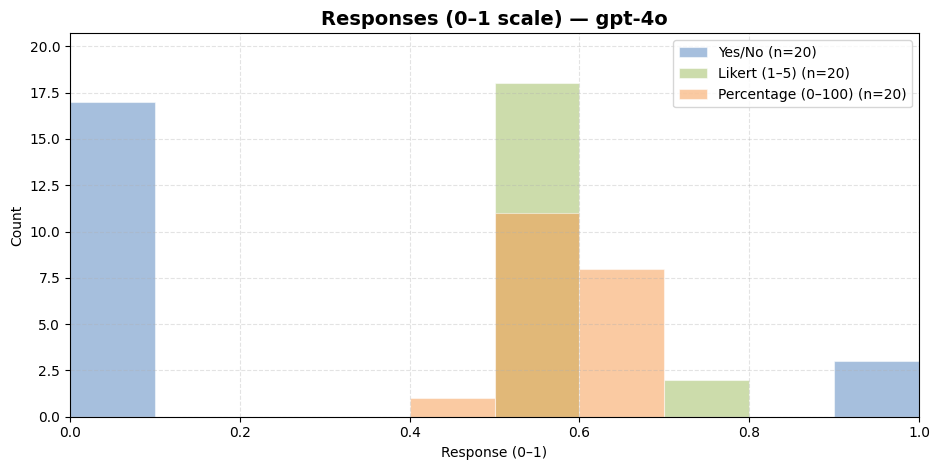}
\includegraphics[width=0.49\textwidth]{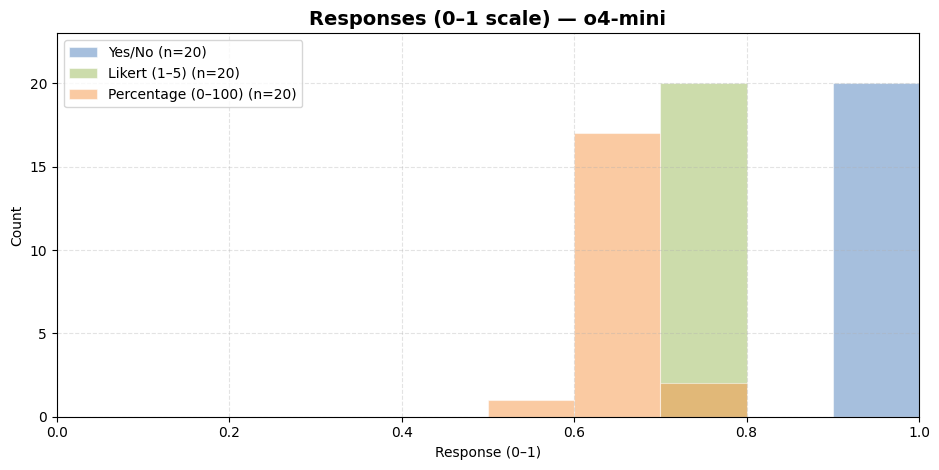}
\caption{Level~1 format sensitivity: distributions for binary, Likert, and probability encodings by model ($n{=}20$ per encoding). Response format affects annotation output, paralleling format effects in survey methodology.}
\label{fig:l1-format}
\end{figure}

We presented the same proposition (``Business model A is better than business model B'') in three encodings: binary Yes/No, 5-point Likert, and probability 0--100. Each encoding was sampled $n{=}20$ times for \texttt{gpt-4o}, \texttt{gpt-4o-mini}, and \texttt{o4-mini}. Figure~\ref{fig:l1-format} overlays the three distributions per model. \texttt{gpt-4o-mini} is near-deterministic ``No'' in binary format yet centers toward ``Yes'' under Likert and probability. \texttt{o4-mini} is comparatively self-consistent across formats. Agreement across formats is low: binarizing Likert and probability at pre-registered thresholds yields small Cohen's $\kappa$, and Wasserstein distances between continuous distributions are large.

This pattern parallels research on response format effects in survey methodology, where the choice between forced-choice and Likert scales systematically affects measured values \parencite{Krosnick1999SurveyResearch}. The encoding choice is not a neutral presentation detail---it is part of the measurement model. If probabilities are used, calibration (Brier score and reliability curves) and pre-specified operating thresholds should accompany any claim.\

\subsubsection*{Level 1: Position and Order Effects (A/B Swap)}
\label{subsubsec:l1-order}

\begin{figure}[htbp]
\centering
\includegraphics[width=0.9\linewidth]{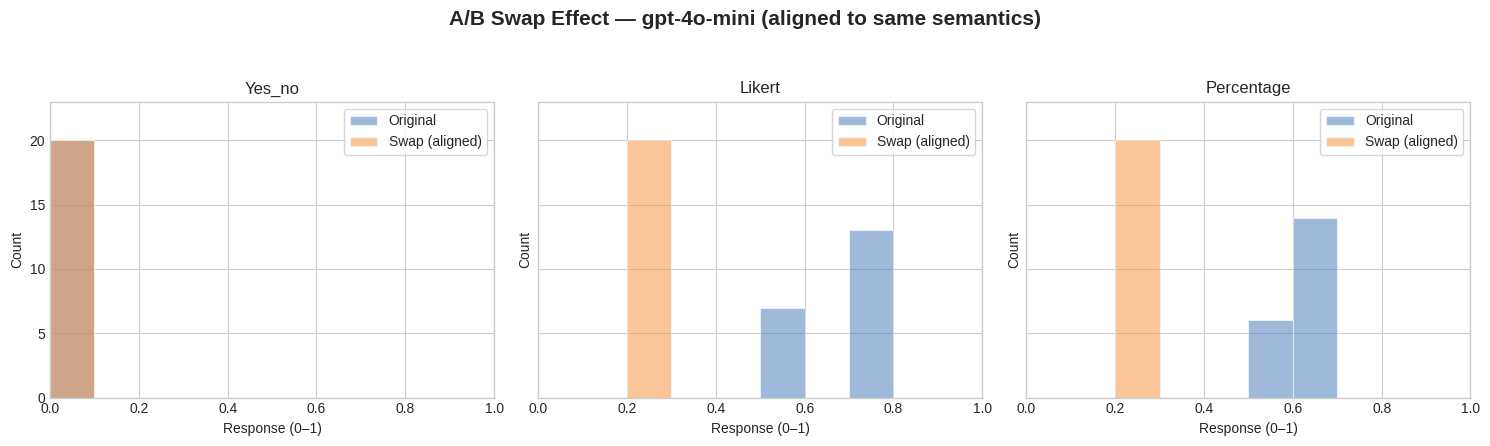}
\includegraphics[width=0.9\linewidth]{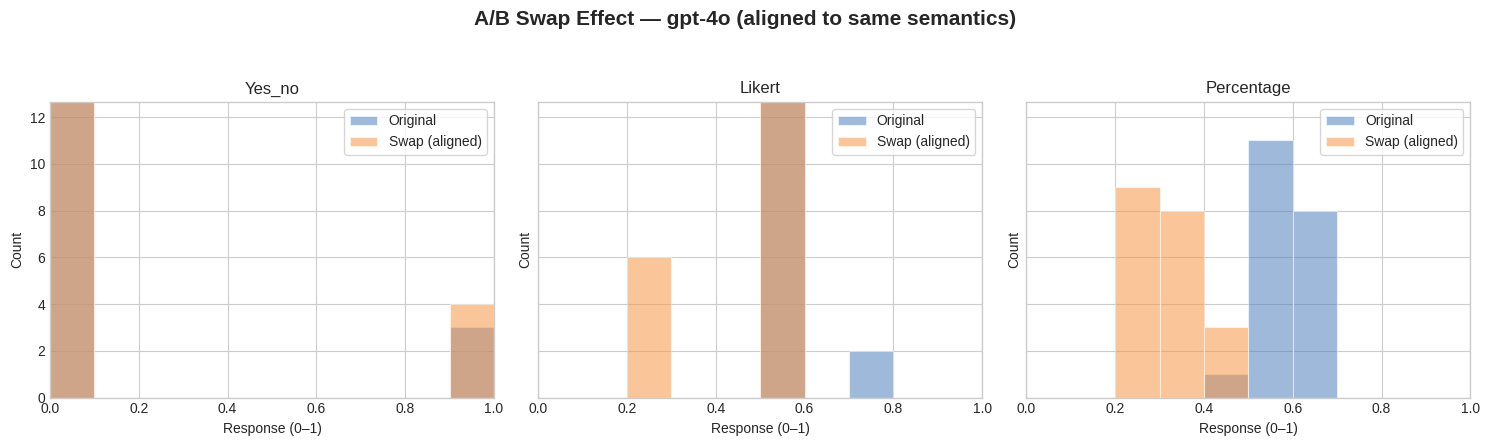}
\includegraphics[width=0.9\linewidth]{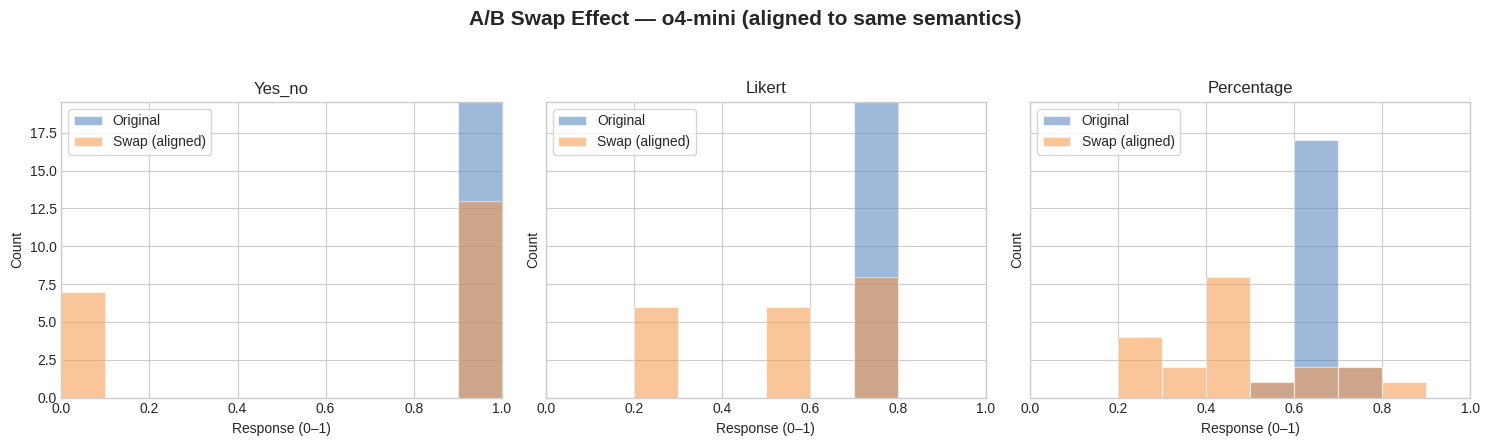}
\caption{Level~1 A/B swap: aligned original vs.\ swap distributions by model ($n{=}20$ per condition). Position effects in LLM annotation parallel order effects documented in survey methodology.}
\label{fig:l1-swap}
\end{figure}

To test position sensitivity, we first ran the base condition with business model A presented first and B second, then re-ran after swapping the textual order while keeping the label set $\{\mathrm{A},\mathrm{B}\}$ unchanged. All three models show some order dependence (Fig.~\ref{fig:l1-swap}). \texttt{gpt-4o-mini} shows the tightest overlap; \texttt{gpt-4o} and \texttt{o4-mini} shift more visibly.

Position and order effects are well documented in survey methodology \parencite{Krosnick1987ResponseOrder}: respondents may favor options presented first (primacy) or last (recency) depending on task characteristics. LLM annotators exhibit analogous sensitivity. For applied work, randomizing option order and reporting invariance checks should be standard practice.

\subsubsection*{Level 2: Separate vs.\ Joint Elicitation}

We used LLMs to code one venture description on ten criteria (0--100): \emph{Innovation \& Value Proposition}, \emph{Product \& Services}, \emph{Market Opportunity \& Strategy}, \emph{Team}, \emph{Execution Plan}, \emph{Financials}, \emph{Structure of Venture Summary}, \emph{Sustainable Competitive Advantage}, \emph{Potential to Invest}, and \emph{Viability}. Each criterion was asked either \emph{separately} (ten independent prompts) or \emph{jointly} (a single schema with ten fields). We drew $n{=}30$ samples per condition for \texttt{GPT-5 reasoning}, \texttt{GPT-5-mini}, \texttt{gpt-4o}, and \texttt{gpt-4o-mini}. We compared separate vs.\ joint scores using Welch's $t$-test and the 1-Wasserstein distance.

Across models, many criteria show statistically significant mode effects (Table~\ref{tab:l2-compact}). Wasserstein magnitudes are generally small to moderate; one large shift appears for \emph{Structure of Venture Summary} with \texttt{gpt-4o}. Multi-field questionnaires are not merely convenient containers: asking together versus separately alters the measurement. This parallels findings on context effects in survey design, where preceding questions influence responses to subsequent ones \parencite{Schuman1981QuestionOrder}. If independence across criteria is important, elicit them separately; if dependence is intended, declare it and keep the fields in a single, typed schema.

\begin{table}[htbp]
\centering
\tiny
\setlength{\tabcolsep}{3.5pt}
\renewcommand{\arraystretch}{0.95}
\begin{tabular}{l *{4}{cc}}
\toprule
& \multicolumn{2}{c}{\textbf{GPT-5 (reason.)}} & \multicolumn{2}{c}{\textbf{GPT-5-mini}} & \multicolumn{2}{c}{\textbf{gpt-4o}} & \multicolumn{2}{c}{\textbf{gpt-4o-mini}} \\
\cmidrule(lr){2-3}\cmidrule(lr){4-5}\cmidrule(lr){6-7}\cmidrule(lr){8-9}
\textbf{Criterion} & \textit{Welch-$p$} & \textit{$W$} & \textit{Welch-$p$} & \textit{$W$} & \textit{Welch-$p$} & \textit{$W$} & \textit{Welch-$p$} & \textit{$W$} \\
\midrule
Team & 0.000 & 3.70 & 0.000 & 11.17 & 0.006 & 2.17 & 0.008 & 2.33 \\
Execution Plan & 0.000 & 8.57 & 0.000 & 2.73 & 0.000 & 9.00 & 0.150 & 2.50 \\
Financials & 0.011 & 1.67 & 0.000 & 6.50 & 0.001 & 3.33 & 0.326 & 0.17 \\
Structure of Venture Summary & 0.000 & 7.37 & 0.000 & 3.77 & 0.000 & 15.33 & 0.000 & 9.00 \\
Innovation \& Value Proposition & 0.000 & 3.23 & 0.016 & 1.73 & 0.000 & 3.67 & 0.001 & 2.83 \\
Market Opportunity \& Strategy & 0.000 & 5.60 & 0.029 & 1.57 & 0.506 & 0.67 & 0.064 & 1.67 \\
Sustainable Competitive Advantage & 0.592 & 0.33 & 0.000 & 7.97 & 0.000 & 5.83 & 0.181 & 1.00 \\
Product \& Services & 0.000 & 9.57 & 0.000 & 8.00 & 0.000 & 7.33 & 0.000 & 8.93 \\
Potential to Invest & 0.000 & 2.57 & 0.751 & 0.63 & 0.000 & 14.00 & 0.772 & 0.91 \\
Viability & 0.385 & 0.50 & 0.000 & 6.10 & 0.005 & 2.50 & 0.000 & 5.43 \\
\bottomrule
\end{tabular}
\tiny \caption{Level~2 dependence: separate vs.\ joint elicitation (30 draws/condition). $W$ = 1-Wasserstein distance (points on 0--100). Elicitation mode affects annotation output, paralleling context effects in survey design.}
\label{tab:l2-compact}
\end{table}

\subsubsection*{Level 2: Re-ordering Criteria}
\label{subsubsec:l2-order}

We re-ordered the joint schema (from high to low) and repeated the $n{=}30$ condition. Some criteria shift upward or downward depending on their position, consistent with position effects in multi-question prompts and weak cross-criterion dependence (Table~\ref{tab:l2-reorder}). This parallels question-order effects documented in survey methodology \parencite{Schuman1981QuestionOrder}. A practical safeguard is to randomize criterion order across draws or include position as a factor in the analysis.

\begin{table}[htbp]
\centering
\tiny
\setlength{\tabcolsep}{3.5pt}
\renewcommand{\arraystretch}{0.95}
\begin{tabular}{l *{4}{cc}}
\toprule
& \multicolumn{2}{c}{\textbf{GPT-5 (reason.)}} & \multicolumn{2}{c}{\textbf{GPT-5-mini}} & \multicolumn{2}{c}{\textbf{gpt-4o}} & \multicolumn{2}{c}{\textbf{gpt-4o-mini}} \\
\cmidrule(lr){2-3}\cmidrule(lr){4-5}\cmidrule(lr){6-7}\cmidrule(lr){8-9}
\textbf{Criterion} & \textit{Welch-$p$} & \textit{$W$} & \textit{Welch-$p$} & \textit{$W$} & \textit{Welch-$p$} & \textit{$W$} & \textit{Welch-$p$} & \textit{$W$} \\
\midrule
Team & 0.000 & 4.17 & 0.000 & 6.07 & 0.000 & 7.83 & 0.000 & 12.33 \\
Execution Plan & 0.000 & 3.20 & 0.921 & 0.40 & 0.000 & 4.17 & 0.049 & 1.67 \\
Financials & 0.000 & 4.50 & 0.099 & 1.63 & 0.000 & 7.17 & 0.000 & 11.83 \\
Structure of Venture Summary & 0.887 & 0.57 & 0.198 & 1.03 & 0.001 & 5.17 & 0.000 & 6.50 \\
Innovation \& Value Proposition & 0.004 & 1.67 & 0.003 & 2.33 & 0.000 & 4.00 & 0.000 & 3.50 \\
Market Opportunity \& Strategy & 0.078 & 1.43 & 0.000 & 3.47 & 0.000 & 7.93 & 0.000 & 6.67 \\
Sustainable Competitive Advantage & 0.177 & 0.80 & 0.965 & 0.50 & 0.000 & 11.23 & 0.000 & 7.33 \\
Product \& Services & 0.017 & 1.40 & 0.000 & 3.57 & 0.000 & 6.93 & 0.000 & 8.50 \\
Potential to Invest & 0.874 & 0.70 & 0.296 & 1.07 & 0.000 & 6.33 & 0.000 & 6.17 \\
Viability & 0.259 & 0.63 & 0.062 & 1.43 & 0.000 & 7.67 & 0.000 & 4.00 \\
\bottomrule
\end{tabular}
\tiny \caption{Level~2 re-ordering: base vs.\ anchored sequence (30 draws/condition). $W$ = 1-Wasserstein distance. Position in the elicitation sequence affects annotation output.}
\label{tab:l2-reorder}
\end{table}

\subsubsection*{Level 3: Free-Form Text as Authoritative Output}

When open text is the annotation output itself (e.g., summaries or extractions), variability reflects both sampling and inference-stack nondeterminism. A simple ensemble of ten summaries of the same transcript may yield clusters with different topical coverage, even under matched instructions \parencite{atil2024non}. This is useful signal when text is the object of study, but risky when the same text is later mined to justify numeric scores. If a number is the target, an ``evidence-then-score'' pipeline (L3 to L2) with clear anchors is safer. If text remains authoritative, variance should be shown explicitly---for example, by reporting dispersion across summaries.

\subsubsection*{Does LLM Reasoning Help?}
\label{subsubsec:l2-reasoning}

\begin{figure}[htbp]
    \centering
    \includegraphics[width=0.5\textwidth]{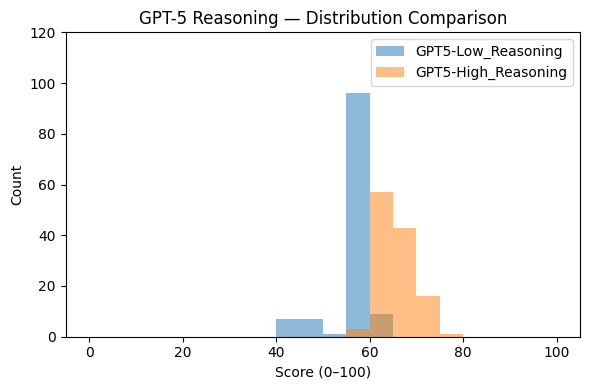}
    \caption{Score distributions under low vs.\ high reasoning (GPT-5 reasoning).\\Requiring explicit reasoning increases annotation variance.}
    \label{fig:l2-reasoning-dist}
\end{figure}

We tested whether requiring extensive reasoning affects annotation outcomes. The model compared two venture descriptions, coding whether business plan A was superior to plan B. We ran two conditions with \texttt{GPT-5 reasoning}: \emph{low reasoning} (minimal chain of thought) and \emph{high reasoning}, each with $n{=}120$ annotations. Figure~\ref{fig:l2-reasoning-dist} shows the score distributions. 
Scores under high reasoning are systematically higher and more dispersed than under low reasoning. With $n_{\text{low}}{=}n_{\text{high}}{=}120$, the mean rises from $54.57$ to $65.00$. Welch test yields $p{<}0.001$ showing significant difference. The 1-Wasserstein distance is $W{=}10.43$ points (Moderate), indicating a broad upward shift of the entire distribution. The high-reasoning distribution is less concentrated, implying greater annotation variance and more differentiated coding when explicit reasoning is required.

\subsection*{Connecting Empirical Findings to Measurement Error}

The patterns documented above are not harmless noise---they translate directly into the measurement error that can invalidate downstream inference. The magnitudes we observe---construct ambiguity shifting annotations by 12--13 points, context effects by 69--85 points, prestige cues by 12--13 points---are far too large and systematic to be treated as random noise. Measurement error of this kind is non-classical: it does not merely attenuate estimated relationships but can distort, exaggerate, or even reverse them in downstream analyses. \textcite{Ludwig2025LLMsEconometric} show that when annotation error $\Delta_r = \hat{V}_r - V_r$ correlates with economic covariates $W_r$, plug-in regression using LLM annotations produces inconsistent estimates. Our findings reveal multiple pathways through which such correlation can arise. 

\subsection*{Implications for Strategy Research}

Our findings establish that LLM annotation should be treated as a measurement problem requiring the same rigor applied to human coding and survey design. The implications span five dimensions that characterize the normative protocol in Section 5. First, researchers should begin with a construct map and a valid, behaviorally anchored rubric following established practices in content analysis \parencite{Krippendorff2018Content}, state the output level (L1/L2/L3) upfront, specify which fields are authoritative, anonymize when identity is not part of the construct, and validate inputs before coding. Second, consistent with survey design principles \parencite{Schuman1981QuestionOrder, haaland2025understanding}, interface and context should be controlled by keeping prompts single-question and tied to the rubric, fixing layout, randomizing option order, avoiding task mixing, and running paraphrase ensembles with planned aggregation. Third, researchers should assume that models differ in what they reward---analogous to rater effects in human coding \parencite{Neuendorf2017Content}---by using neutral annotator roles, allowing ``cannot judge'' responses, preferring evidence-then-score pipelines for numeric targets, and treating persistent cross-model disagreement as a signal to revisit the construct or rubric. Fourth, output constraining and extraction should enforce valid labels through decoding constraints, typed JSON schemas with validation for L2, and per-label probability logging with calibration reporting when decisions depend on thresholds. Fifth, system and aggregation procedures should expose variance through sampling replicates at non-zero temperature, use prompt ensembles and multiple model families as parallel coders requiring reliability assessment, report agreement with bootstrap intervals, employ reliability-aware aggregation when agreement is low, and maintain pinned audit sets with version tracking. These implications lead directly to practice. The next section turns these five dimensions into concrete defaults and a checklist for designing and reporting LLM-based annotation.

\section*{A Variance-Aware Annotation Protocol}
\label{sec:variance-aware-protocol}

The protocol is organized around the five sources of variance and by annotation output type (Levels 1--3 and pipelines). For each, we specify what to design, what to test, and what to report. The protocol is modular so researchers can ``rightsize'' its use, matching rigor to stakes while keeping procedures auditable and replicable.

\subsection*{Design Primitives and Notation}

We formalize the sampling design using three primitives that parallel multi-rater reliability designs in content analysis. The \textit{prompt ensemble} $P$ specifies the number of meaning-preserving paraphrases of the coding question, typically $P \in \{3,4,5\}$, analogous to training multiple coders with slightly different codebook interpretations. The \textit{sampling replicates} $S$ specifies the number of independent draws per prompt, typically $S \in \{5,10,20\}$, capturing within-configuration variance that would be hidden by deterministic single-shot annotation. The \textit{model ensemble} $M$ specifies the number of independent model families, typically $M \in \{1,2,3\}$, analogous to using multiple independent coders to assess intercoder reliability. 
For item $i$, prompt variant $u \in \{1,\dots,P\}$, sample $s \in \{1,\dots,S\}$, and model $m \in \{1,\dots,M\}$, let $y_{ius}^{(m)}$ denote the annotation output. Aggregation proceeds in stages: within $(u,m)$ across samples $s$; across prompts $u$ within model $m$; then across models $m$ to yield the final item-level annotation $\hat{y}_i$. Researchers should distinguish \emph{authoritative} fields (schema-constrained outputs used in analysis) from \emph{non-authoritative} fields (rationales, notes). For L2, they should declare whether criteria are \emph{independent} (coded separately) or \emph{dependent} (e.g., budget shares summing to 100). They should also state the aggregation target: single canonical label, distribution over labels, or ensemble of summaries.

\subsection*{Codebook Development: From Construct to Prompt}
\label{subsec:theory-first}

Our approach is \emph{theory-first}---the same principle that guides codebook development in traditional content analysis. The target construct should be translated into codable criteria and a rubric that an LLM can apply with fidelity. Validity is supported by a cumulative program of evidence (content, response process, relations to other variables) rather than by any single procedure \parencite{Messick1995ValidityAssessment}. Researchers should state the construct, boundaries, and dimensionality. They should specify whether the measure--construct link is reflective or formative and justify indicator choice and aggregation accordingly \parencite{Jarvis2003ConstructIndicatorsReview,Diamantopoulos2001FormativeIndex}. Misclassification biases inference in strategy settings. They should also state four design choices up front: (1) Output level: L1/L2/L3 or pipeline. (2) Decision scope: pointwise/pairwise/listwise/setwise. (3) Evidence binding: span-grounded quotes vs.\ none; declare authoritative fields. (4) Tool use: retrieval/validators allowed or not; constraints for each tool. They should use behaviorally anchored rating scales keyed to observable cues and counter-examples \parencite{Smith1963RetranslationAnchors,Latham1977BOS}. \emph{Every} point on the scale should be anchored with concrete examples and the anchors should be aligned with the level and scope selected. Researchers should: (1) ask one question at a time tied to the rubric; (2) randomize option order in pairwise/listwise settings; (3) constrain outputs to the admissible label set or schema; and (4) keep any free-text rationale \emph{non-authoritative} unless Level~3 is declared. Before scaling, researchers should collect content evidence (expert Q-sort/CVI; brief cognitive interviews) and relations evidence (small MTMM-style slice with human coders) \parencite{Campbell1959MTMM,Cronbach1955ConstructValidity}. If probabilities exist, they should be kept for calibration and uncertainty reporting.

\subsection*{Defaults by Output Level (L1/L2/L3)}

Table~\ref{tab:annotation-design} provides level-specific defaults for output constraints, sampling budgets $(P,S,M)$, aggregation rules, agreement metrics, and calibration/uncertainty reporting. These defaults balance rigor with feasibility and should be adjusted based on construct complexity and stakes.

\begin{table}[htbp]
\centering
\tiny
\caption{Annotation Level $\times$ Recommended Design Choices}
\label{tab:annotation-design}
\begin{tabularx}{\textwidth}{
>{\raggedright\arraybackslash}p{2.6cm}
>{\raggedright\arraybackslash}X
>{\raggedright\arraybackslash}X
>{\raggedright\arraybackslash}p{2.6cm}
>{\raggedright\arraybackslash}X
>{\raggedright\arraybackslash}X
>{\raggedright\arraybackslash}X}
\toprule
\textbf{Annotation Level} & \textbf{Typical Use} & \textbf{Output Constraint} & \textbf{Sampling (P,S,M)} & \textbf{Aggregation} & \textbf{Agreement / Validity} & \textbf{Calibration \& Uncertainty}\\
\midrule
\textbf{Level 1: Single-Label} & Binary/multiclass labels; pairwise A vs.\ B & \texttt{max\_tokens}=1; deterministic label map; stop tokens; $T{=}0$--$0.3$ & Minimal: $P{=}3$, $S{=}10$, $M{=}1$; Rec.: $P{=}3$--$5$, $S{=}20$, $M{=}2$ & Majority; DS/GLAD when coders disagree & Cohen/Fleiss $\kappa$ (95\% CI); human benchmark if feasible & Keep per-label probs; Brier/log-loss; reliability curves; bootstrap CIs\\
\addlinespace
\textbf{Level 2: Multi-Label (Typed Schema)} & Multi-criteria coding (e.g., novelty, feasibility) & JSON schema with enums/ranges; cross-field constraints; reject-on-fail \& resample & Minimal: $P{=}3$, $S{=}5$, $M{=}1$; Rec.: $P{=}3$--$5$, $S{=}10$--$15$, $M{=}2$ & Slot-wise: median/trim for numeric; vote for categorical; joint validity check & Weighted $\kappa$/ICC per slot; Kendall's $W$ across criteria & Calibrate cut-points vs.\ human anchors; show spread with CIs\\
\addlinespace
\textbf{Level 3: Free-Form Text (Authoritative)} & Span extraction or bounded summary as the measure & Length/structure caps; span-grounding; no score parsing & Minimal: $P{=}3$, $S{=}5$, $M{=}1$; Rec.: $P{=}3$--$4$, $S{=}10$, $M{=}2$ & Publish ensemble dispersion; pre-registered selection if one text required & Dispersion across summaries; coverage of rubric elements & If L3 feeds L2, revert to L2 calibration; otherwise report ensemble variance\\
\addlinespace
\textbf{Pipeline (Meta)} & Extract $\rightarrow$ verify $\rightarrow$ code $\rightarrow$ synthesize & Per-stage schemas/constraints; DAG declared & Minimal per stage: $P{=}2$, $S{=}3$, $M{=}1$; Rec.: $P{=}3$, $S{=}5$--$10$, $M{=}2$ & Aggregate per stage, then compose; propagate uncertainty & Stage-wise metrics + end-to-end; sensitivity to stage failure & Calibrate critical stages; drift audit per stage; propagated CIs (bootstrap)\\
\bottomrule
\end{tabularx}
\end{table}

\subsection*{Reliability Engineering: The Five Sources of Variance}

For each of the five sources introduced in Section 3, we specify what it looks like in practice, core controls in the protocol, and where to log/report. This subsection translates the variance framework into operational guidance.
\subsubsection*{Source 1: Construct \& Rubric Specification}
Our \emph{theory-first} approach requires that the target construct is captured by a valid measure and translated into codable criteria and a rubric that an LLM can apply with fidelity. 
\textbf{Core controls.} Freeze the construct map $\rightarrow$ write BARS-style anchors for every scale point $\rightarrow$ pick L1/L2/L3 accordingly;  declare independence/dependence in L2; set decision scope (point/pair/list/set). Mark authoritative fields; keep rationales non-authoritative unless L3 is the measure. Anonymize identities not in the construct. Validate transcripts or ASR/OCR outputs before coding.
\textbf{Where to report.} Human$\leftrightarrow$LLM agreement on calibration slice; MTMM tables if applicable; schema definitions; examples of anchors and counter-anchors. Document any redaction or anonymization procedures.
\subsubsection*{Source 2: Interface \& Context Control}
This set of choices comprises prompt and layout. It countermeasures order/position and length effects,  paraphrase brittleness, single- vs.\ multi-question prompts and identity or prestige cues (halo effects).
\textbf{Core controls.} One question per prompt; fixed layout; randomize option order; equalize verbosity when comparing items; place rubric and key evidence adjacent to the query; anonymize entities when not part of the construct. Build a prompt ensemble ($P \geq 3$) of meaning-preserving paraphrases.
\textbf{Where to report.} $\kappa$/weighted $\kappa$, flip rates, and layout invariance checks (e.g., A/B swap tests). Document prompt templates and randomization procedures.
\subsubsection*{Source 3: Model Preference \& Rater Bias}
These choices countermeasure alignment-induced deference to stated beliefs, persona and framing dependence, family or self-preference, and rationale--label mismatch from unfaithful chain-of-thought. 
\textbf{Core controls.} Neutral persona by default; allow ``cannot judge'' responses; use evidence-then-score pipelines (L3 spans $\rightarrow$ L2 codes) when numbers are the target; constrain L1 to a single token; include canary items where superficial cues conflict with truth to detect deference patterns.
\textbf{Where to report.} Cross-model $\kappa$ and any abstain/tie rates; rationale consistency spot-checks. If persona is part of the construct, document and assess robustness to persona removal.
\subsubsection*{Source 4: Output Constraining \& Extraction}
This set of choices comprises decoding constraints, deterministic token$\rightarrow$label mapping, JSON-schema validation with bounded retries, probability logging, and post-hoc calibration.
\textbf{Core controls.} L1: \texttt{max\_tokens=1} with a fixed label map. L2: typed JSON with range checks and bounded retries; reject-on-fail \& resample. Log per-label probabilities when available; apply temperature or isotonic calibration on a held-out set if decisions depend on thresholds.
\textbf{Where to report.} Invalid-output rate, exact match/$\kappa$ per slot, Brier/log-loss with CIs, and calibrated operating thresholds if used.
\subsubsection*{Source 5: System \& Aggregation Robustness}
These choices countermeasure cross-model disagreement, provider and version drift, inference nondeterminism and numerical precision or device effects. It applies aggregation rules and calibration, and preregistration and audit sets.
\textbf{Core controls.} Use replicates $S$, paraphrases $P$, and at least two independent families $M$ when feasible. Pin provider, model name, version/date, precision, batch, and device; keep an audit set and re-run it on version changes. Preregister aggregation (majority vs.\ Dawid--Skene/GLAD) and calibration procedures.
\textbf{Where to report.} $\kappa$/Krippendorff's $\alpha$/ICC with bootstrap CIs; DS/GLAD vs.\ majority outcomes if applicable; version/precision manifests; drift audit tables comparing pre/post version metrics.

\subsection*{Sampling, Aggregation, and Cross-Model Triangulation}

The protocol operationalizes variance control through systematic sampling and aggregation procedures that parallel multi-rater reliability assessment in content analysis \parencite{Krippendorff2018Content}.
\textbf{Prompt ensemble and extraction.} Build $P\!\ge\!3$ meaning-preserving prompt templates tied to the rubric, with one question per template. Randomize option order for L1/L2 tasks and record permutations. Enforce output constraints appropriate to the level: \texttt{max\_tokens=1} with a deterministic token-to-label map for L1, typed JSON with schema validation for L2, and declared structure for L3. Store authoritative fields, prompt IDs, option orders, schema versions, decoding settings, and per-label probabilities where available.
\textbf{Replication and within-configuration aggregation.} For variance estimation, draw $S$ replicates at $T\!\approx\!1$ per prompt-model pair; for final annotation, use a single constrained draw at $T\!=\!0$ only after aggregation rules are locked. Aggregate within each prompt-model combination using majority voting for nominal labels or median/trimmed mean for numeric outputs, then aggregate across prompts to obtain model-level annotations. For numeric slots, z-score by prompt before averaging to remove prompt-specific scale effects.
\textbf{Cross-model triangulation.} Use at least $M\!\ge\!2$ independent model families when stakes allow, treating each family as an independent coder with potentially different biases. Apply the same prompt ensemble and sampling budget across all models to ensure comparability. For baseline aggregation, use majority voting for nominal labels and median for numeric outputs. When cross-model agreement is modest ($\kappa < 0.6$), employ noise-aware aggregation methods such as Dawid--Skene \parencite{Dawid1979ObserverErrorEM} or GLAD \parencite{Whitehill2009WhoseVoteCounts}, which jointly estimate annotator reliability and true labels. Report both per-model annotations and final aggregates, along with inter-model agreement ($\kappa$/ICC) with bootstrap confidence intervals. For pairwise or listwise tasks, convert annotations to pairwise wins and fit Bradley--Terry or Plackett--Luce models with randomized presentation order.
\textbf{Documentation requirements.} Record prompts and IDs, option orders, schema versions, decoding settings, environment specifications (provider, model family and name, version/date, numerical precision, batch size, device), per-slot probabilities, and both within-prompt and cross-prompt aggregates.

\subsection*{Aggregation and Downstream Estimation}
\label{subsec:aggregation-estimation}

When LLM annotations are used in downstream regressions, aggregation serves not only to improve reliability but also to strengthen estimation. \textcite{Ludwig2025LLMsEconometric} show that combining LLM annotations with validation data can produce more precise estimates than relying on validation data alone, with larger gains when annotation reliability is higher. By reducing annotation variance through $P$–$S$–$M$ ensembles, our protocol increases the value of limited validation samples. High and systematically correlated annotation error leads to inconsistent estimates, whereas smaller, approximately classical error mainly attenuates coefficients and can be corrected using standard methods. Our variance-reduction procedures shift annotation error toward this latter case. For estimation problems in which LLM annotations enter regression analyses, researchers should apply the full protocol and report reliability metrics with confidence intervals. A modest validation sample—typically 5–10\% of the corpus annotated by experts using the same codebook—is sufficient to implement debiasing procedures that both correct bias and improve precision. In this sense, the cost of ensemble annotation is offset by the smaller validation samples required for credible inference.

\subsection*{Calibration, Audit Sets, and Drift}

\textbf{Calibration.} When per-label probabilities are available, assess calibration using reliability diagrams and proper scoring rules (Brier score, log-loss) on a held-out set. Apply post-hoc calibration (temperature scaling, isotonic regression) if miscalibration is detected and decisions depend on probability thresholds. Report pre- and post-calibration metrics.
\textbf{Audit sets.} Maintain a time-boxed audit set---a fixed sample of items with known properties (e.g., human consensus labels or canary items with expected patterns). Re-run the audit set whenever provider, model version, precision, or decoding settings change. Compare agreement and calibration metrics pre/post change; pause or roll back if reliability degrades materially.
\textbf{Drift monitoring.} Pin provider (e.g., OpenAI, Anthropic), model name (e.g., gpt-4o), version/date, numerical precision (FP16/BF16/FP32), batch size, and device. Document any changes and their impact on audit set performance. For long-running studies, schedule periodic drift checks (e.g., quarterly).
\textbf{Hardware/precision effects.} Seemingly identical runs can diverge under low-precision inference (FP16/BF16 vs.\ FP32) or across hardware/batch configurations due to floating-point rounding cascades. When using local models or controlling inference hardware, document precision settings and device specifications. For API-based models, acknowledge that precision may vary and is typically not user-controllable; drift audits become especially important. Full technical details on FP16/BF16/TF32 effects are provided in Appendix D.

\subsection*{Agreement, Validity, and Uncertainty Reporting}

Fix the \emph{unit} (item/pair/list) and \emph{slot types}. Use chance-corrected agreement for nominal labels (Cohen/Fleiss $\kappa$) \parencite{Cohen1960Kappa}, weighted $\kappa$ or ICC for ordinal/numeric \parencite{fleiss1973equivalence}, Kendall's $W$ for multi-criteria concordance, and EM/F1 for set-valued fields. Bootstrap over items for CIs; use cluster bootstrap for grouped data \parencite{Efron1994Bootstrap}. For Level~3 text as the measure, report dispersion across summary ensembles (e.g., cosine similarity spread) and coverage of rubric elements. If Level~3 feeds Level~2, revert to L2 slot-wise metrics.

\subsection*{Human-in-the-Loop Thresholds and Preregistration}

Pre-register triggers and actions for human escalation. For nominal labels, cross-prompt or cross-model agreement below $\kappa{=}0.4$ should trigger escalation or design revision. Near-tie cases warrant review: for L1, items where the top-two log-odds gap is small; for L2, scores near decision cut-points; for L3, low ensemble similarity across summaries. Schema validation failures or pipeline stage errors should short-circuit directly to human review. Use blinded dual review with adjudication for escalated cases, and report overturn rates alongside human--LLM agreement using the same metrics applied to the main corpus.

\subsection*{Reporting Standards: A Checklist for LLM Annotation Studies}

Table~\ref{box:checklist} provides a distilled checklist for designing, conducting, and reporting LLM annotation studies. This checklist operationalizes the five-source framework and can be used both for preregistration and for review.\footnote{The checklist is analogous to established reporting standards in other methodological domains---STROBE for observational studies, PRISMA for systematic reviews, and CONSORT for randomized trials.} Journals adopting this standard can include checklist compliance as part of their submission requirements.

\begin{table}[htbp]
\centering
\caption{Reporting Checklist for LLM Annotation Studies}
\label{box:checklist}
\tiny

\begin{tabular}{@{}m{0.22\linewidth}@{} m{0.78\linewidth}@{}}

\textbf{Pre-Study Design (Codebook Development)} &
\begin{itemize}[leftmargin= 20pt, itemsep=0pt, topsep=0pt, parsep=0pt]
\item Define construct with behavioral anchors; choose L1/L2/L3 and scope (point/pair/list/set)
\item Specify authoritative vs.\ non-authoritative fields; declare L2 independence/dependence
\item Build prompt ensemble ($P \geq 3$) with fixed layout and randomized option order
\item Select sampling budget $(P,S,M)$ appropriate to construct complexity
\item Preregister aggregation rules (majority, DS/GLAD, median/trim) and calibration procedures
\item Establish audit set with expected patterns or human consensus labels
\end{itemize} \\

\textbf{During Data Collection (Reliability Engineering)} &
\begin{itemize}[leftmargin= 20pt, itemsep=0pt, topsep=0pt, parsep=0pt]
\item Log prompts, prompt IDs, option orders, and schema versions
\item Log models (provider/name/version/date/precision/batch/device)
\item Log decoding settings ($T$, top-$p$, \texttt{max\_tokens}, stop tokens)
\item Store per-label probabilities when available; store authoritative field outputs
\item Validate outputs against schema; log invalid-output rates and resample counts
\item Run audit set at study start; re-run on any model/precision change
\end{itemize} \\

\textbf{Reporting and Transparency} &
\begin{itemize}[leftmargin= 20pt, itemsep=0pt, topsep=0pt, parsep=0pt]
\item Report agreement: Cohen's $\kappa$/weighted $\kappa$/ICC with bootstrap 95\% CIs
\item Report calibration: Brier score, log-loss, reliability curves (pre- and post-calibration if applied)
\item Report cross-model agreement and flip rates; document any abstain/tie rules
\item Report drift checks: audit set metrics pre/post version changes
\item Provide prompt templates, schema definitions, and aggregation code in appendices or supplementary materials
\item Acknowledge limitations: scope conditions where LLM annotation may be inappropriate; residual uncertainty after variance controls
\end{itemize} \\

\textbf{Human-in-the-Loop Escalation (if applicable)} &
\begin{itemize}[leftmargin= 20pt, itemsep=0pt, topsep=0pt, parsep=0pt]
\item Preregister triggers: agreement floors (e.g., $\kappa < 0.4$), near-tie margins, schema/stage failures
\item Document human review workflow: blinded dual review with adjudication
\item Report overturn rates and human--LLM agreement using same metrics
\end{itemize} \\

\textbf{Downstream Estimation (if annotations enter regression)} &
\begin{itemize}[leftmargin= 20pt, itemsep=0pt, topsep=0pt, parsep=0pt]
\item Report validation sample size and selection procedure
\item Document debiasing procedure applied \parencite{Ludwig2025LLMsEconometric}
\item Report both na\"ive and corrected parameter estimates with standard errors
\end{itemize}

\end{tabular}

\end{table}

Table~\ref{tab:five-dims} synthesizes the five variance dimensions with failure signatures, minimal diagnostics, and controls/reporting requirements. This framework, combined with the level-specific defaults in Table~\ref{tab:annotation-design} and the checklist in Box~\ref{box:checklist}, provides a complete protocol for variance-aware LLM annotation in strategy research.

\begin{table}[htbp]
\centering
\caption{Five variance dimensions: signals, diagnostics, controls, reporting}
\label{tab:five-dims}
\tiny
\begin{tabular}{p{2.5cm}p{4.0cm}p{4.5cm}p{5.0cm}}
\toprule
\textbf{Dimension} & \textbf{Failure signatures} & \textbf{Minimal diagnostics (quick A/Bs)} & \textbf{Controls \& reporting} \\
\midrule
\textbf{(1) Construct \& rubric} &
Vague construct; missing behavioral anchors; mixing rationales into scores; undeclared dependence across L2 fields; scope mismatch; identity/leakage in items. &
Expert content review (Q--sort/CVI) on anchors; small human$\leftrightarrow$LLM calibration slice; L1 vs L2 vs L3 ablation to detect level mis-spec; transcript spot-checks; name/redaction probes. &
Construct map $\rightarrow$ behaviorally anchored rubric; declare L1/L2/L3 and scope; mark authoritative vs non-authoritative fields; anonymize/redact non-construct cues; validate item integrity; report MTMM and human$\leftrightarrow$LLM agreement. \\
\addlinespace
\textbf{(2) Interface \& context} &
A/B flips on order swap; longer or more polished text wins; mid-context evidence ignored; identity cues shift labels (halo effects); paraphrase brittleness. &
Order swap test; equal-length rewrite test; ``lost-in-the-middle'' placement test; identity ablation; paraphrase ensemble and $\kappa$ across paraphrases. &
Fixed layout; single-question prompts; option randomization; equalize verbosity; place rubric near the query; anonymize entities when not part of construct; report $\kappa$/$\kappa_{w}$ and flip rates. \\
\addlinespace
\textbf{(3) Model preference \& rater bias} &
Persona/framing flips; belief-matching (sycophancy); within-family ``wins''; rationale contradicts label (unfaithful composition). &
Persona flip test (neutral vs role); belief-flip probes; family panel comparison; rationale--label consistency audit on a sample. &
Neutral persona; allow ``cannot judge''; prefer evidence-then-score (L3$\rightarrow$L2) over free-form rationales; constrain L1 to \texttt{max\_tokens=1}; use canary items; report cross-model $\kappa$. \\
\addlinespace
\textbf{(4) Output constraining \& extraction} &
``Right answer, wrong score'' from parsing; invalid/ambiguous tokens; schema fill errors; uncalibrated probabilities; threshold flips near cut-points. &
Invalid-output rate on a schema; parser audits with seeded edge cases; replicate runs to estimate near-tie instability; reliability diagrams and Brier/ECE; threshold sweep. &
Constrain L1 to valid labels; deterministic token$\rightarrow$label map; JSON/grammar validation with bounded retries; log class probabilities; post-hoc calibration on a hold-out; publish EM/$\kappa$ and Brier with CIs. \\
\addlinespace
\textbf{(5) System \& aggregation} &
Cross-model disagreement; pre/post version shifts; FP16/BF16/TF32 flips; majority vs DS/GLAD deltas; unstable CIs. &
Pinned audit set re-run pre/post version; multi-family panel; precision/device toggles; compare majority vs DS/GLAD; bootstrap CIs. &
Replicates $S$, paraphrases $P$, families $M$; pin provider/name/version/date/precision/device; preregister aggregation and calibration; report $\kappa$/$\alpha$/ICC with bootstrap CIs; keep an audit set for drift. \\
\bottomrule
\end{tabular}
\end{table}

The detailed implementation runbook, including pseudocode for prompt ensembles, sampling loops, aggregation functions, and calibration procedures, is provided in Appendix A.

\section*{Discussion and Conclusion}
\label{sec:discussion}

Our results show that seemingly minor design choices can shift LLM annotation output by 12--85 percentage points, with uncertainty bands near 24 points, even in simple (Level~1) coding settings. These are not negligible fluctuations. They are measurement problems tied to construct specification, interface and context, model preferences, output constraints, and system choices. Treating LLM annotation as deterministic misstates how annotation output is produced and potentially engenders inference errors. Treating it as a measurement instrument enables reliable output and honest acknowledgment of uncertainty. Practical demonstrations of the variance-aware protocol applied to organizational settings (grant screening) and educational contexts (MBA case analysis) are provided in Appendix E. 

Our contribution extends their \textcite{Carlson2025LLMsAnnotate}'s  foundational framework by operationalizing sensitivity analysis into a systematic protocol with specific diagnostics, budgets, and reporting standards. Where they establish \emph{what} decisions matter (method selection, model selection, prompt engineering, cost considerations, validation), we specify \emph{how} to make those decisions in a variance-aware manner: behaviorally anchored rubrics that translate constructs into codable criteria; $P$--$S$--$M$ sampling budgets that expose and aggregate across sources of variation; noise-aware aggregation methods (Dawid--Skene, GLAD) that account for differential annotator reliability; and reporting standards that make procedures transparent and auditable. The volatility magnitudes we document underscore why operationalization matters. When 90\% of annotation outcomes are determined by model choice rather than substantive content, conclusions hinge on API selection rather than strategic phenomena. Variance-aware annotation converts ad hoc LLM use into portable, auditable measurement infrastructure that supports cumulative science. Published studies using LLM annotations without variance controls should be interpreted with caution. When constructs are underspecified, prompts uncontrolled, and single models used without replication, reported effects may reflect measurement artifacts rather than substantive phenomena. Reviewers and editors should require: (i) explicit construct maps and rubrics with behavioral anchors; (ii) documentation of prompt ensembles, sampling budgets, and model families; (iii) agreement metrics with confidence intervals; (iv) drift checks on audit sets; (v) acknowledgment of scope conditions where LLM annotation may be inappropriate. Our protocol does not eliminate all variance---nor should it, as legitimate construct complexity and item difficulty introduce meaningful variation. Rather, it surfaces variance, reduces what can be reduced through design, and reports the rest with explicit uncertainty. This shifts LLM-based annotation from ad hoc practice to portable, auditable measurement.

Our empirical results sharpen the boundary conditions identified in Section 2 . Following \textcite{Ludwig2025LLMsEconometric}, we distinguish between prediction problems, where text is used to forecast outcomes, and estimation problems, where LLM annotations enter inferential analyses. For prediction, our protocol improves the stability of out-of-sample performance across models and prompts, assuming no training data leakage. For estimation, the protocol reduces but does not eliminate measurement error; when annotations are used in hypothesis testing, validation data are needed to correct residual bias, as even small correlated errors can invalidate inference. Descriptive and exploratory uses require variance reduction and reliability reporting, but not necessarily formal validation.
These differences imply a graduated approach to validation: descriptive applications require variance reduction, predictive studies require the full protocol with out-of-sample evaluation, and causal or hypothesis-testing studies require both validation samples and debiasing procedures, with reporting of naïve and corrected estimates. 

Journals should establish reporting standards for LLM annotation studies analogous to existing standards for experiments (CONSORT), observational studies (STROBE), and systematic reviews (PRISMA). Authors should confirm adherence to the checklist in Table~\ref{box:checklist} in their submission letter or methods section. Our protocol can serve as a basis for journal guidelines. Just as journals require power analyses for experiments or robustness checks for regressions, they should require variance documentation for LLM-based measurements. Editors can request protocol compliance during desk review; reviewers can use the checklist to assess methodological rigor.

The reproducibility challenges that have affected psychology, medicine, and other fields arose partly from underpowered studies, flexible analysis, and inadequate reporting. LLM annotation without variance controls risks introducing analogous problems: results that depend on arbitrary prompt choices, model selection, or sampling configurations rather than substantive phenomena. By requiring variance-aware protocols, preregistration, and transparent reporting, journals can prevent LLM-based measurement from becoming a new source of irreproducibility in strategy research.

Several directions warrant further investigation. Building shared audit sets and calibration slices for common strategy constructs---stakeholder orientation, strategic foresight, innovation, organizational ambidexterity, dynamic capabilities---would enable cross-study comparison and cumulative validation. Community-maintained benchmarks analogous to shared datasets in machine learning could accelerate protocol standardization and provide reference points for reliability assessment, allowing researchers to report not only internal reliability but also agreement with established reference annotations.

Our protocol assumes LLM annotations are the primary measurement mode with human review reserved for escalation cases, but alternative hybrid designs may achieve better cost-quality trade-offs. Workflows where LLMs handle high-volume initial coding while humans adjudicate borderline or disagreement cases might combine the scale advantages of LLM annotation with the validity assurance of expert human judgment. Investigating optimal division of labor, escalation thresholds, and feedback loops between human and LLM coders remains an open question. We provide default budgets by annotation level, but optimal allocation may depend on construct complexity, item heterogeneity, corpus size, and available resources. Simulation studies varying prompt ensemble size, replication depth, and model panel breadth while measuring downstream inference quality could guide budget decisions and identify when additional prompt variants matter more than additional model families or when marginal returns to replication diminish.

Adopting a variance-aware protocol turns LLM annotation from ad hoc practice into portable measurement infrastructure. Substantively, constructs that were once costly to code at scale, become tractable without diluting standards. Methodologically, shared anchors, schemas, and checklists shift effort from prompt tinkering to reproducible procedures, enabling replication, extension, and meta-analysis. When strategy researchers measure variance, reduce what they can, and report what remains, LLM-based annotation becomes a credible scientific instrument rather than a source of untracked noise. The protocols we propose are not barriers to innovation but foundations for cumulative knowledge. By treating LLM annotation as a measurement problem and designing accordingly, strategy scholars can harness the speed and scale of AI while maintaining the rigor that enables inference and theory development.

Our goal is not to discourage the use of LLMs in strategy research but to elevate its practice. The magnitudes of variance we document underscore the stakes: without procedural safeguards, LLM-based annotations risk becoming a new source of irreproducibility in the field. With variance-aware protocols, preregistration, and transparent reporting, LLM annotation can extend the empirical toolkit of strategy research while preserving the epistemic standards that underpin scientific progress. The choice is not between speed and rigor, but between implicit variance and explicit measurement.
\begingroup
\doublespacing
\printbibliography
\endgroup

\clearpage
\newpage

\appendix
\setcounter{page}{1}

\singlespacing

\begin{center}
    \LARGE{Online Appendix: Variance-Aware LLM Annotation for Strategy Research: Sources, Diagnostics, and a Protocol for Reliable Measurement}
\end{center}

\doublespacing

\section*{Appendix A: Implementation Runbook for Variance-Aware LLM Annotation}
\label{app:A}
\label{app:protocol-runbook}

This appendix provides the complete operational implementation guide referenced in Section 5, including project structure, execution workflows, audit set construction, and reporting protocols. It provides step-by-step operational instructions for implementing the variance-aware protocol described in Section 5. It contains roles and responsibilities, project manifest structure, repository layout, execution steps, and a paper-facing methods table template.

\subsection*{Roles, Responsibilities, and Deliverables Within the Research Team}

LLM-based annotation studies typically require coordination across multiple roles:

\begin{itemize}[leftmargin=*,topsep=3pt,itemsep=2pt]
\item \textbf{Principal Investigator (PI)}: Owns construct map, rubric design, level selection (L1/L2/L3), and preregistered protocol. Approves final methods table and materials archive.
\item \textbf{Research Engineer (RE)}: Owns environment manifests, version pinning, schema enforcement, run orchestration, log collection, and drift monitoring. Implements sampling loops and aggregation functions.
\item \textbf{Research Analyst (RA)}: Owns artifact preparation (item cleaning, redaction, validation), prompt paraphrase generation, audit set curation, quality spot-checks, and escalation packet assembly.
\item \textbf{Human Reviewers/Adjudicators}: Engaged only on escalated items under preregistered triage policy. Conduct blinded dual review with adjudication for ties.
\end{itemize}

\subsection*{Project Manifest (Versioned with Repository)}

Create \texttt{manifest.yaml} capturing everything needed to reproduce the annotation. This file is read by orchestration scripts and archived with materials.

\begin{verbatim}
project:
  title: "LLM-as-Annotator: <Study short name>"
  run_id: "<YYYY-MM-DD>_<short-tag>"
  data_root: "data/"
  out_root: "runs/<run_id>/"

design:
  level: <1|2|3|pipeline>   # annotation level (see Table~\ref{tab:levels})
  scope: <pointwise|pairwise|listwise|setwise>
  rubric_id: "rubrics/v1.json"
  label_map_id: "labels/v1.json"
  schema_id: "schemas/<level>_v1.json"
  prompts_id: "prompts/<construct>_P3.toml"
  randomize_options: true
  P: 3                      # prompt ensemble size
  S: 20                     # samples per prompt
  M: 2                      # model families

environment:
  providers:
    - {name: "OpenAI", model: "gpt-4o", version: "2025-08",
       precision: "default", device: "cloud", notes: "pinned"}
    - {name: "Anthropic", model: "claude-sonnet-4", version: "2025-08",
       precision: "default", device: "cloud", notes: "pinned"}
  decoding:
    temperature: {estimation: 1.0, final: 0.0}
    top_p: 1.0
    max_tokens: 1   # for categorical L1
  seeds: {collection: 1234, shuffling: 5678}

inputs:
  items_path: "data/items.jsonl"
  audit_set_path: "data/audit.jsonl"
  gold_items_path: "data/gold.jsonl"  # optional

outputs:
  logs_dir: "runs/<run_id>/logs/"
  dumps_dir: "runs/<run_id>/raw/"
  aggregates_dir: "runs/<run_id>/agg/"
  methods_table_tex: "runs/<run_id>/methods_table.tex"
  materials_bundle: "runs/<run_id>/materials.zip"

triage:
  policy_id: "policies/triage_v1.md"
  reviewers_roster: "human/reviewers.csv"
\end{verbatim}

\subsection*{Repository Layout (Conventional, Minimal)}

\begin{verbatim}
.
|-- rubrics/         # construct map and behavioral anchors
|-- labels/          # deterministic token<->label maps
|-- schemas/         # JSON/EBNF schemas for L2/L3/pipeline
|-- prompts/         # P paraphrases with IDs and randomization rules
|-- policies/        # preregistered triage policy
|-- data/
|   |-- items.jsonl        # source artifacts to judge
|   |-- audit.jsonl        # frozen audit set
|   `-- gold.jsonl         # optional human consensus labels
|-- runs/<run_id>/
|   |-- logs/              # per-call logs with probabilities
|   |-- raw/               # raw outputs per (prompt, model)
|   |-- agg/               # aggregated decisions and escalations
|   `-- materials.zip      # archive for publication
`-- scripts/
    |-- validate.py        # pre-flight manifest validation
    |-- collect.py         # sampling loop orchestration
    |-- aggregate.py       # within/across prompt/model aggregation
    `-- export.py          # methods table and materials bundle
\end{verbatim}

\subsection*{Pre-Flight Checklist}

Before full data collection:
\begin{enumerate}[leftmargin=*,topsep=3pt,itemsep=2pt]
\item \textbf{Validate manifests}: Run \texttt{scripts/validate.py --manifest manifest.yaml} to ensure all IDs resolve (rubric, labels, schema, prompts) and $P,S,M$ are specified.
\item \textbf{Schema dry-run}: Test schema validator on 10 dummy outputs per level/slot; confirm reject-on-fail is active.
\item \textbf{Audit set freeze}: Verify \texttt{data/audit.jsonl} covers edge cases; store hash for audit trail.
\item \textbf{Minimal pilot}: Run 20 items $\times$ 1 prompt $\times$ 1 model at estimation temperature; inspect raw dumps for schema cleanliness and option randomization.
\end{enumerate}

\subsection*{Audit Set Construction: Design Principles and Best Practices}
\label{subsec:audit-set-construction}

\subsubsection*{Purpose and Function}

Audit sets serve as frozen benchmarks for detecting annotation drift when models, providers, or decoding parameters change. Unlike validation sets (used for calibration) or gold-standard slices (used to measure accuracy against human consensus), audit sets monitor \emph{temporal stability}. The core principle: re-running the same audit items with the same prompts should yield similar distributions if the annotation process has not drifted.

\subsubsection*{Size Recommendations}

Minimum viable audit set sizes depend on annotation level and label cardinality:

\begin{itemize}[topsep=2pt,itemsep=1pt]
\item \textbf{L1 binary/categorical (small $K$):} 30--50 items provide sufficient power to detect $|\Delta\kappa| > 0.05$ with reasonable confidence.
\item \textbf{L1 ordinal or multi-class ($K > 5$):} 50--100 items to ensure adequate representation of all label classes.
\item \textbf{L2 multi-field:} 50--100 items $\times$ number of fields. For example, if evaluating 5 criteria, audit set should contain 50--100 complete item annotations.
\item \textbf{L3 extractive:} 30--100 items, depending on corpus heterogeneity. Longer texts may require fewer items; short texts more.
\end{itemize}

Recommended default: \textbf{100 items} for most use cases, stratified across label distribution and item characteristics.

\subsubsection*{Selection Strategy}

\paragraph{Representativeness.}
Audit sets must reflect the corpus distribution along key dimensions:
\begin{itemize}[topsep=2pt,itemsep=1pt]
\item \textbf{Label distribution:} Stratified sampling ensures all labels appear in proportions similar to full corpus.
\item \textbf{Text characteristics:} Include range of document lengths, genres, or domains if corpus is heterogeneous.
\item \textbf{Temporal spread:} If corpus spans time periods, sample across periods to avoid cohort effects.
\end{itemize}

\paragraph{Edge cases and boundary items.}
Deliberately include:
\begin{itemize}[topsep=2pt,itemsep=1pt]
\item \textbf{Prototypical examples:} Clear instances of each label class that should be consistently annotated.
\item \textbf{Near-boundary items:} Cases where human coders might disagree or where label assignment is non-obvious. These are sensitive to model drift.
\item \textbf{Construct edge cases:} Items that test rubric boundaries (e.g., very high or very low scores).
\end{itemize}

Ratio recommendation: 60\% prototypical, 30\% near-boundary, 10\% edge cases.

\paragraph{Human consensus labels (optional but recommended).}
If resources permit, obtain human consensus labels for audit set items through dual-coding with adjudication. This enables:
\begin{itemize}[topsep=2pt,itemsep=1pt]
\item Absolute accuracy checks (not just drift detection).
\item Assessment of whether LLM annotations regress toward or away from human consensus over time.
\end{itemize}

If human labels are unavailable, track \emph{relative stability}: distributions should remain consistent even if absolute accuracy is unknown.

\subsubsection*{Freezing and Versioning}

\paragraph{Immutability.}
Once constructed, audit sets must never be modified. Any changes (adding items, correcting labels, updating text) invalidate temporal comparisons. If audit set needs revision, create a new version with distinct identifier and discard historical comparisons.

\paragraph{Version control.}
Store audit set with:
\begin{itemize}[topsep=2pt,itemsep=1pt]
\item Unique identifier: \texttt{audit\_v1\_YYYY-MM-DD}
\item SHA-256 hash of file contents for tamper detection
\item Metadata: creation date, corpus source, sampling procedure, human labels (if any)
\end{itemize}

\paragraph{Persistence.}
Archive audit sets with study materials. Published papers should report audit set size, composition strategy, and hash for reproducibility.

\subsubsection*{Drift Monitoring Protocol}

\paragraph{Trigger conditions for re-running audit set.}
Re-run audit set whenever any of the following change:
\begin{itemize}[topsep=2pt,itemsep=1pt]
\item Provider or model family (e.g., switch from GPT-4o to Claude-Sonnet-4)
\item Model version (e.g., GPT-4o updated by OpenAI)
\item Decoding precision (e.g., FP32 to FP16)
\item Inference hardware/device (relevant for local models)
\item Temperature or other decoding parameters
\end{itemize}

\paragraph{Decision rules.}
Preregister thresholds for acceptable drift. Recommended defaults:
\begin{itemize}[topsep=2pt,itemsep=1pt]
\item $|\Delta\kappa| < 0.05$: Pass (continue with new configuration)
\item $0.05 \le |\Delta\kappa| < 0.10$: Warning (investigate causes, consider recalibration)
\item $|\Delta\kappa| \ge 0.10$: Fail (do not proceed; diagnose or rollback)
\end{itemize}

For continuous metrics (ICC, Brier score): similar thresholds.

\paragraph{Logging.}
Maintain drift log in project manifest:
\begin{verbatim}
drift_audits:
  - date: 2025-10-15
    trigger: "Model updated to gpt-4o-2025-10"
    baseline_kappa: 0.68
    new_kappa: 0.65
    delta_kappa: -0.03
    decision: "PASS"
  - date: 2025-11-01
    trigger: "Switched to claude-sonnet-4"
    baseline_kappa: 0.68
    new_kappa: 0.52
    delta_kappa: -0.16
    decision: "FAIL - rolled back to gpt-4o"
\end{verbatim}

\subsubsection*{Example: Audit Set for Strategic Clarity Annotation}

\paragraph{Corpus:} 500 earnings call transcripts, annotated for strategic clarity (5-point ordinal scale).

\paragraph{Audit set design:}
\begin{itemize}[topsep=2pt,itemsep=1pt]
\item Size: 100 transcripts
\item Sampling: Stratified by clarity score (20 per score level) and company size (50 large-cap, 30 mid-cap, 20 small-cap)
\item Boundary items: 30 transcripts with scores near decision thresholds (e.g., between "2" and "3")
\item Edge cases: 10 transcripts with unusual characteristics (highly technical jargon, non-English phrases, very short/long)
\item Human labels: Dual-coded by two domain experts with adjudication; $\kappa_{\text{human}} = 0.72$
\end{itemize}

\paragraph{Versioning:}
\begin{itemize}[topsep=2pt,itemsep=1pt]
\item Identifier: \texttt{strategic\_clarity\_audit\_v1\_2025-08-01}
\item SHA-256: \texttt{a3f2...b9c1}
\item Stored in: \texttt{data/audit\_sets/strategic\_clarity\_v1.jsonl}
\end{itemize}

\paragraph{Drift monitoring:}
\begin{itemize}[topsep=2pt,itemsep=1pt]
\item Baseline (August 2025): $\kappa$ with human consensus = 0.61
\item October 2025 (model update): $\kappa$ = 0.58 ($|\Delta\kappa| = 0.03$, PASS)
\item November 2025 (precision change): $\kappa$ = 0.53 ($|\Delta\kappa| = 0.08$, WARNING, recalibrated temperature)
\end{itemize}

\subsubsection*{Summary}

Audit sets are not optional; they are essential infrastructure for detecting when annotation processes drift. By freezing a representative sample at study outset and monitoring it systematically, researchers can ensure that LLM annotations remain stable even as underlying technologies evolve. Audit set construction should be documented in preregistration and reported transparently in methods sections.

\subsection*{Execution Workflow}

\subsubsection*{Data Collection}

For each item $x$, for each prompt variant $p^{(u)}$ ($u \in \{1,\dots,P\}$), for each model family $m$ ($m \in \{1,\dots,M\}$):
\begin{enumerate}[leftmargin=*,topsep=3pt,itemsep=2pt]
\item Draw $S$ independent samples at estimation temperature ($T{\approx}1$) with constrained decoding and randomized option order.
\item Persist authoritative fields, per-label log-probabilities (if available), prompt ID, option permutation, seed, and timestamp.
\item Write raw dumps to \texttt{runs/<run\_id>/raw/} in one file per $(u,m)$ combination.
\end{enumerate}

\subsubsection*{Within-Prompt Aggregation}

For each $(u,m)$ combination:
\begin{enumerate}[leftmargin=*,topsep=3pt,itemsep=2pt]
\item Collapse $S$ samples to a single decision per slot using appropriate aggregation:
  \begin{itemize}[topsep=2pt,itemsep=1pt]
  \item Categorical slots: majority vote
  \item Ordinal/numeric slots: median or trimmed mean
  \item Set-valued slots: union with frequency weights
  \end{itemize}
\item Retain per-slot uncertainty objects (vote distributions, quartiles, ensemble variance).
\end{enumerate}

\subsubsection*{Across-Prompts and Across-Models Aggregation}

\begin{enumerate}[leftmargin=*,topsep=3pt,itemsep=2pt]
\item Aggregate across prompts $u$ for each model $m$ to obtain one model-level decision per item. For numeric slots, z-score by prompt before robust averaging.
\item Aggregate across models $m$ to the study-level decision per item:
  \begin{itemize}[topsep=2pt,itemsep=1pt]
  \item Baseline: majority vote (categorical) or median (numeric)
  \item When agreement is low ($\kappa < 0.4$): apply Dawid--Skene or GLAD to learn rater reliabilities; store posteriors alongside hard labels
  \end{itemize}
\end{enumerate}

\subsubsection*{Triage and Human Review}

\begin{enumerate}[leftmargin=*,topsep=3pt,itemsep=2pt]
\item Auto-generate \texttt{runs/<run\_id>/agg/escalations.csv} listing items meeting preregistered triggers (low agreement, near-ties, schema failures).
\item Materialize reviewer packets in \texttt{runs/<run\_id>/agg/review\_kits/}: item text, rubric excerpt, blinded to LLM output.
\item RA files escalations; two blinded reviewers render independent decisions; adjudicator resolves ties.
\item RE merges human outcomes into aggregated table and updates exports.
\end{enumerate}

\subsubsection*{Exports}

\begin{enumerate}[leftmargin=*,topsep=3pt,itemsep=2pt]
\item Render paper-facing methods table (template in \S\ref{subsec:methods-table-template}) to \texttt{methods\_table.tex}.
\item Archive prompts, schemas, manifest, and de-identified sample of items/outputs into \texttt{materials.zip}.
\end{enumerate}

\subsection*{Drift Monitoring (Operational Cadence)}

On any provider/model/version/precision change:
\begin{enumerate}[leftmargin=*,topsep=3pt,itemsep=2pt]
\item Re-run frozen audit set with unchanged prompts/schemas.
\item Log deltas in agreement metrics ($\kappa$, ICC) and calibration (Brier score).
\item Apply preregistered decision rule: continue if $|\Delta\kappa| < 0.05$, otherwise pause for diagnosis or rollback.
\item Update manifest with new version/date and drift audit results.
\end{enumerate}

\subsection*{Paper-Facing Methods Table Template}
\label{subsec:methods-table-template}

\begin{table}[H]
\centering
\caption{Methods Table Template for LLM Annotation Studies (Author-Fillable)}
\label{tab:methods-template}
\begin{tabular}{p{0.27\linewidth} p{0.69\linewidth}}
\toprule
\textbf{Element} & \textbf{Specification} \\
\midrule
Construct \& Measure & Construct definition; dimensionality; reflective vs.\ formative; rubric source/anchors. \\
Annotation Design & Level (L1/L2/L3/pipeline); scope (pointwise/pairwise/listwise/setwise); evidence binding; authoritative vs.\ non-authoritative fields. \\
Constraints \& Schema & Label map; output grammar/JSON; max tokens; invalid-output policy (reject-on-fail with bounded retries). \\
Sampling Plan & $P$ prompts (paraphrase ensemble); $S$ samples/prompt (estimation); $M$ model families (triangulation). \\
Decoding \& Pinning & Temperature (estimation/final), top-$p$, stop tokens; provider/model/version/date/precision; device/hardware notes. \\
Aggregation & Within-prompt rule (majority/median/trim); across-prompts; across-models; noise-aware model if used (DS/GLAD). \\
Agreement Metrics & Slot-wise metrics (Cohen's $\kappa$/weighted $\kappa$/ICC/F1) and unit (item/pair/ranking); CI method (bootstrap/cluster). \\
Calibration & Held-out split; method (temperature/isotonic scaling); pre/post Brier score or reliability curves. \\
Triage Policy & Triggers (pointer to preregistration); human workflow (blinding, adjudication); \% escalated/\% overturned. \\
Drift Audits & Audit-set composition; cadence; pass/fail thresholds; rollback/rescore policy. \\
Materials & Bundle contents (prompts, schemas, manifest, de-identified sample); run date; repository hash. \\
\bottomrule
\end{tabular}
\end{table}

\subsection*{Checklist for Authors and Reviewers}

This appendix provides an expanded, preregistration-ready version of the distilled checklist presented in Box~\ref{box:checklist} in the main text. The checklist operationalizes the five-source variance framework for both planning (preregistration) and reporting (submission). Authors should use this expanded version during study design and adapt Box~\ref{box:checklist} for inclusion in their Methods section or cover letter.

\label{subsec:expanded-checklist}

This checklist operationalizes the protocol for preregistration (planning) and submission (reporting).

\subsubsection*{For Preregistration (Planning \& Ex Ante Specification)}

\begin{enumerate}[leftmargin=*,topsep=3pt,itemsep=2pt]
  \item \textbf{Design choices and scope}
  \begin{itemize}[topsep=2pt,itemsep=1pt]
    \item[$\square$] Evaluation level (L1/L2/L3/pipeline) and scope (pointwise/pairwise/listwise/setwise) selected from Table~\ref{tab:levels}.
    \item[$\square$] Construct definition (boundaries, indicators, dimensionality) and reflective vs.\ formative specification.
    \item[$\square$] Decision grain (ranking/grading/scoring) and unit of analysis (item/pair/list).
  \end{itemize}

  \item \textbf{Rubric, labels, and task form}
  \begin{itemize}[topsep=2pt,itemsep=1pt]
    \item[$\square$] Analytic rubric with behaviorally anchored levels and explicit edge cases.
    \item[$\square$] Label set $\mathcal{Y}$ precisely specified (including ordering if ordinal); gold-probe items for drift checks.
  \end{itemize}

  \item \textbf{Prompt templates and layout control}
  \begin{itemize}[topsep=2pt,itemsep=1pt]
    \item[$\square$] One-question paraphrased prompt ensemble ($P\!\ge\!3$) with stable layout and randomized option order.
    \item[$\square$] Verbatim system/user templates stored and versioned.
  \end{itemize}

  \item \textbf{Output constraints and extraction}
  \begin{itemize}[topsep=2pt,itemsep=1pt]
    \item[$\square$] Constrained decoding to schema/label map (e.g., \texttt{max\_tokens=1} for L1; JSON schemas for L2/L3).
    \item[$\square$] Deterministic token$\leftrightarrow$label map; invalid-output policy (reject-on-fail with bounded resample) prespecified.
    \item[$\square$] Free-text rationales (if collected) treated as non-authoritative and stored separately.
  \end{itemize}

  \item \textbf{Models, decoding, and environment pinning}
  \begin{itemize}[topsep=2pt,itemsep=1pt]
    \item[$\square$] Provider, family, model name, version/date/precision pinned for each judge.
    \item[$\square$] Decoding settings (temperature, top-$p$, stop tokens) prespecified for estimation and final passes.
  \end{itemize}

  \item \textbf{Sampling, aggregation, and agreement plan}
  \begin{itemize}[topsep=2pt,itemsep=1pt]
    \item[$\square$] Sampling budgets: prompts $P$, samples per prompt $S$, model families $M$ (from Table~\ref{tab:annotation-design}).
    \item[$\square$] Within-prompt and across-prompt aggregation rules (vote/median/trim; prompt-wise standardization for L2 numeric slots).
    \item[$\square$] Agreement/validity metrics matched to slot type (EM/$\kappa$/ICC/F1); CI method (bootstrap/cluster).
  \end{itemize}

  \item \textbf{Calibration and uncertainty}
  \begin{itemize}[topsep=2pt,itemsep=1pt]
    \item[$\square$] Held-out split and method (temperature/isotonic scaling); operating cut-points for decisions.
    \item[$\square$] Logging plan for per-label probabilities to support Brier/log-loss and reliability curves.
  \end{itemize}

  \item \textbf{Human-in-the-loop triage \& review}
  \begin{itemize}[topsep=2pt,itemsep=1pt]
    \item[$\square$] Triage triggers by level (margin floors, low inter-prompt/model agreement, schema/stage failures).
    \item[$\square$] Roles \& blinding: two independent reviewers and adjudicator; decision log and disposition codes.
    \item[$\square$] Sampling of non-escalated items for quality control (if applicable).
  \end{itemize}

  \item \textbf{Drift watch and audit set}
  \begin{itemize}[topsep=2pt,itemsep=1pt]
    \item[$\square$] Stable audit set composition/size; cadence; version/precision pinning; rollback policy.
    \item[$\square$] Re-run policy upon provider/model/version changes; criteria for pause/diagnose/rescore.
  \end{itemize}

  \item \textbf{Project manifest and artifacts}
  \begin{itemize}[topsep=2pt,itemsep=1pt]
    \item[$\square$] \texttt{manifest.yaml} includes: level/scope; IDs for rubric, label map, schema, prompts; flags (randomize\_options); $P,S,M$; providers and decoding; paths for data/outputs.
    \item[$\square$] Directory layout for reproducible runs (\texttt{data/}, \texttt{runs/<run\_id>/}, etc.).
  \end{itemize}
\end{enumerate}

\subsubsection*{For Submission (Reporting \& Archiving)}

\begin{enumerate}[leftmargin=*,topsep=3pt,itemsep=2pt]
  \item \textbf{Methods table \& departures from defaults}
  \begin{itemize}[topsep=2pt,itemsep=1pt]
    \item[$\square$] One-page methods table (template above) covering: construct/rubric; level \& constraints; $P,S,M$; decoding; aggregation; agreement/validity metrics with 95\% CIs; calibration; drift plan; provider/model/version/precision with run date.
    \item[$\square$] Deviations from Table~\ref{tab:levels} defaults stated and justified; preregistration identifier provided.
  \end{itemize}

  \item \textbf{Agreement, validity, and uncertainty}
  \begin{itemize}[topsep=2pt,itemsep=1pt]
    \item[$\square$] Metrics per slot type/level (EM/$\kappa$ for nominal; weighted $\kappa$/ICC for ordinal; per-slot EM/F1 for structured; stage-wise + end-to-end for pipelines) with CI method documented.
    \item[$\square$] Content-validity checks via rubric alignment (schema adherence, cross-field checks).
    \item[$\square$] Calibration results: pre/post Brier or log-loss; reliability curves; operating cut-points.
  \end{itemize}

  \item \textbf{Human-in-the-loop outcomes}
  \begin{itemize}[topsep=2pt,itemsep=1pt]
    \item[$\square$] Report \% escalated, overturn rate, clustering (by prompt/model/criterion/stage); human--human and human--LLM agreement with 95\% CIs.
    \item[$\square$] Document how human decisions were integrated into aggregates.
  \end{itemize}

  \item \textbf{Drift monitoring}
  \begin{itemize}[topsep=2pt,itemsep=1pt]
    \item[$\square$] Drift-audit manifest (audit items, dates, versions/precision); any recalibration or threshold updates; actions taken.
  \end{itemize}

  \item \textbf{Materials and reproducibility}
  \begin{itemize}[topsep=2pt,itemsep=1pt]
    \item[$\square$] Archive: prompts (with IDs), schemas, label maps, manifest, decoder settings, seeds/permutations, per-slot probabilities; minimal de-identified sample.
    \item[$\square$] Rendered methods table and \texttt{materials.zip} with artifacts.
  \end{itemize}
\end{enumerate}

\section*{Appendix B: Empirical Design, Stimuli, and Full Results}
\label{app:B}
\label{app:empirical-details}

This appendix provides complete details for the empirical demonstrations in Section 4, including verbatim stimuli, prompt variants, model specifications, and extended statistical results.

\subsection*{Business Model Descriptions (Verbatim)}

\begin{quote}
\textbf{Business Model A:} Targets parents of young children who want to dress their kids in trendy, comfortable clothing made from safe, eco-friendly materials. Offers an online boutique for children's wear focusing on trendy designs and eco-friendly materials, with a subscription service for regular delivery of age-appropriate clothes. Main activities include selecting eco-friendly children's wear, maintaining an online boutique, managing a subscription service, coordinating regular deliveries, and handling customer service inquiries.
\end{quote}

\begin{quote}
\textbf{Business Model B:} Targets environmentally conscious millennials who value sustainability in their clothing choices and are willing to spend extra for ethically made clothes using organic or recycled materials. Offers a range of sustainable clothing items including everyday essentials, workout gear, and fashionable outfits, all designed with minimalistic and timeless style. Main activities include sourcing sustainable fabrics, designing clothes, manufacturing, selling directly through an online platform, and partnering with environmental organizations for tree planting initiatives.
\end{quote}

\subsection*{Prompt Variants for Semantic-Preserving Rephrasing}
\label{subsec:paraphrase-variants}

\begin{table}[H]
\centering
\caption{Semantic-Preserving Prompt Reformulations (Section~\ref{subsubsec:prompt-brittleness})}
\label{tab:paraphrase-variants}
\begin{tabular}{p{5cm}p{11cm}}
\toprule
\textbf{Variation Type} & \textbf{Reformulated Question} \\
\midrule
Original & ``From your viewpoint as an investor in this startup, which \textbf{business model} is more \textbf{likely to succeed}?" \\
Pronoun substitution & ``From your viewpoint as an investor in \textbf{it}, which \textbf{one} is more likely to succeed?" \\
Synonym substitution & ``From your viewpoint as an investor in this startup, which business model is more \textbf{probable to succeed}?" \\
Nominalization & ``From your viewpoint as an investor in this startup, which business model \textbf{has the greater likelihood of success}?" \\
Clause restructuring & ``\textbf{Which business model is more likely to succeed} from your viewpoint as an investor in this startup?" \\
\bottomrule
\end{tabular}
\end{table}

\subsection*{Implementation Details}

\paragraph{Models and time window.}
All empirical analyses use production API models accessed through provider SDKs. We pin model names and calendar window but do not control underlying hardware or inference stack. Models used:
\begin{itemize}[topsep=2pt,itemsep=1pt]
\item OpenAI: \texttt{gpt-4o}, \texttt{o4-mini}, \texttt{GPT-5 reasoning}, \texttt{GPT-5-mini}
\item Anthropic: \texttt{Claude-3.5-Sonnet}, \texttt{Claude-Sonnet-4}
\item Google: \texttt{Gemini-2.0-Flash}, \texttt{Gemini-2.5-Pro}
\end{itemize}
Level~1 analyses (construct ambiguity, context, rephrasing, sampling, endorsement, cross-model) executed August 2025. Level~1 format and Level~2 analyses executed October 2025.

\paragraph{Decoding and output constraints.}
Unless noted, all conditions use temperature $T{=}1$. Other parameters at provider defaults. We instruct models to output single admissible labels:
\begin{itemize}[topsep=2pt,itemsep=1pt]
\item Binary forced choice (L1): "Answer only 'A' or 'B'" with \verb|max_tokens=1|
\item Likert (5-point): "Answer only with integer 1--5" with \verb|max_tokens=1|
\item Probability (0--100): "Return single integer 0--100" with small token budget
\end{itemize}
Deterministic token-to-label map applied based on first non-whitespace token. Each API call treated as independent draw.

\paragraph{Logged probabilities.}
For binary $\{\mathrm{A},\mathrm{B}\}$ outcomes we request per-token log-probabilities and reconstruct per-label log-probabilities by summing over label tokens. Analyses compare induced Bernoulli distributions using $t$-tests and ANOVA. For percentage/probability responses we use Welch's $t$-test and 1-Wasserstein distance.

\paragraph{Sampling design and replications.}
Level~1 manipulations: $n{=}20$ repeated evaluations per condition and model. Level~2 manipulations: $n{=}30$ repeated evaluations per condition and model. Artifacts (business-model descriptions) held fixed; only prompts, instructions, and models vary.

\paragraph{Order and position effects.}
For Level~1 order sensitivity (Section~\ref{subsubsec:l1-order}): First run $n{=}20$ evaluations with baseline order (A first, then B), then swap presentation order while keeping label set $\{\mathrm{A},\mathrm{B}\}$ unchanged. Run another $n{=}20$ evaluations and compare distributions, aligning probabilities as $p(A)$ vs.\ $1{-}p(A)$.

For Level~2 position (Section~\ref{subsubsec:l2-order}): Ten criteria (0--100 scale) presented in base order vs.\ re-ordered sequence (high to low). For each model, $n{=}30$ samples per order and criterion; compare using Welch's $t$-test and 1-Wasserstein distance.

\paragraph{Sampling variance experiment.}
Section~\ref{subsubsec:sampling-variance} uses \texttt{gpt-4o} on baseline investor prompt. Repeatedly query $N{=}100$ times under identical conditions. Log-probabilities show $P(A)$ ranges from 0.68 to 0.92. Replicate at $T{=}0$; 24-point range persists due to hardware nondeterminism.

\paragraph{Cross-model disagreement.}
Section~\ref{subsubsec:cross-model} compares six models on 20 binary comparisons. Each model produces 20 $\{\mathrm{A},\mathrm{B}\}$ choices (four API failures discarded), yielding 120 usable observations. Compute per-model distributions, test differences by one-way ANOVA, run pairwise $t$-tests, quantify inter-model agreement with Cohen's $\kappa$.

\paragraph{Level~2 separate vs.\ joint and reasoning.}
Score one venture description on ten criteria (0--100 scale): Innovation \& Value Proposition, Product \& Services, Market Opportunity \& Strategy, Team, Execution Plan, Financials, Structure of Venture Summary, Sustainable Competitive Advantage, Potential to Invest, Viability. Separate condition: each criterion in own prompt. Joint condition: all ten in single prompt with tabular schema. For each of four models, $n{=}30$ samples per condition and criterion; compare using Welch's $t$-test and 1-Wasserstein distance.

Reasoning manipulation (Section~\ref{subsubsec:l2-reasoning}) uses \texttt{GPT-5 reasoning}: low-reasoning (brief justification) vs.\ high-reasoning (explicit step-by-step). Each condition sampled $n{=}120$ times; compare 0--100 scores with Welch's $t$-test and 1-Wasserstein distance.

\paragraph{Statistical tests and aggregation.}
Omnibus tests: ANOVA with Tukey's HSD at $\alpha{=}0.05$. Pairwise comparisons: Welch's $t$-test. Distributional effect size: 1-Wasserstein distance. Agreement: Cohen's $\kappa$ (nominal), weighted $\kappa$ (ordinal), with bootstrap CIs.

\subsection*{Agreement Statistics Definition}

Cohen's $\kappa$ \parencite{Cohen1960Kappa} for two-rater categorical outcomes and weighted $\kappa$ for ordinal labels:
\[
\kappa \;=\; \frac{P_o - P_e}{1 - P_e}, \qquad
P_o \;=\; \sum_{i=1}^{K} p_{ii}, \qquad
P_e \;=\; \sum_{i=1}^{K} p_{i\cdot}\,p_{\cdot i},
\]
where $P_o$ is observed agreement and $P_e$ is chance agreement from marginal label frequencies. Weighted $\kappa$ uses distance weights for ordinal disagreements. Bootstrap CIs computed over items with 10,000 resamples.

\subsection*{Extended Results: Level 2 Re-ordering}

Table~\ref{tab:l2-reorder-full} provides complete results for the Level~2 criterion re-ordering experiment (Section~\ref{subsubsec:l2-order}).

\begin{table}[H]
\centering
\tiny
\caption{Level 2 re-ordering: base vs.\ anchored sequence (30 draws/condition)}
\label{tab:l2-reorder-full}
\setlength{\tabcolsep}{3.5pt}
\renewcommand{\arraystretch}{0.95}
\begin{tabular}{l *{4}{cc}}
\toprule
& \multicolumn{2}{c}{\textbf{GPT-5 (reason.)}} & \multicolumn{2}{c}{\textbf{GPT-5-mini}} & \multicolumn{2}{c}{\textbf{gpt-4o}} & \multicolumn{2}{c}{\textbf{gpt-4o-mini}} \\
\cmidrule(lr){2-3}\cmidrule(lr){4-5}\cmidrule(lr){6-7}\cmidrule(lr){8-9}
\textbf{Criterion} & \textit{Welch-$p$} & \textit{Wasserstein} & \textit{Welch-$p$} & \textit{Wasserstein} & \textit{Welch-$p$} & \textit{Wasserstein} & \textit{Welch-$p$} & \textit{Wasserstein} \\
\midrule
Team & 0.000 & 4.17 & 0.000 & 6.07 & 0.000 & 7.83 & 0.000 & 12.33 \\
Execution Plan & 0.000 & 3.20 & 0.921 & 0.40 & 0.000 & 4.17 & 0.049 & 1.67 \\
Financials & 0.000 & 4.50 & 0.099 & 1.63 & 0.000 & 7.17 & 0.000 & 11.83 \\
Structure of Venture Summary & 0.887 & 0.57 & 0.198 & 1.03 & 0.001 & 5.17 & 0.000 & 6.50 \\
Innovation \& Value Proposition & 0.004 & 1.67 & 0.003 & 2.33 & 0.000 & 4.00 & 0.000 & 3.50 \\
Market Opportunity \& Strategy & 0.078 & 1.43 & 0.000 & 3.47 & 0.000 & 7.93 & 0.000 & 6.67 \\
Sustainable Competitive Advantage & 0.177 & 0.80 & 0.965 & 0.50 & 0.000 & 11.23 & 0.000 & 7.33 \\
Product \& Services & 0.017 & 1.40 & 0.000 & 3.57 & 0.000 & 6.93 & 0.000 & 8.50 \\
Potential to Invest & 0.874 & 0.70 & 0.296 & 1.07 & 0.000 & 6.33 & 0.000 & 6.17 \\
Viability & 0.259 & 0.63 & 0.062 & 1.43 & 0.000 & 7.67 & 0.000 & 4.00 \\
\bottomrule
\end{tabular}
\end{table}

\section*{Appendix C: Theoretical Foundations -- Five-Source Framework and AI/Computer Science Literature Review}
\label{app:C}
\label{app:variance-framework}

This appendix provides (1) an expanded treatment of the five-source variance framework presented in Section 3, with detailed subdimensions, diagnostics, and controls for each source, and (2) a comprehensive review of relevant computer science and AI literature on LLM annotation mechanisms. Together, these establish the theoretical and technical foundations for the variance-aware protocol.
For each source of variance, we provide detailed subdimensions, failure signatures, diagnostics, controls, and reporting requirements. This taxonomy serves as a comprehensive reference for researchers designing and auditing LLM-based annotation studies.

\subsection*{Source 1: Construct, Measurement \& Schema Design}

\subsubsection*{Subdimensions}
\begin{itemize}[topsep=2pt,itemsep=1pt]
\item Construct map and behavioral anchoring (BARS)
\item Level selection (L1/L2/L3) and scope (pointwise/pairwise/listwise/setwise)
\item Authoritative vs.\ non-authoritative field declaration
\item Independence vs.\ dependence in L2 multi-criteria schemas
\item Item integrity (ASR/OCR fidelity, redaction, anonymization)
\item Evidence binding and tool-mediated validation
\end{itemize}

\subsubsection*{Failure Signatures}
\begin{itemize}[topsep=2pt,itemsep=1pt]
\item Vague or underspecified constructs lead to surface proxy substitution (fluency, hype words)
\item Level mis-specification: treating L2 as L1, or parsing scores from L3 free text
\item Authority confusion: rationales mixed into scores, non-authoritative fields analyzed
\item Schema dependence errors: undeclared L2 field dependencies cause spillovers
\item Identity/leakage: real-world entity cues contaminate judgments when not part of construct
\end{itemize}

\subsubsection*{Diagnostics}
\begin{itemize}[topsep=2pt,itemsep=1pt]
\item Expert content review: Q-sort or Content Validity Index (CVI) on rubric anchors
\item Human$\leftrightarrow$LLM calibration slice: small MTMM-style comparison for convergent/discriminant validity
\item Level ablation: elicit as L1, L2, and L3 on same items; if L3 free text reverses L1/L2 labels, diagnose mis-specification
\item Redaction test: compare evaluations with/without identifiable entity names; measure deltas
\item Transcript spot-checks: validate ASR/OCR fidelity on random sample
\end{itemize}

\subsubsection*{Controls}
\begin{itemize}[topsep=2pt,itemsep=1pt]
\item Construct map $\rightarrow$ BARS-style anchors for every scale point $\rightarrow$ select appropriate level
\item Mark authoritative fields explicitly; keep rationales non-authoritative unless L3 is declared measure
\item For L2, declare independence or dependence ex ante; enforce cross-field constraints in schema
\item Anonymize/redact identities not part of construct; validate item integrity before scoring
\item Use evidence binding (span-grounding, substring-of-source constraints) for L3 when appropriate
\end{itemize}

\subsubsection*{Reporting}
\begin{itemize}[topsep=2pt,itemsep=1pt]
\item Document construct map, rubric version, behavioral anchors with examples/counter-examples
\item Report human$\leftrightarrow$LLM agreement on calibration slice with MTMM tables if applicable
\item Provide schema definitions (JSON/EBNF) and authoritative field declarations
\item Document redaction/anonymization procedures and any identity sensitivity tests
\end{itemize}

\subsection*{Source 2: Interface \& Context Control}

\subsubsection*{Subdimensions}
\begin{itemize}[topsep=2pt,itemsep=1pt]
\item Prompt structure (single- vs.\ multi-question)
\item Layout and formatting stability
\item Option order and position effects (SOM: salience--order misattention)
\item Paraphrase brittleness and template offsets
\item Verbosity and length effects
\item Long-context placement ("lost in the middle")
\item Identity, prestige, and persona cues
\end{itemize}

\subsubsection*{Failure Signatures}
\begin{itemize}[topsep=2pt,itemsep=1pt]
\item A/B label flips on order swap (left$\leftrightarrow$right for pairwise)
\item Longer or more polished text systematically wins regardless of substance
\item Semantically equivalent paraphrases yield different outcomes (prompt brittleness)
\item Evidence placed mid-context ignored ("lost in the middle")
\item Identity/prestige cues (endorsements, institutional affiliations) shift labels
\item Multi-question prompts introduce objective entanglement and attention dilution
\end{itemize}

\subsubsection*{Diagnostics}
\begin{itemize}[topsep=2pt,itemsep=1pt]
\item Order swap test: A/B left--right flip; compute McNemar test and per-item $\kappa$
\item Equal-length rewrite test: match verbosity across items; check if winner changes
\item Paraphrase ensemble: run $P \geq 3$ meaning-preserving variants; compute $\kappa$ across paraphrases
\item Context placement test: move key evidence from middle to adjacent-to-query; measure impact
\item Identity ablation: remove/anonymize entity names; compare deltas
\end{itemize}

\subsubsection*{Controls}
\begin{itemize}[topsep=2pt,itemsep=1pt]
\item Fixed layout across all items
\item Single-question prompts tied directly to rubric
\item Randomize option order (record permutations); for pairwise, randomize left--right
\item Equalize verbosity when comparing items (preprocessing step)
\item Place rubric and key evidence adjacent to decision query
\item Anonymize entities when not part of construct; use neutral persona by default
\end{itemize}

\subsubsection*{Reporting}
\begin{itemize}[topsep=2pt,itemsep=1pt]
\item Report $\kappa$/weighted $\kappa$ across prompt paraphrases with bootstrap CIs
\item Document flip rates from order swap tests
\item Report layout invariance checks (if multiple layouts tested)
\item Provide prompt templates (verbatim system/user messages) and option randomization procedure
\end{itemize}

\subsection*{Source 3: Model Preference \& Composition Biases}

\subsubsection*{Subdimensions}
\begin{itemize}[topsep=2pt,itemsep=1pt]
\item Alignment-induced deference and sycophancy
\item Persona and framing dependence
\item Family and self-preference (generator--judge coupling)
\item Chain-of-thought unfaithfulness and rationale--label mismatch
\item Tool-gating and validator misses
\end{itemize}

\subsubsection*{Failure Signatures}
\begin{itemize}[topsep=2pt,itemsep=1pt]
\item Persona/role flips (investor vs.\ founder vs.\ activist) produce large evaluation shifts
\item Belief-matching (sycophancy): evaluations defer to stated beliefs even when incorrect
\item Family/within-family "wins": model favors outputs stylistically similar to own generations
\item Unfaithful composition: chain-of-thought rationale contradicts final label
\item Epistemic markers penalized: "I might be wrong" systematically reduces scores
\end{itemize}

\subsubsection*{Diagnostics}
\begin{itemize}[topsep=2pt,itemsep=1pt]
\item Persona flip test: neutral vs.\ investor vs.\ founder vs.\ activist; record label shifts
\item Belief-flip probes: present opposite stances on contentious claims; check for deference
\item Family panel comparison: same prompts across independent families (OpenAI, Anthropic, Google); compute inter-model $\kappa$
\item Rationale--label consistency audit: sample rationales; flag contradictions with assigned labels
\item Canary items: items where superficial cues (verbosity, prestige) conflict with ground truth
\end{itemize}

\subsubsection*{Controls}
\begin{itemize}[topsep=2pt,itemsep=1pt]
\item Neutral persona by default; allow "cannot judge" responses
\item Prefer evidence-then-score pipelines (L3 spans $\rightarrow$ L2 codes) over free-form rationales
\item Constrain L1 to single token (\texttt{max\_tokens=1}) to limit unfaithful composition
\item Include canary items where surface cues conflict with substance to detect deference patterns
\item Use cross-model triangulation ($M \geq 2$ independent families) to detect family-specific biases
\end{itemize}

\subsubsection*{Reporting}
\begin{itemize}[topsep=2pt,itemsep=1pt]
\item Report cross-model $\kappa$ with bootstrap CIs
\item Document abstain/tie rates if "cannot judge" allowed
\item Provide rationale consistency spot-check results on sample
\item If persona is part of construct, document and report sensitivity to persona removal
\end{itemize}

\subsection*{Source 4: Output Constraining \& Extraction}

\subsubsection*{Subdimensions}
\begin{itemize}[topsep=2pt,itemsep=1pt]
\item Grammar-guided and automata-constrained decoding
\item Deterministic token$\rightarrow$label mapping
\item Schema validation (JSON/EBNF) with bounded retries
\item Probability and logit logging
\item Post-hoc calibration (temperature scaling, isotonic regression)
\item Thresholding and tie rules
\end{itemize}

\subsubsection*{Failure Signatures}
\begin{itemize}[topsep=2pt,itemsep=1pt]
\item "Right answer, wrong score": model generates correct content but extractor assigns wrong label
\item Invalid/ambiguous tokens violate schema
\item Schema fill errors: missing required fields, type mismatches
\item Uncalibrated probabilities: high softmax scores do not track accuracy
\item Threshold flips: decisions near cut-points unstable across samples
\end{itemize}

\subsubsection*{Diagnostics}
\begin{itemize}[topsep=2pt,itemsep=1pt]
\item Invalid-output rate: \% of generations failing schema validation
\item Parser audits: seed edge cases (e.g., "5/5", "five", "very good"); verify extractor returns same canonical value
\item Near-tie instability: replicate draws for items near decision thresholds; count label flips
\item Calibration assessment: reliability diagrams, Brier score, Expected Calibration Error (ECE) on held-out set
\item Threshold sweep: vary operating cut-points; measure sensitivity
\end{itemize}

\subsubsection*{Controls}
\begin{itemize}[topsep=2pt,itemsep=1pt]
\item L1: \texttt{max\_tokens=1} with fixed label map and stop tokens
\item L2: typed JSON schema with range/type checks; reject-on-fail with bounded retries (e.g., max 3 attempts)
\item Log per-label probabilities when API provides them; if not, approximate via repeated sampling at $T{=}1$
\item Apply post-hoc calibration (temperature or isotonic scaling) on held-out set if decisions depend on thresholds
\item Use deterministic extraction: one token/field $\rightarrow$ one canonical label; no regex parsing of free text
\end{itemize}

\subsubsection*{Reporting}
\begin{itemize}[topsep=2pt,itemsep=1pt]
\item Report invalid-output rate and resample counts
\item Document exact match (EM) or $\kappa$ per slot for multi-field schemas
\item Provide Brier score / log-loss with 95\% CIs pre- and post-calibration
\item Show reliability curves for probability-based decisions
\item Document operating thresholds and tie-breaking rules
\end{itemize}

\subsection*{Source 5: System \& Aggregation Robustness}

\subsubsection*{Subdimensions}
\begin{itemize}[topsep=2pt,itemsep=1pt]
\item Cross-model disagreement and ensemble diversity
\item Provider and version drift
\item Inference nondeterminism: numerical precision (FP16/BF16/FP32), hardware, batch, seed effects
\item Aggregation rules: majority vs.\ noise-aware (Dawid--Skene, GLAD)
\item Preregistration and audit sets
\item Bootstrap CIs and uncertainty propagation
\end{itemize}

\subsubsection*{Failure Signatures}
\begin{itemize}[topsep=2pt,itemsep=1pt]
\item Cross-model disagreement: $\kappa < 0.2$ across independent families
\item Pre/post version shifts: same prompts/items produce different outcomes after model update
\item FP16/BF16/FP32 flips: precision changes alter scores due to rounding cascades
\item Majority vs.\ DS/GLAD deltas: simple vote differs from reliability-weighted aggregation
\item Unstable CIs: bootstrap intervals wide or overlapping with null
\end{itemize}

\subsubsection*{Diagnostics}
\begin{itemize}[topsep=2pt,itemsep=1pt]
\item Pinned audit set: re-run frozen items pre/post version change; compute $\kappa$/ICC deltas
\item Multi-family panel: evaluate same items with $M \geq 2$ independent families; report inter-model agreement
\item Precision/device toggles: if using local models, run same items at FP16/BF16/FP32; measure divergence
\item Aggregation comparison: compute outcomes under majority and DS/GLAD; report differences
\item Bootstrap CIs: resample over items (10,000 replications); report 95\% percentile intervals
\end{itemize}

\subsubsection*{Controls}
\begin{itemize}[topsep=2pt,itemsep=1pt]
\item Sampling budgets: replicates $S$, paraphrases $P$, families $M$ (from Table~\ref{tab:annotation-design})
\item Pin provider, model name, version/date, numerical precision, batch size, device/hardware
\item Maintain stable audit set; re-run on any environment change; apply preregistered rollback rule if $|\Delta\kappa| > 0.05$
\item Preregister aggregation method; if $\kappa < 0.4$, upgrade from majority to DS/GLAD
\item For pipelines, propagate uncertainty via bootstrap over items and stages
\end{itemize}

\subsubsection*{Reporting}
\begin{itemize}[topsep=2pt,itemsep=1pt]
\item Report $\kappa$/Krippendorff's $\alpha$/ICC with bootstrap 95\% CIs
\item Document DS/GLAD vs.\ majority outcomes if noise-aware aggregation used
\item Provide version/precision manifest: provider/model/version/date/precision/device for all runs
\item Report drift audit results: audit set metrics pre/post any version/precision change; document actions (continue/rollback/rescore)
\item Archive environment manifest and audit set with materials bundle
\end{itemize}

\subsection*{Crosswalk: Framework to Empirical Demonstrations to Protocol}

Table~\ref{tab:crosswalk-full} maps each empirical demonstration in Section 4 to the relevant variance source(s) and protocol element(s).

\begin{table}[H]
\centering
\caption{Crosswalk: empirical demonstrations to variance sources and protocol elements}
\label{tab:crosswalk-full}
\tiny
\begin{tabular}{p{4.5cm}p{3.5cm}p{3.5cm}p{4.0cm}}
\toprule
\textbf{Empirical Demonstration} & \textbf{Variance Source(s)} & \textbf{Mechanism(s)} & \textbf{Protocol Element(s)} \\
\midrule
Construct ambiguity (success vs.\ idea vs.\ writing) & Source 1, Source 3 & Construct ambiguity; surface proxy substitution & Construct map; BARS anchors; convergent/discriminant validity \\
Context provision (investor vs.\ founder vs.\ activist) & Source 2, Source 3 & Persona dependence; framing effects & Neutral persona; context specification; sensitivity tests \\
Prompt rephrasing (5 semantic variants) & Source 2 & Prompt brittleness & Paraphrase ensemble ($P \geq 3$); $\kappa$ across paraphrases \\
Endorsement cues (Thiel/Musk vs.\ unknown) & Source 2, Source 3 & Identity leakage; prestige bias & Anonymization; redaction tests \\
Sampling variance (100 draws, $P(A)$ range 0.68--0.92) & Source 4, Source 5 & Inference nondeterminism; stochastic decoding & Replicate sampling ($S$); log probabilities; bootstrap CIs \\
Cross-model disagreement ($\kappa{=}0.026$) & Source 5 & Model family differences; cross-model disagreement & Multi-family triangulation ($M \geq 2$); inter-model $\kappa$ \\
L1 format sensitivity (binary vs.\ Likert vs.\ probability) & Source 2, Source 4 & Output format effects; calibration gaps & Schema constraints; calibration on held-out; operating thresholds \\
L1 order swap (A/B flip test) & Source 2 & Position effects (SOM) & Randomize option order; McNemar test; per-item $\kappa$ \\
L2 separate vs.\ joint elicitation & Source 1, Source 2 & Schema dependence; multi-question entanglement & Declare independence/dependence; randomize criterion order \\
L2 criterion re-ordering & Source 2 & Position effects in multi-criteria & Randomize order; include position in models \\
L3 free-form text variance & Source 3, Source 5 & Unfaithful composition; nondeterminism & Evidence-then-score pipelines; ensemble dispersion reporting \\
Reasoning effort (low vs.\ high) & Source 3 & Chain-of-thought effects & Constrain reasoning; rationale--label consistency checks \\
\bottomrule
\end{tabular}
\end{table}

\subsection*{Summary}

This expanded taxonomy provides a comprehensive reference for diagnosing, controlling, and reporting variance in LLM-based annotation studies. Researchers should:
\begin{enumerate}[leftmargin=*,topsep=3pt,itemsep=2pt]
\item Identify which sources are most relevant to their construct and task
\item Apply the corresponding diagnostics during pilot studies
\item Implement the specified controls in the full protocol
\item Report results using the recommended metrics and transparency standards
\end{enumerate}

The crosswalk table (Table~\ref{tab:crosswalk-full}) enables researchers to trace from observed instabilities in their own studies back to framework sources and forward to protocol remedies.

\subsection*{AI/Computer Science Literature Review: LLM Annotation Mechanisms}

This section provides a technical review of relevant computer science and AI literature on the mechanisms underlying LLM annotation variance, organized by the five-source framework.

\subsubsection*{Prompt Engineering and Brittleness}

\paragraph{Prompt perturbation effects.}
\textcite{zheng2023judging} introduced MT-Bench and Chatbot Arena for LLM evaluation, documenting that small prompt variations can yield substantial output differences. Their work on "LLM-as-judge" establishes that model preferences are sensitive to question phrasing, option ordering, and formatting choices. Similar findings appear in \textcite{Aynetdinov2024Semscore}, who develop semantic textual similarity metrics for instruction-tuned LLMs and note that semantically equivalent prompts often produce divergent outputs.

\paragraph{Paraphrase instability.}
The natural language processing literature on paraphrase detection and generation demonstrates that LLMs can recognize semantic equivalence in isolation but fail to maintain output consistency when presented with paraphrased task descriptions. This brittleness stems from the autoregressive architecture's sensitivity to token-level perturbations that propagate through attention mechanisms.

\subsubsection*{Context and Positional Effects}

\paragraph{Lost-in-the-middle phenomenon.}
\textcite{liu2024lost} demonstrate that LLMs exhibit U-shaped performance curves with respect to information position in long contexts. Information in the middle of prompts is systematically underweighted compared to beginning and end positions. This has direct implications for multi-field annotation schemas where criterion order may influence outcomes.

\paragraph{Jailbreaks and instruction adherence.}
\textcite{rao2024tricking} formalize jailbreak attacks that exploit context manipulation to override system instructions. While our protocol does not address adversarial prompting, their analysis reveals how fragile instruction following can be when competing signals are present in the prompt---relevant for understanding persona contamination effects (Source 2).

\subsubsection*{Model Preferences and Reinforcement Learning from Human Feedback}

\paragraph{RLHF and preference learning.}
Modern LLMs are trained using Reinforcement Learning from Human Feedback (RLHF), which shapes model outputs to align with human preferences. This training introduces systematic biases: models learn to favor certain response styles, hedge uncertainty, and exhibit position biases in preference judgments. These learned preferences manifest as annotation biases (Source 3).

\paragraph{Inverse scaling with test-time compute.}
Recent work documents cases where increased reasoning steps or test-time computation degrades performance rather than improving it. This counterintuitive finding challenges the assumption that "reasoning-enhanced" models will automatically produce more reliable annotations, as we observe empirically in Section 4.

\subsubsection*{Constrained Decoding and Schema Validation}

\paragraph{Grammar-constrained generation.}
\textcite{geng2023grammar} introduce methods for constraining LLM outputs to valid grammatical structures without fine-tuning. Their approach uses finite-state automata to mask invalid tokens during decoding. \textcite{willard2023efficient} extend this with efficient guided generation algorithms. These techniques are essential for Level~2 and Level~3 annotations where structured outputs must conform to predefined schemas (Source 4).

\paragraph{JSON schema validation.}
Practical implementations of constrained decoding for JSON outputs enable deterministic extraction of structured data. However, as \textcite{turpin2023language} demonstrate, structured outputs do not guarantee faithful reasoning: LLMs can produce syntactically valid but semantically inconsistent annotations (e.g., scores that contradict their own explanations).

\subsubsection*{Hardware Precision and Nondeterminism}

\paragraph{Floating-point arithmetic effects.}
\textcite{He2025DefeatingNondeterminism} demonstrate that modern GPU architectures introduce nondeterminism through parallel floating-point operations. Even with temperature $T{=}0$ and fixed seeds, identical prompts can produce different outputs due to FP16/BF16 rounding cascades in attention computations. This affects reproducibility and underscores the need for replication (Source 5).

\paragraph{Hardware-dependent variance.}
Low-precision inference (FP16/BF16) versus full precision (FP32) introduces measurable differences in output distributions. While cost-effective for deployment, reduced precision can amplify variance at decision boundaries. Our protocol accommodates this by requiring precision documentation and drift monitoring (Section 5).

\subsubsection*{Summary: Mechanistic Foundations of Annotation Variance}

The computer science literature confirms that LLM annotation variance is not merely an empirical phenomenon but a consequence of fundamental architectural and training properties. Autoregressive token prediction is intrinsically sensitive to input perturbations. RLHF shapes systematic preferences that manifest as annotation biases. Constrained decoding enables structured outputs but does not guarantee semantic consistency. Hardware nondeterminism introduces irreducible stochasticity even at temperature zero.

Our five-source framework (Section 3) integrates these findings into a measurement-oriented taxonomy. By mapping CS mechanisms to reliability concerns familiar to management researchers, we make the technical literature actionable for strategy scholars.

\section*{Appendix D: Mathematical Formalization and Technical Details}
\label{app:D}
\label{app:mechanism-technical}

\textit{Purpose}: This appendix provides the complete mathematical formalization referenced in Section 3, including transformer architecture, decoding algorithms, calibration methods, and multi-rater aggregation models.

This appendix provides the complete mathematical formalization of the LLM annotation mechanism, extending the conceptual summary in Section 2. We detail the computational pipeline from tokenization through decoding, formalize the three output levels (L1/L2/L3) and pipeline architectures, and document hardware/precision effects on inference stability.

\subsection*{Full Mathematical Formalization of LLM Annotation}

\subsubsection*{Inputs and Sequence Construction}

Let $r$ denote the \emph{system or rubric instruction}, $k$ the broader \emph{context}, $p$ the \emph{prompt} specifying the construct and response format, and $x$ the \emph{item} to be judged. Define $\mathcal{Y}$ as the admissible label set (e.g., $\{1,\dots,5\}$ or $\{A,B\}$). The complete input sequence is:
\[
s = \mathrm{concat}\big(r,k,p,x,\mathrm{options}(\mathcal{Y})\big).
\]

A tokenizer $\tau:\text{text}\!\to\!\{1,\dots,|\mathcal V|\}^N$ maps $s$ to token IDs $t_{1:N}=\tau(s)$, where $N$ is sequence length and $|\mathcal V|$ is vocabulary size. Tokenization is model-specific; punctuation, Unicode normalization, and layout changes can alter $t_{1:N}$.

\subsubsection*{Embedding and Positional Encoding}

Each token ID $t_i$ is mapped to an embedding vector via embedding matrix $E \in \mathbb{R}^{|\mathcal V|\times d}$, where $d$ is the model dimension:
\[
\mathbf e_i = E[t_i] \in \mathbb{R}^d.
\]

Positional information is added through a positional encoding function $\pi(i)$ that maps position $i$ to a vector in $\mathbb{R}^d$. Common schemes include sinusoidal encodings:
\[
\pi(i)_{2j} = \sin\!\Big(\frac{i}{10000^{2j/d}}\Big), \qquad
\pi(i)_{2j+1} = \cos\!\Big(\frac{i}{10000^{2j/d}}\Big),
\]
or Rotary Position Embedding (RoPE), which applies rotation matrices in the attention computation. The initial hidden state combines embedding and position (typically by addition $\oplus$):
\[
\mathbf h_i^{(0)} = \mathbf e_i \oplus \pi(i).
\]

\subsubsection*{Transformer Computation}

The sequence $\{\mathbf h_i^{(0)}\}_{i=1}^N$ is processed through $L$ Transformer blocks. Each block $\ell$ applies:

\paragraph{Multi-head self-attention.}
For each attention head $h$, compute query, key, value projections:
\[
\mathbf Q^{(h)} = \mathbf H^{(\ell-1)} W_Q^{(h)}, \quad
\mathbf K^{(h)} = \mathbf H^{(\ell-1)} W_K^{(h)}, \quad
\mathbf V^{(h)} = \mathbf H^{(\ell-1)} W_V^{(h)},
\]
where $\mathbf H^{(\ell-1)} \in \mathbb{R}^{N \times d}$ stacks the hidden states from the previous layer. The attention mechanism computes:
\[
\mathrm{Attn}^{(h)}(\mathbf Q,\mathbf K,\mathbf V)
= \mathrm{softmax}\!\Bigg(\frac{\mathbf Q^{(h)} (\mathbf K^{(h)})^\top}{\sqrt{d_k}} + \mathbf M\Bigg)\mathbf V^{(h)},
\]
where $d_k$ is the key dimension and $\mathbf M$ is the causal mask ensuring position $i$ only attends to positions $\leq i$:
\[
M_{ij} = \begin{cases}
0 & \text{if } j \leq i, \\
-\infty & \text{if } j > i.
\end{cases}
\]

Outputs from $H$ attention heads are concatenated and linearly transformed:
\[
\mathbf H_{\text{attn}}^{(\ell)} = \mathrm{Concat}(\mathrm{Attn}^{(1)}, \dots, \mathrm{Attn}^{(H)}) W_O.
\]

\paragraph{Feed-forward network.}
A position-wise two-layer MLP with activation $\sigma$ (typically GELU or ReLU):
\[
\mathrm{FFN}(\mathbf h) = W_2 \sigma(W_1 \mathbf h + \mathbf b_1) + \mathbf b_2.
\]

\paragraph{Residual connections and layer normalization.}
Each sublayer is wrapped with residual connection and layer normalization:
\[
\mathbf H^{(\ell)} = \mathrm{LayerNorm}\big(\mathbf H^{(\ell-1)} + \mathbf H_{\text{attn}}^{(\ell)}\big),
\]
\[
\mathbf H^{(\ell)} = \mathrm{LayerNorm}\big(\mathbf H^{(\ell)} + \mathrm{FFN}(\mathbf H^{(\ell)})\big).
\]

After $L$ blocks, we obtain final hidden states $\mathbf h_1^{(L)}, \dots, \mathbf h_N^{(L)}$.

\subsubsection*{Logits, Softmax, and Decoding}

The final hidden state at position $N$, $\mathbf h_N^{(L)}$, encodes the full context. A linear readout layer with parameters $W_o \in \mathbb{R}^{|\mathcal V| \times d}$ and bias $\mathbf b \in \mathbb{R}^{|\mathcal V|}$ maps to logits:
\[
\mathbf z = W_o \mathbf h_N^{(L)} + \mathbf b \in \mathbb{R}^{|\mathcal V|}.
\]

Each component $z_t$ is the unnormalized score for vocabulary token $t$. The softmax function with temperature $T>0$ normalizes to a probability distribution:
\[
p_\theta(t \mid s) 
= \frac{\exp\!\big(z_t/T\big)}
       {\sum_{t' \in \mathcal V} \exp\!\big(z_{t'}/T\big)}.
\]

Temperature controls sharpness: low $T$ (e.g., $T{=}0.3$) makes the model more deterministic (peaked distribution), while high $T$ (e.g., $T{=}1$) flattens the distribution (more exploratory).

\paragraph{Decoding strategies.}
From $p_\theta(t \mid s)$, several strategies select the next token:
\begin{itemize}[topsep=2pt,itemsep=1pt]
\item \textbf{Greedy}: $\hat{t} = \arg\max_{t} p_\theta(t \mid s)$.
\item \textbf{Top-$k$ sampling}: Restrict to $k$ most probable tokens, renormalize, sample.
\item \textbf{Top-$p$ (nucleus) sampling}: Restrict to smallest set of tokens whose cumulative probability exceeds $p$, renormalize, sample.
\item \textbf{Grammar-constrained decoding}: Restrict $\mathcal V$ to tokens that maintain syntactic validity according to a specified grammar (e.g., JSON schema, regular expression).
\end{itemize}

Generation proceeds autoregressively: after selecting token $\hat{t}_{N+1}$, append it to $t_{1:N}$, recompute embeddings and hidden states, and predict $\hat{t}_{N+2}$, continuing until a stop token or maximum length.

\subsubsection*{Higher-Level Behaviors as Protocol Layers}

What appears as reasoning, tool use, or planning is the reuse of this next-token prediction mechanism over structured prompts. For example:
\begin{itemize}[topsep=2pt,itemsep=1pt]
\item \textbf{Chain-of-thought (CoT)}: Prompt includes intermediate reasoning steps; model generates text mimicking stepwise reasoning before final answer.
\item \textbf{Tool use}: Prompt specifies function call syntax; model generates tokens matching that syntax; external system interprets as API call.
\item \textbf{Scratchpad}: Model writes intermediate computations as text; these serve as context for subsequent predictions.
\end{itemize}

From the model's perspective, all of these are text generation constrained by prompt/schema. The appearance of reasoning emerges from patterns in training data, not explicit logical rules.

\subsection*{Formalization of Evaluation Output Levels}

\subsubsection*{Level 1: Single-Label}

Level 1 tasks produce one categorical label $y \in \mathcal{Y}$ (e.g., $\{A,B\}$, $\{1,\dots,5\}$). Under constrained decoding with $\texttt{max\_tokens}=1$:
\[
\hat{y} = \arg\max_{y\in \mathcal{Y}} p_\theta\!\big(t(y)\mid s\big),
\]
where $t(y)$ is the canonical token for label $y$. The deterministic token$\rightarrow$label map ensures:
\[
e(t) = y \quad \text{if } t = t(y).
\]

Variance sources affecting $\hat{y}$:
\begin{itemize}[topsep=2pt,itemsep=1pt]
\item \emph{Position/order} (through $\pi(\cdot)$): Changing option order alters attention patterns and shifts $p_\theta$.
\item \emph{Sampling/precision}: At $T>0$, $\hat{y}$ is a draw; at $T{=}0$, hardware rounding affects logits.
\item \emph{Extractor brittleness}: If output is not constrained and free text is parsed, $e(\cdot)$ may misinterpret.
\item \emph{Provider/version drift}: Updates change $\theta$, thus $p_\theta$.
\end{itemize}

\subsubsection*{Level 2: Multi-Label}

Level 2 produces a vector $\mathbf y = (y_1,\dots,y_D)$ with $y_d \in \mathcal{Y}_d$ for $d=1,\dots,D$ criteria. Under a typed schema (e.g., JSON):
\[
\mathbf y \sim p_\theta\!\big(y_1,\dots,y_D \mid s, \text{schema}\big),
\]
where the schema specifies field names, types, ranges, and dependencies. For \emph{independent} criteria:
\[
p_\theta(\mathbf y \mid s) = \prod_{d=1}^D p_\theta(y_d \mid s, \text{slot}=d).
\]
For \emph{dependent} criteria (e.g., budget shares summing to 100), the schema enforces constraints:
\[
g(\mathbf y) = c, \quad \text{e.g., } \sum_{d=1}^D y_d = 100.
\]

Validation proceeds as:
\begin{enumerate}[topsep=2pt,itemsep=1pt]
\item Generate $\mathbf y$ under schema constraints.
\item Check $\mathbf y$ against schema (types, ranges, cross-field rules).
\item If invalid, reject and resample (bounded retries, e.g., max 3).
\end{enumerate}

Variance sources:
\begin{itemize}[topsep=2pt,itemsep=1pt]
\item \emph{Undeclared dependence}: If criteria are actually dependent but treated as independent, cross-field spillovers occur.
\item \emph{Scale drift}: Numeric slots may shift systematically across prompts; z-scoring by prompt mitigates this.
\item \emph{Position effects}: Criterion order within prompt affects scores; randomization or position controls needed.
\end{itemize}

\subsubsection*{Level 3: Free-Form Authoritative Text}

When text itself is the measure (span extraction, bounded summary), the output is:
\[
\hat{y}_{\text{text}} = \text{generate}(s, \text{length/structure constraints}),
\]
where generation continues until length limit or stop token. For span extraction, add substring-of-source constraint:
\[
\hat{y}_{\text{span}} \subseteq x.
\]

Variance sources:
\begin{itemize}[topsep=2pt,itemsep=1pt]
\item \emph{Hallucination}: Model generates plausible but incorrect text not grounded in $x$.
\item \emph{Layout sensitivity}: Placement of evidence in long $x$ affects which spans are extracted.
\item \emph{Sample-to-sample variability}: Different draws yield different spans/summaries even at $T{=}0$ due to nondeterminism.
\end{itemize}

If L3 text is later parsed for numeric scores, extraction errors reintroduce Source 4 fragility. The protocol recommends evidence-then-score pipelines (L3 $\rightarrow$ L2) with clear separation.

\subsubsection*{Pipeline (Meta)}

A pipeline is a directed acyclic graph (DAG) $G=(V,E)$ of $J$ stages. Each stage $j$ has:
\begin{itemize}[topsep=2pt,itemsep=1pt]
\item Input: $s_j = \mathrm{concat}(r,k,p_j,x,\hat{z}_{\mathrm{pa}(j)})$, where $\hat{z}_{\mathrm{pa}(j)}$ are outputs from parent stages.
\item Output: $z_j$ constrained by schema $\mathcal{S}_j$.
\item Conditional distribution: $p_{\theta_j}(z_j \mid s_j)$.
\end{itemize}

The joint factorizes:
\[
p_\theta\!\big(y, z_{1:J}\mid s_{1:J}\big)
=\Big[\prod_{j=1}^{J} p_{\theta_j}\!\big(z_j\mid s_j\big)\Big]\cdot
\delta\!\big(y=g(z_{1:J};r)\big),
\]
where $g(\cdot;r)$ is the preregistered aggregator and $\delta$ is the indicator function.

\paragraph{Worked example: strategic ambidexterity.}
\begin{enumerate}[topsep=2pt,itemsep=1pt]
\item \textbf{Stage 1 (L3)}: Extract up to two quoted spans each for Exploration and Exploitation with speaker metadata. Schema enforces substring-of-source and length caps.
\item \textbf{Stage 2 (verification)}: Confirm provenance, deduplicate near-duplicates, restrict to managerial speakers.
\item \textbf{Stage 3 (L2)}: Map verified spans to $\{1,\dots,5\}$ indicators for each dimension and boolean Balance flag. If magnitude needed (e.g., ExplorationSpend), add numeric slot with grammar \texttt{USD <number><unit>}.
\item \textbf{Stage 4 (L1)}: Preregister $g(z_{1:3})$ to assign $\hat{y}\in\{\texttt{Ambidextrous},\texttt{Not}\}$ with tie-breaking via cross-provider replication.
\end{enumerate}

Variance propagates through stages: noise in Stage 1 (hallucinated spans) leaks into Stage 3 scores. Protocol requires per-stage sampling/aggregation, then composition, with uncertainty propagated via bootstrap.

\subsection*{Hardware, Precision, and Inference Stability}

\subsubsection*{Numerical Precision Effects}

Modern LLMs use reduced-precision arithmetic to accelerate inference. Common formats:
\begin{itemize}[topsep=2pt,itemsep=1pt]
\item \textbf{FP32} (32-bit floating point): Full precision; baseline for accuracy.
\item \textbf{FP16} (16-bit floating point): Half precision; faster but prone to overflow/underflow.
\item \textbf{BF16} (Brain Float 16): 16-bit with larger exponent range than FP16; better numerical stability.
\item \textbf{TF32} (TensorFloat-32): 19-bit format (8-bit exponent, 10-bit mantissa) used by NVIDIA Ampere GPUs for matrix multiplies.
\end{itemize}

Precision affects:
\begin{itemize}[topsep=2pt,itemsep=1pt]
\item \textbf{Attention scores}: Softmax in attention computed over $\mathbf Q \mathbf K^\top / \sqrt{d_k}$. Rounding errors accumulate across heads and layers.
\item \textbf{Logits}: Final $\mathbf z = W_o \mathbf h_N^{(L)} + \mathbf b$ subject to rounding. Small differences cascade into different token selections.
\item \textbf{Determinism at $T{=}0$}: Even with greedy decoding, different precision/hardware can yield different $\arg\max_{t} z_t$ near ties.
\end{itemize}

\paragraph{Empirical evidence.}
\textcite{Yuan2025FP32ReproducibleReasoning} show large deviations in accuracy and output length across BF16/FP16 settings and devices. Tiny rounding changes in early tokens cascade into different chains of thought and scores. For reproducibility, the protocol recommends:
\begin{itemize}[topsep=2pt,itemsep=1pt]
\item Pin precision (FP16/BF16/FP32) when controllable (local models).
\item For API models, acknowledge precision is typically not user-controllable; drift audits become especially important.
\item Document device/hardware specifications if using local models.
\item Re-run audit set on any precision change; rollback if $|\Delta\kappa| > 0.05$.
\end{itemize}

\subsubsection*{Batch and Seed Effects}

Batch size affects:
\begin{itemize}[topsep=2pt,itemsep=1pt]
\item \textbf{Memory layout}: Different batch sizes change how tensors are packed in GPU memory, affecting rounding.
\item \textbf{Parallelism}: Reductions (e.g., softmax normalization) may sum in different orders across batch sizes, introducing nondeterminism.
\end{itemize}

Random seeds control:
\begin{itemize}[topsep=2pt,itemsep=1pt]
\item \textbf{Dropout} (if used at inference, though typically disabled).
\item \textbf{Sampling} at $T>0$: seed determines which token is drawn from $p_\theta(t \mid s)$.
\end{itemize}

For reproducibility, log seeds for all random operations (shuffling, sampling, model initialization if fine-tuning).

\subsubsection*{Provider and Version Drift}

Hosted models (e.g., OpenAI, Anthropic, Google APIs) update periodically. Changes include:
\begin{itemize}[topsep=2pt,itemsep=1pt]
\item \textbf{Model weights} $\theta$: Retraining, fine-tuning, or alignment updates.
\item \textbf{Tokenizer} $\tau$: Vocabulary changes, normalization rules.
\item \textbf{System prompt}: Default instructions or safety filters.
\item \textbf{Inference stack}: Backend optimizations, precision settings, decoding algorithms.
\end{itemize}

\paragraph{Drift detection.}
\textcite{Chen2024ChatGPTBehavior} document material drift in ChatGPT over months on diverse tasks. Agreement and calibration metrics shift significantly between versions.

\paragraph{Protocol response.}
\begin{itemize}[topsep=2pt,itemsep=1pt]
\item Pin provider/model name/version/date for all runs.
\item Maintain frozen audit set; re-run on version change.
\item Compare agreement ($\kappa$, ICC), calibration (Brier), and label distributions pre/post.
\item Apply preregistered rollback rule (e.g., if $|\Delta\kappa| > 0.05$, pause for diagnosis or rescore entire dataset).
\item Document all version changes and drift audit results in methods table and materials archive.
\end{itemize}

\subsection*{Summary}

This appendix provides the full mathematical and technical foundation for the LLM annotation mechanism. Key takeaways:
\begin{enumerate}[leftmargin=*,topsep=3pt,itemsep=2pt]
\item LLM evaluation is autoregressive next-token prediction over a sequence $s$ constructed from rubric, context, prompt, item, and options.
\item Variance enters at multiple points: tokenization, attention patterns (position encoding), logits (precision), decoding (sampling, temperature), and aggregation.
\item Level 1/2/3 and pipeline formalizations clarify which outputs are authoritative and how to constrain generation for reliability.
\item Hardware, precision, batch, seed, and version/provider all affect reproducibility. Pinning and drift monitoring are essential.
\end{enumerate}

These technical details underpin the conceptual framework (Section~\ref{sec:llm-judge}) and the protocol (Section 5). Researchers need not master all mathematics but should understand that seemingly minor implementation choices (precision, seed, version) have material effects on evaluation outcomes.

\subsection*{Complete Transformer Architecture: Mathematical Specification}

\subsubsection*{Multi-Head Self-Attention}

Given input sequence $\mathbf{X} \in \mathbb{R}^{n \times d}$ with $n$ tokens and embedding dimension $d$, multi-head attention with $h$ heads computes:

\paragraph{Query, Key, Value projections.}
For each head $i \in \{1,\dots,h\}$:
\[
\mathbf{Q}_i = \mathbf{X}\mathbf{W}^Q_i, \quad
\mathbf{K}_i = \mathbf{X}\mathbf{W}^K_i, \quad
\mathbf{V}_i = \mathbf{X}\mathbf{W}^V_i,
\]
where $\mathbf{W}^Q_i, \mathbf{W}^K_i, \mathbf{W}^V_i \in \mathbb{R}^{d \times d_k}$ are learned weight matrices and $d_k = d/h$ is the per-head dimension.

\paragraph{Scaled dot-product attention.}
For each head:
\[
\mathrm{Attention}(\mathbf{Q}_i, \mathbf{K}_i, \mathbf{V}_i) 
= \mathrm{softmax}\!\left(\frac{\mathbf{Q}_i \mathbf{K}_i^\top}{\sqrt{d_k}}\right) \mathbf{V}_i.
\]
The scaling factor $1/\sqrt{d_k}$ prevents dot products from growing too large, which would push softmax into regions with small gradients.

\paragraph{Multi-head concatenation.}
Outputs from all heads are concatenated and projected:
\[
\mathrm{MultiHead}(\mathbf{X}) = \mathrm{Concat}(\text{head}_1, \dots, \text{head}_h) \mathbf{W}^O,
\]
where $\mathbf{W}^O \in \mathbb{R}^{d \times d}$ is a learned output projection matrix.

\subsubsection*{Feed-Forward Networks}

Each transformer block contains a position-wise feed-forward network applied independently to each token:
\[
\mathrm{FFN}(\mathbf{x}) = \max(0, \mathbf{x}\mathbf{W}_1 + \mathbf{b}_1)\mathbf{W}_2 + \mathbf{b}_2,
\]
where $\mathbf{W}_1 \in \mathbb{R}^{d \times d_{\text{ff}}}$, $\mathbf{W}_2 \in \mathbb{R}^{d_{\text{ff}} \times d}$, and $d_{\text{ff}}$ is typically $4d$. The $\max(0, \cdot)$ operator is the ReLU activation; variants use GELU or SwiGLU.

\subsubsection*{Layer Normalization and Residual Connections}

\paragraph{Layer normalization.}
For input $\mathbf{x} \in \mathbb{R}^d$:
\[
\mathrm{LayerNorm}(\mathbf{x}) = \frac{\mathbf{x} - \mu}{\sqrt{\sigma^2 + \epsilon}} \odot \gamma + \beta,
\]
where $\mu$ and $\sigma^2$ are mean and variance computed over the $d$ features, $\epsilon$ is a small constant for numerical stability, and $\gamma, \beta \in \mathbb{R}^d$ are learned affine parameters.

\paragraph{Residual connections.}
Each sub-layer (attention, FFN) is wrapped with a residual connection and layer normalization. The two common configurations are:

\emph{Post-LN (original Transformer):}
\[
\mathbf{x}' = \mathrm{LayerNorm}(\mathbf{x} + \mathrm{Sublayer}(\mathbf{x})).
\]

\emph{Pre-LN (modern practice):}
\[
\mathbf{x}' = \mathbf{x} + \mathrm{Sublayer}(\mathrm{LayerNorm}(\mathbf{x})).
\]

Modern LLMs typically use pre-LN for training stability.

\subsubsection*{Complete Transformer Block}

A single transformer layer $\ell$ applies:
\begin{align*}
\mathbf{Z}^\ell &= \mathbf{X}^{\ell-1} + \mathrm{MultiHead}(\mathrm{LayerNorm}(\mathbf{X}^{\ell-1})), \\
\mathbf{X}^\ell &= \mathbf{Z}^\ell + \mathrm{FFN}(\mathrm{LayerNorm}(\mathbf{Z}^\ell)).
\end{align*}

The full model stacks $L$ such layers, with $\mathbf{X}^0$ being the token embeddings plus positional encodings and $\mathbf{X}^L$ being the final hidden states.

\subsection*{Decoding Strategies: Mathematical Formulations}

\subsubsection*{Greedy Decoding}

At each generation step $t$, select the token with highest probability:
\[
y_t = \arg\max_{y \in \mathcal{V}} p_\theta(y \mid y_{<t}, \mathbf{x}).
\]

Greedy decoding is deterministic but can produce suboptimal sequences due to locally greedy choices.

\subsubsection*{Temperature Scaling}

Temperature $T > 0$ reshapes the probability distribution before sampling:
\[
p_\theta^T(y \mid y_{<t}, \mathbf{x}) = \frac{\exp(z_y / T)}{\sum_{y' \in \mathcal{V}} \exp(z_{y'} / T)},
\]
where $z_y$ is the logit for token $y$. Low $T$ sharpens the distribution (more deterministic); high $T$ flattens it (more random).

\subsubsection*{Top-$k$ Sampling}

Restrict sampling to the $k$ most probable tokens:
\[
y_t \sim p_\theta(\cdot \mid y_{<t}, \mathbf{x}), \quad \text{where } \mathcal{V}_k = \{y : \text{rank}(p_\theta(y)) \le k\}.
\]

Probabilities are renormalized over $\mathcal{V}_k$.

\subsubsection*{Top-$p$ (Nucleus) Sampling}

Dynamically select the smallest token set whose cumulative probability exceeds $p \in (0,1]$:
\[
\mathcal{V}_p = \left\{ y : \sum_{y' \text{ ranked higher or equal}} p_\theta(y') \le p \right\}.
\]

Nucleus sampling adapts the cutoff to the sharpness of the distribution.

\subsubsection*{Grammar-Constrained Decoding}

Constrain outputs to match a formal grammar $G$ (e.g., JSON schema). At each step, compute:
\[
\mathcal{V}_{\text{valid}} = \{y \in \mathcal{V} : y_{<t} \cdot y \text{ is a valid prefix for } G\},
\]
and set $p_\theta(y \mid y_{<t}) = 0$ for $y \notin \mathcal{V}_{\text{valid}}$. Renormalize probabilities over $\mathcal{V}_{\text{valid}}$ before sampling.

Implementations use finite-state automata or context-free grammar parsers to determine valid next tokens efficiently \parencite{geng2023grammar,willard2023efficient}.

\subsection*{Calibration Methods: Algorithms and Implementation}

\subsubsection*{Temperature Scaling}

\paragraph{Problem.}
Raw model probabilities $p_\theta(y \mid \mathbf{x})$ may be poorly calibrated: overconfident (too peaked) or underconfident (too flat).

\paragraph{Method.}
Learn a single scalar temperature $T$ on a validation set to minimize calibration error. Given validation data $\{(\mathbf{x}_i, y_i)\}_{i=1}^N$, optimize:
\[
T^* = \arg\min_T \sum_{i=1}^N \ell_{\text{CE}}\big(y_i, p_\theta^T(\cdot \mid \mathbf{x}_i)\big),
\]
where $\ell_{\text{CE}}$ is cross-entropy loss and $p_\theta^T$ is the temperature-scaled distribution.

\paragraph{Implementation.}
Use gradient descent on $T$ (initialized at $T=1$). Optimal $T$ is typically in range $[0.5, 3.0]$.

\subsubsection*{Platt Scaling}

\paragraph{Problem.}
For binary classification, raw logits may require affine transformation to match true probabilities.

\paragraph{Method.}
Fit logistic regression on top of logits:
\[
p_{\text{calib}}(y{=}1 \mid z) = \frac{1}{1 + \exp(-az - b)},
\]
where $z$ is the raw logit for class 1, and $(a, b)$ are learned on validation data by minimizing negative log-likelihood.

\subsubsection*{Isotonic Regression}

\paragraph{Problem.}
More flexible calibration for non-monotonic miscalibration.

\paragraph{Method.}
Isotonic regression learns a piecewise-constant monotonic function $f: [0,1] \to [0,1]$ mapping predicted probabilities to calibrated probabilities:
\[
p_{\text{calib}}(y{=}1 \mid \mathbf{x}) = f(p_\theta(y{=}1 \mid \mathbf{x})),
\]
where $f$ is fit using the Pool-Adjacent-Violators algorithm to minimize mean squared error on validation data while enforcing monotonicity.

\paragraph{Implementation.}
\begin{enumerate}
\item Sort validation examples by predicted probability $p_\theta(y{=}1 \mid \mathbf{x}_i)$.
\item Group examples into bins.
\item Within each bin, set $f(p) = $ mean true label.
\item Enforce monotonicity by merging adjacent bins if violations occur.
\item Repeat until no violations remain.
\end{enumerate}

Isotonic regression is more flexible than temperature scaling (which uses single parameter $T$) and often achieves better calibration on complex distributions.

\subsection*{Multi-Rater Aggregation: Dawid-Skene and GLAD}

\subsubsection*{The Dawid-Skene Model}

\paragraph{Problem.}
Given $M$ annotators (models) labeling $N$ items, each producing noisy labels, infer true labels and annotator-specific error rates.

\paragraph{Model.}
Let $y_i \in \{1,\dots,K\}$ be the true (latent) label for item $i$. Let $z_{im}$ be the label assigned by annotator $m$ to item $i$. The Dawid-Skene model assumes:
\begin{itemize}
\item Prior: $p(y_i = k) = \pi_k$ for all items (can be relaxed).
\item Conditional independence: Given true label $y_i$, annotators produce labels independently.
\item Annotator confusion matrices: $\theta_m^{jk} = p(z_{im} = j \mid y_i = k)$ is the probability that annotator $m$ assigns label $j$ when the true label is $k$.
\end{itemize}

\paragraph{Inference via EM.}
The EM algorithm iterates:

\emph{E-step:} Compute posterior distribution over true labels:
\[
p(y_i = k \mid \{z_{im}\}_m, \{\theta_m\}_m, \pi) 
\propto \pi_k \prod_{m=1}^M \theta_m^{z_{im} k}.
\]

\emph{M-step:} Update parameters:
\[
\pi_k = \frac{1}{N} \sum_{i=1}^N p(y_i = k \mid \text{data}),
\]
\[
\theta_m^{jk} = \frac{\sum_{i : z_{im} = j} p(y_i = k \mid \text{data})}{\sum_{i} p(y_i = k \mid \text{data})}.
\]

\paragraph{When to use.}
Dawid-Skene is appropriate when you have multiple annotators (models) labeling the same items and suspect systematic annotator-specific biases. It outperforms majority voting when annotators have different error patterns.

\subsubsection*{GLAD: Generative Model of Labels, Abilities, and Difficulties}

\paragraph{Problem.}
Extend Dawid-Skene to account for both annotator ability and item difficulty.

\paragraph{Model.}
GLAD models the probability that annotator $m$ correctly labels item $i$:
\[
p(z_{im} = y_i) = \frac{1}{1 + \exp(-\alpha_m \beta_i)},
\]
where:
\begin{itemize}
\item $\alpha_m \in \mathbb{R}$ is annotator $m$'s ability (positive = high ability).
\item $\beta_i \in \mathbb{R}$ is item $i$'s difficulty (positive = easy to label).
\end{itemize}

\paragraph{Inference.}
GLAD uses EM or variational inference to jointly estimate $\{\alpha_m\}_m$, $\{\beta_i\}_i$, and latent true labels $\{y_i\}_i$.

\paragraph{Comparison with Dawid-Skene.}
\begin{itemize}
\item Dawid-Skene: Annotator confusion matrices ($K \times K$ parameters per annotator); no item-specific difficulty.
\item GLAD: Scalar ability per annotator + scalar difficulty per item; assumes symmetric errors.
\end{itemize}

GLAD is simpler (fewer parameters) and explicitly models item difficulty, making it suitable when some items are inherently harder to annotate than others. Dawid-Skene is more flexible for capturing asymmetric confusion patterns.

\paragraph{Implementation.}
Both models are implemented in packages such as \texttt{crowdkit} (Python) and can be applied to LLM annotation data by treating each model as an "annotator."

\paragraph{When to use in LLM annotation.}
Use Dawid-Skene or GLAD when:
\begin{itemize}
\item Cross-model agreement is low ($\kappa < 0.4$).
\item You suspect systematic model-specific biases (e.g., one model consistently overestimates scores).
\item Items vary in difficulty (some are near decision boundaries).
\end{itemize}

These methods aggregate annotations more intelligently than simple majority voting by weighting models according to their estimated reliability.

\section*{Appendix E: Extensions, Examples, and Applications}
\label{app:E}
\label{app:extensions}

This appendix provides practical demonstrations of the variance-aware protocol applied to real-world organizational and educational settings, including strategic clarity annotation, entrepreneurial pitch evaluation, grant screening, and MBA case analysis.

\subsection*{Alternative Agreement and Calibration Models}

Beyond the baseline majority vote and simple calibration methods discussed in Section 5, several advanced models can improve aggregation and uncertainty quantification when LLM "raters" exhibit heterogeneous reliability.

\subsubsection*{Dawid--Skene Model}

The Dawid--Skene (DS) model \parencite{Dawid1979ObserverErrorEM} treats each rater (model, prompt variant) as having a fixed confusion matrix $\pi^{(m)}$ describing the probability of assigning label $j$ when the true label is $i$. Given $N$ items, $M$ raters, and observed labels $\{y_i^{(m)}\}$, the model estimates:
\begin{itemize}[topsep=2pt,itemsep=1pt]
\item Item difficulties: $p(z_i)$ for latent true label $z_i$.
\item Rater confusion matrices: $\pi_{ij}^{(m)} = p(y^{(m)} = j \mid z = i)$.
\end{itemize}

Estimation via Expectation-Maximization (EM):
\begin{enumerate}[topsep=2pt,itemsep=1pt]
\item \textbf{E-step}: Compute posterior $p(z_i \mid \{y_i^{(m)}\}, \{\pi^{(m)}\})$.
\item \textbf{M-step}: Update $\pi^{(m)}$ and $p(z_i)$ to maximize log-likelihood.
\end{enumerate}

Iterate until convergence. Final output: posterior distribution over $z_i$ for each item, and reliability matrix for each rater.

\paragraph{When to use DS.}
When inter-rater agreement is modest ($\kappa < 0.6$) and you have $M \geq 3$ raters with varying reliabilities. DS learns which models/prompts are more reliable and upweights them.

\subsubsection*{GLAD (Generative model of Labels, Abilities, and Difficulties)}

GLAD \parencite{Whitehill2009WhoseVoteCounts} extends DS by modeling item difficulty explicitly. Each item $i$ has difficulty $\beta_i$ and each rater $m$ has ability $\alpha_m$. The probability rater $m$ labels item $i$ correctly is:
\[
p(\text{correct} \mid \alpha_m, \beta_i) = \frac{1}{1 + \exp(-\alpha_m \beta_i)}.
\]

Estimation again via EM, producing:
\begin{itemize}[topsep=2pt,itemsep=1pt]
\item Item difficulties $\{\beta_i\}$.
\item Rater abilities $\{\alpha_m\}$.
\item Posterior over true labels $\{z_i\}$.
\end{itemize}

\paragraph{When to use GLAD.}
When items vary substantially in difficulty (e.g., some business models clearly superior, others borderline) and you want to identify which items are driving disagreement.

\subsubsection*{Isotonic Regression for Calibration}

Post-hoc calibration via isotonic regression fits a monotonic mapping from predicted probabilities to empirical frequencies on a held-out set. Given pairs $(p_i, y_i)$ where $p_i$ is model probability and $y_i \in \{0,1\}$ is outcome:
\begin{enumerate}[topsep=2pt,itemsep=1pt]
\item Sort by $p_i$.
\item Fit isotonic (monotonically increasing) function $f: [0,1] \rightarrow [0,1]$ minimizing squared error.
\item For new predictions $p$, use $f(p)$ as calibrated probability.
\end{enumerate}

Isotonic regression is more flexible than temperature scaling (which uses single parameter $T$) and often achieves better calibration on complex distributions.

\subsection*{End-to-End Case Examples}

\subsubsection*{Example 1: Evaluating Strategic Clarity in Earnings Calls}

\paragraph{Construct.} Strategic clarity: the extent to which executive communications articulate clear, actionable strategic priorities.

\paragraph{Level.} L2 (multi-label): Four criteria---Goal Specificity, Resource Commitment, Competitive Positioning, Temporal Framing---each scored 1--5 (ordinal), treated as independent.

\paragraph{Rubric.} BARS anchors for each criterion:
\begin{itemize}[topsep=2pt,itemsep=1pt]
\item \emph{Goal Specificity 5}: "States measurable targets with deadlines."
\item \emph{Goal Specificity 1}: "Generic aspirations with no metrics."
\item (Similar anchors for other criteria.)
\end{itemize}

\paragraph{Items.} 200 earnings call transcripts, ASR-validated, redacted for company names.

\paragraph{Protocol.}
\begin{itemize}[topsep=2pt,itemsep=1pt]
\item $P=4$ paraphrases, $S=10$ samples/prompt, $M=2$ families (OpenAI gpt-4o, Anthropic claude-sonnet-4).
\item JSON schema with four ordinal slots; reject-on-fail with max 3 retries.
\item Randomize criterion order across samples.
\item Aggregate: median per criterion within $(u,m)$; then across $u$; then across $m$.
\end{itemize}

\paragraph{Results.}
\begin{itemize}[topsep=2pt,itemsep=1pt]
\item Weighted $\kappa$ (per criterion) ranges 0.52--0.68 across models.
\item Audit set re-run after 3 months: $|\Delta\kappa| = 0.03$ (within tolerance).
\item Human calibration slice ($n{=}30$): ICC(2,1) = 0.61 (moderate agreement with human consensus).
\end{itemize}

\paragraph{Reporting.} Methods table documents rubric, schema, sampling plan, agreement metrics with bootstrap CIs, audit set results. Materials bundle includes prompts, schema JSON, de-identified sample of 10 transcripts.

\subsubsection*{Example 2: Ranking Entrepreneurial Pitches on Novelty}

\paragraph{Construct.} Novelty: degree to which a venture's value proposition differs from existing offerings.

\paragraph{Level.} L1 (pairwise): For each pair of pitches, select more novel one (A or B).

\paragraph{Rubric.} BARS with 5 novelty levels; pairwise prompt asks "Which pitch is more novel?" with randomized left--right order.

\paragraph{Items.} 50 pitch blurbs (2-paragraph descriptions); $\binom{50}{2} = 1225$ pairs (subsampled to 200 for feasibility).

\paragraph{Protocol.}
\begin{itemize}[topsep=2pt,itemsep=1pt]
\item $P=3$ paraphrases, $S=20$ samples/prompt, $M=2$ families.
\item Constrained to single token A or B; \texttt{max\_tokens=1}.
\item Randomize left--right order; record permutations.
\item Aggregate pairwise wins; fit Bradley--Terry model for global ranking.
\end{itemize}

\paragraph{Results.}
\begin{itemize}[topsep=2pt,itemsep=1pt]
\item Cohen's $\kappa$ across models = 0.48 (moderate).
\item A/B swap test: 12\% flip rate (items near decision boundary).
\item Bradley--Terry ranking: top 10 pitches consistent across models; middle 30 show substantial rank variance.
\end{itemize}

\paragraph{Reporting.} Methods table documents pairwise design, sampling, order randomization, swap test results, and Bradley--Terry estimates with CIs.

\subsection*{Organizational Practice: Using LLM-as-Annotator in Live Settings}

Beyond research, organizations increasingly use LLM-based annotation for operational decisions (e.g., screening grant applications, triaging customer feedback, scoring innovation proposals). The protocol adapts for live settings:

\subsubsection*{Differences from Research Protocol}

\begin{itemize}[leftmargin=*,topsep=3pt,itemsep=2pt]
\item \textbf{Throughput priority}: Organizations may tolerate lower $P,S,M$ for speed. Minimum viable: $P{=}2$, $S{=}5$, $M{=}1$ with mandatory human review for all high-stakes items.
\item \textbf{Continuous drift monitoring}: Unlike research (one-time data collection), live systems require scheduled audit set runs (e.g., weekly) with automated alerts on $|\Delta\kappa| > 0.05$.
\item \textbf{Feedback loop}: Human overturns feed back into calibration. Track overturn patterns by criterion/model; retrain or reweight models if systematic biases emerge.
\item \textbf{Transparency to stakeholders}: Applicants/customers may request explanations. L3 rationales (non-authoritative) can be shared; primary decision based on constrained L1/L2 output.
\end{itemize}

\subsubsection*{Example: Grant Screening at a Foundation}

A foundation receives 5,000 grant applications annually. Human reviewers can deeply assess 500; the rest screened by LLM.

\paragraph{Protocol.}
\begin{itemize}[topsep=2pt,itemsep=1pt]
\item \textbf{Level}: L2 (six criteria: Alignment, Feasibility, Impact, Novelty, Team, Budget).
\item \textbf{Sampling}: $P{=}2$, $S{=}5$, $M{=}1$ (cost constraint).
\item \textbf{Triage}: Top 500 by aggregate score $\rightarrow$ full human review. Bottom 4000 by score $\rightarrow$ human spot-check 5\%. Middle 500 $\rightarrow$ dual human review on 100 (20\%) to calibrate.
\item \textbf{Drift monitoring}: Weekly audit set ($n{=}50$ historical applications with consensus labels); alert if $|\Delta\kappa| > 0.05$.
\item \textbf{Feedback}: Human overturns logged by criterion; quarterly review identifies if LLM systematically underweights certain criteria; adjust rubric or weights accordingly.
\end{itemize}

\paragraph{Results (Year 1).}
\begin{itemize}[topsep=2pt,itemsep=1pt]
\item LLM--human agreement on calibration set: ICC = 0.54 (moderate).
\item Overturn rate on spot-checks: 8\% (mostly borderline cases).
\item Drift alerts: 2 in 12 months; both resolved by rubric clarification (no model change needed).
\item Efficiency gain: Reduced human screening workload by 80\%; redirected effort to in-depth review of finalists.
\end{itemize}

\subsection*{Teaching and Pedagogy: LLM-as-Annotator in the Classroom}

Strategy instructors increasingly use LLMs to provide feedback on student work (e.g., case analyses, strategy memos). The protocol adapts for educational settings:

\subsubsection*{Key Considerations}

\begin{itemize}[leftmargin=*,topsep=3pt,itemsep=2pt]
\item \textbf{Formative vs.\ summative}: For formative feedback (low stakes), single-model L2 with detailed L3 rationales suffices. For summative assessment (grades), require human review and multi-model triangulation.
\item \textbf{Learning objectives}: Rubric should map to learning objectives; BARS anchors communicate expectations to students.
\item \textbf{Transparency}: Share rubric and sample LLM evaluations with students; demystify AI feedback.
\item \textbf{Bias awareness}: Monitor for systematic biases (e.g., length/verbosity effects favoring certain writing styles); adjust rubric or use human adjudication for equity.
\end{itemize}

\subsubsection*{Example: MBA Strategy Case Analysis}

Instructor assigns Porter's Five Forces analysis of a case. 120 students submit 3-page memos.

\paragraph{Protocol (Formative Feedback).}
\begin{itemize}[topsep=2pt,itemsep=1pt]
\item \textbf{Level}: L2 (five criteria: Completeness, Depth, Clarity, Evidence Use, Recommendations).
\item \textbf{Sampling}: $P{=}1$ (single prompt), $S{=}3$, $M{=}1$ (time constraint).
\item \textbf{Output}: Median score per criterion + brief L3 rationale (non-authoritative) for each.
\item \textbf{Human review}: Instructor spot-checks 10\% (12 memos) for calibration; adjusts rubric if LLM systematically misaligns.
\end{itemize}

\paragraph{Student reception.}
\begin{itemize}[topsep=2pt,itemsep=1pt]
\item Positive: Immediate feedback (within 1 hour of submission).
\item Concerns: Some students felt length/polish rewarded over substance. Instructor added explicit rubric anchor: "Brevity with depth valued over verbosity."
\end{itemize}

\paragraph{Instructor insights.}
\begin{itemize}[topsep=2pt,itemsep=1pt]
\item LLM feedback freed instructor time for office hours and in-depth discussions.
\item Rubric refinement improved over iterations; by third case, LLM--instructor agreement reached ICC = 0.68.
\end{itemize}

\subsection*{Summary}

This appendix demonstrates the protocol's flexibility across contexts:
\begin{itemize}[leftmargin=*,topsep=3pt,itemsep=2pt]
\item \textbf{Research}: Full protocol with high $P,S,M$, preregistration, and transparency for publication.
\item \textbf{Organizational practice}: Streamlined protocol with continuous drift monitoring and feedback loops.
\item \textbf{Education}: Formative applications with lower rigor; summative applications require human review.
\end{itemize}

The core principles---construct clarity, constrained outputs, replication, agreement metrics, drift monitoring---apply universally. Researchers, practitioners, and educators should "rightsize" the protocol to match stakes and resources while maintaining transparency and accountability.

\end{document}